\documentclass[aps,prd,preprint,floatfix,12pt,showpacs,showkeys,tightenlines,nofootinbib,longbibliography,a4paper]{revtex4-1}
\usepackage{soul}
\usepackage[table]{xcolor}
\usepackage{amsfonts}
\usepackage{amsmath}
\usepackage{array}
\usepackage[markup=nocolor]{changes}
\usepackage{graphicx}
\usepackage{hyperref}
\usepackage{multirow}
\usepackage{diagbox} 
\usepackage{commath}
\usepackage{url}
\usepackage{xspace}
\usepackage{setspace}
\usepackage{graphics}
\usepackage[parfill]{parskip}
\usepackage{xcolor}
\usepackage{tikz}
\usepackage{amssymb}
\usepackage{pifont}
\usepackage{pgfplots}
\newcommand{\xmark}{\ding{55}}%
\hypersetup{pdfstartview=FitV,colorlinks=true,linkcolor=blue,citecolor=blue, filecolor=black,urlcolor=blue,linktoc=page}
\definecolor{color1}{HTML}{cc9966}
\definecolor{color2}{HTML}{ff0000}
\definecolor{color3}{HTML}{dddd00}
\definecolor{color4}{HTML}{33ff00}
\definecolor{color5}{HTML}{ff00dd}
\definecolor{color6}{HTML}{0044dd}
\definecolor{color7}{HTML}{8822cc}
\definecolor{color8}{HTML}{00ffff}
\definecolor{color9}{HTML}{aa5511}
\definecolor{color10}{HTML}{9999cc}
\definecolor{color11}{HTML}{225522}
\definecolor{color12}{HTML}{ffa322}
\definecolor{sigstrcolor1}{HTML}{ee0000}
\definecolor{sigstrcolor2}{HTML}{00aaaa}
\definecolor{sigstrcolor3}{HTML}{550099}
\definecolor{sigstrcolor4}{HTML}{2255ff}
\definecolor{sigstrcolor5}{HTML}{22cc00}
\definecolor{sigstrcolor6}{HTML}{ff9900}
\definecolor{isocolor0}{HTML}{4fcf36}
\definecolor{isocolor12m}{HTML}{75cf36}
\definecolor{isocolor12}{HTML}{75cf36}
\definecolor{isocolor1}{HTML}{b4c400}
\definecolor{isocolor1m}{HTML}{b4c400}
\definecolor{isocolor32m}{HTML}{dddd00}
\definecolor{isocolor32}{HTML}{dddd00}
\definecolor{isocolor2}{HTML}{ddb800}
\definecolor{isocolor2m}{HTML}{ddb800}
\newcommand{\ld}{\lambda}
\newcommand{\be}{\begin{equation}}
\newcommand{\ee}{\end{equation}}
\newcommand{\bea}{\begin{eqnarray}}
\newcommand{\eea}{\end{eqnarray}}
\newcommand{\eqn}[2]{\begin{equation} \label{eq:#1} #2 \end{equation}}
\newcommand\Tstrut{\rule{0pt}{2.6ex}}      

\newcommand{\tr}{\mathrm{Tr}}
\definecolor{lime}{HTML}{A6CE39}
\DeclareRobustCommand{\orcidicon}{\hspace{-1mm}
	\begin{tikzpicture}
	\draw[lime, fill=lime] (0,0) 
	circle [radius=0.16] 
	node[white] at (-0.007,-0.007) {{\fontfamily{qag}\selectfont \tiny \,ID}};
	\draw[white, fill=white] (-0.067,0.095) 
	circle [radius=0.005];
	\end{tikzpicture}
	\hspace{-3mm}
}

\foreach \x in {A, ..., Z}{\expandafter\xdef\csname orcid\x\endcsname{\noexpand\href{https://orcid.org/\csname orcidauthor\x\endcsname}
			{\noexpand\orcidicon}}
}
\usepackage{calc}
\newlength{\depthofsumsign}
\newcommand{\nsum}[1][1.4]{
    \mathop{%
        \raisebox
            {-#1\depthofsumsign+1\depthofsumsign}
            {\scalebox
                {#1}
                {$\displaystyle\sum$}%
            }
    }
}
\makeatletter
\newcommand*{\DivideLengths}[2]{%
  \strip@pt\dimexpr\number\numexpr\number\dimexpr#1\relax*65536/\number\dimexpr#2\relax\relax sp\relax
}
\makeatother

\begin{document}
\title{Next-to-Leading Order Unitarity Fits in the  Extended Georgi-Machacek Model}
\author{Debtosh Chowdhury\orcidA{}}
\email{\textcolor{black}{E-mail:} debtoshc@iitk.ac.in}
\author{Poulami Mondal\orcidB{}}
\email{\textcolor{black}{E-mail:} poulami.mondal@tifr.res.in}
\author{Subrata Samanta\orcidC{}}
\altaffiliation{Corresponding author}
\email{\textcolor{black}{e-mail:} samantaphys@gmail.com}
\affiliation{Department of Physics, Indian Institute of Technology Kanpur, Kanpur 208016, India}

\begin{abstract}
\noindent
We compute one-loop corrections to all $2\to2$ bosonic scattering amplitudes in the generalized two-triplet scalar extension of the Standard Model and place next-to-leading order unitarity bounds on the scalar quartic couplings of the Georgi-Machacek (GM) and the extended Georgi-Machacek (eGM) models. Further, we derive the bounded-from-below (BFB) conditions on the scalar quartic couplings demanding the stability of the scalar potential in the field subspaces. We find that, in the GM and eGM models, the BFB conditions with all combinations of three non-zero scalar fields provide a very good approximation of the all field BFB conditions while being computationally more efficient. With these improved theoretical constraints, we present results for the GM and eGM models from global fits to the latest Higgs signal strength measurements at the $13$ TeV Large Hadron Collider. We observe that the global fit disfavors the regions where $\kappa_V > 1.05$, $\kappa_V < 0.95$, and $\kappa_f > 1.05$, $\kappa_f< 0.92$ at a $95.4\%$ probability for both models. We obtain an upper limit on the absolute values of the scalar quartic couplings to be $1.91\:(3.0)$ in the GM (eGM) model. We find that the absolute mass differences between the heavy Higgs bosons are less than $410$ GeV and $520$ GeV in the GM and eGM models, respectively, if their individual masses are restricted to be below $1.1$ TeV.
\end{abstract}
\date{\today}
\maketitle
\begin{spacing}{0.5}
\tableofcontents
\end{spacing}
\section{Introduction}\label{sec:intro}
Discovery of the Higgs boson~\citep{ATLAS:2012yve,CMS:2012qbp} has already played a significant role in understanding the electroweak symmetry breaking (EWSB) mechanism and its connection to the generation of the masses of elementary particles in the Standard Model (SM). However, the origin of EWSB has not yet been settled. Results from the latest ATLAS and CMS Run 2 data allow for a deviation within $\sim 10\%$ in the Higgs signal strength values~\citep{ATLAS:2016neq,ATLAS:2020rej,ATLAS:2020qcv,ATLAS:2020fcp,ATLAS:2020fzp,CMS:2020xwi,ATLAS:2020bhl,CMS:2021ugl,CMS:2021kom,ATLAS:2021qou,ATLAS:2022yrq,CMS:2022ahq,CMS:2022kdi,CMS:2022uhn,CMS:2022dwd,ATLAS:2022vkf,ATLAS:2022ooq,ATLAS:2022tnm,CMS:2023tfj}. Thus the data does not preclude the existence of additional multiplets (doublet or higher) in the scalar sector of the SM. Such higher multiplets leave an imprint on the EWSB mechanism that can be detected via the SM-like Higgs couplings to the vector bosons. Among these higher multiplets, scalar $SU(2)_L$ triplet extensions of the SM offer various interesting phenomenological signatures at the Large Hadron Collider (LHC)~\citep{Georgi:1985nv,Chanowitz:1985ug,Gunion:1989ci,Gunion:1990dt,Chiang:2012cn,Englert:2013zpa,Hartling:2014zca,Garcia-Pepin:2014yfa,Hartling:2014aga,Chiang:2015amq,Chiang:2018cgb,Banerjee:2019gmr,deLima:2021llm,Mondal:2022xdy,Chen:2022zsh,Kundu:2022bpy,Chen:2023ins,Chen:2023bqr,Chakraborti:2023mya,deLima:2024uwc,Crivellin:2024uhc} and future colliders~\cite{10.21468/SciPostPhysProc.8.111,Bizon:2024juq,Stylianou:2023xit,Torndal:2023mmr,Celada:2024mcf,Aime:2022flm,MuonCollider:2022xlm,Forslund:2022xjq}. For example, unlike complex doublet models, complex triplet models predict both singly and doubly-charged scalars in their spectrum. These doubly-charged scalars couple to massive vector boson pairs at tree-level and open up the possibility that the SM-like Higgs boson’s couplings to $WW$ and $ZZ$ could be larger than those in the SM~\citep{deLima:2021llm,deLima:2024uwc}. The vacuum expectation values (VEVs) of these triplets are constrained from the SM-like Higgs signal strength data and, therefore, affect the mass spectra of heavier Higgs bosons in the model. Additionally, the presence of doubly-charged Higgs bosons in triplet models leads to phenomenologically interesting collider signatures. For example, this doubly-charged Higgs boson can decay into a pair of same-sign $W$ bosons and hence provides further constraints on triplet VEV~\citep{CMS:2017fhs,CMS:2020etf,ATLAS:2018ceg,ATLAS:2023sua}. Furthermore, $B$-physics and electroweak precision observables (EWPO) put indirect constraints~\citep{Ciuchini:2013pca,Hartling:2014aga,deBlas:2016ojx,deBlas:2022hdk} on the extended scalar sector of the beyond SM (BSM). Among the EWPO, the $\rho$ parameter, defined as, $\rho=M_W^2/(M_Z^2\cos^2\theta_W)$,
where $\theta_W$ is the weak-mixing angle, plays a significant role in constraining the structure and parameters in BSM models. The global fit analyses~\citep{ParticleDataGroup:2022pth} suggest that, $\rho=1.00038 \pm 0.00020.$

Minimal triplet scalar extension of the SM needs one complex triplet scalar and one real triplet scalar under $SU(2)_L$ with hypercharge, $Y=1$ and $Y=0$, respectively, such that the $\rho$ parameter at the tree-level remains unity, owing to the consequence of an approximate global $SU(2)_V$ symmetry called custodial symmetry (CS). One such extension is the Georgi-Machacek (GM) model~\citep{Georgi:1985nv,Chanowitz:1985ug}, which preserves the CS at the tree-level, as in the SM, due to the explicit global $SU(2)_L\otimes SU(2)_R$ symmetry on the scalar potential. While the CS at the tree-level is motivated by the stringent constraints from precision electroweak and Higgs observables, the requirement of global $SU(2)_L\otimes SU(2)_R$ symmetry is not necessary. Also, the global $SU(2)_L\otimes SU(2)_R$ symmetry needs to be broken to renormalize the GM model~\cite{Gunion:1990dt,Englert:2013zpa,Hartling:2014aga,Blasi:2017xmc,Chiang:2018xpl,Keeshan:2018ypw}. Recently, Ref.~\cite{Kundu:2021pcg} introduced a triplet extended scalar sector of the SM without requiring the global $SU(2)_L \otimes SU(2)_R$ symmetry in the potential while the CS remains intact at the tree-level, dubbed as the extended GM (eGM) model. In the eGM model, mixing among the members of different custodial multiplets lifts the tree-level mass degeneracy. These mass splittings open up various new decay modes that can be probed at the LHC and other future colliders.

Prior to the SM Higgs discovery, perturbative unitarity bounds were used to place an upper limit on the SM Higgs boson mass~\citep{Dicus:1973gbw,Lee:1977yc, Lee:1977eg,PASSARINO1985231,Luscher:1988gc, PhysRevLett.62.1232, Marciano:1989ns, PhysRevD.40.2880,PASSARINO199031,PhysRevLett.64.1215,Lendvai:1991gd,PhysRevD.45.3112,Maher:1993vj,Durand:1993vn}. Dicus and Mathur~\cite{Dicus:1973gbw}, and  Lee, Quigg, and Thacker~\citep{Lee:1977yc, Lee:1977eg} established an upper limit of SM quartic coupling, $\lambda\leq8\pi/3,$ demanding the unitarity of $2\to2$ partial-wave scattering amplitudes. A number of authors extended this work to one-loop level and derived an upper bound on the running coupling, $\lambda\lesssim 2-2.5$, in a weakly interacting SM Higgs scenario~\citep{PASSARINO1985231,PhysRevLett.62.1232, PhysRevD.40.2880,PASSARINO199031,PhysRevLett.64.1215,Lendvai:1991gd,PhysRevD.45.3112}. A similar methodology was used to test the reliability of the perturbative calculations at the two-loop level and showed that an increase of one order in perturbation theory does not revise the above-stated limit on the running coupling $\lambda$~\citep{Maher:1993vj,Durand:1993vn}. Therefore, perturbative unitarity delineates the region in which perturbative calculations can be trusted, as performed in most phenomenological studies. Perturbative unitarity constraints have been used to bound large quartic couplings as well as constrain exotic Higgs masses in the two-Higgs doublet model (2HDM) with a softly broken $Z_2$ symmetry~\citep{Grinstein:2015rtl,Cacchio:2016qyh,Chowdhury:2017aav}. Clearly, it is imperative to study the constraints arising from the perturbative unitarity of the $2\to2$ partial-wave scattering amplitudes in a weakly interacting Higgs sector of GM and eGM models.

In this paper, we compute, for the first time, one-loop corrections to all the $2\to 2$ Higgs boson and longitudinal vector boson scattering amplitudes in the generalized two-triplet scalar extension of the SM and recast these results for the GM and eGM models. Specifically, we consider $\phi^4$-like interactions which are enhanced, $\mathcal{O}(\lambda_i\lambda_j/16\pi^2)$, in the limit $s\gg |\lambda_i|v^2\gg M_W^2$, $s\gg |\mu_i|v$. Our results improve the tree-level perturbative unitarity bounds on  quartic couplings, previously presented in the literature~\citep{Aoki:2007ah,Hartling:2014zca,Chen:2023ins}. We further derive the 3-field bounded-from-below (BFB) conditions on the scalar quartic couplings. We find that the 3-field BFB conditions are a very good approximation of the full 13-field BFB conditions~\cite{Hartling:2014zca,Moultaka2020} in both models while being computationally more efficient. We perform global fits to the latest Run 2 LHC data on Higgs signal strengths, with improved theoretical constraints, including next-to-leading order (NLO) unitarity and 3-field BFB conditions. Finally, we investigate the impact of these improved theoretical bounds on the global fit in the GM and eGM models.

 Our analysis improves the theoretical bounds in the study of Ref.~\cite{Chiang:2018cgb,Chen:2022zsh} for the GM model and includes the latest Run 2 LHC data on the Higgs signal strengths. We show that a large part of the parameter space in both the GM and eGM models is ruled out by NLO unitarity bounds, which were otherwise viable from the tree-level unitarity conditions. In addition, the inclusion of the latest Run 2 LHC data significantly refines the upper limit of the triplet VEV. Among the EWPO, the most dominant constraint comes from the Peskin-Takauchi $T$ parameter (or the $\rho$ parameter). However, as discussed in~\cite{Gunion:1990dt,Drauksas:2023isd,Chowdhury:2026rus}, we need four input parameters to renormalize GM and eGM models in contrast to the standard three input parameters ($\alpha_{\rm em}, M_Z,$ and $G_F$) that are needed to renormalize the SM. As a result, at the one-loop, $\rho$ parameter is no longer a model prediction in the custodial symmetric triplet extensions of the SM. Instead, it must be treated as an independent input in the renormalization procedure. The global electroweak fit value of the $\rho$ parameter~\citep{deBlas:2016ojx,ParticleDataGroup:2022pth} determines the finite part of the custodial symmetry breaking parameter, which is absent at the tree-level. Hence, the value of the $\rho$ parameter or the Peskin-Takeuchi $T$ parameter does not provide a useful constraint on such triplet extensions of the SM.

The rest of this paper is organized as follows: The model is defined in Section \ref{sec:model}. Bounded-from-below conditions and NLO unitarity constraints are discussed in Section \ref{sec:bfb} and Section \ref{sec:uni}, respectively. We explain our global fit set-up in Section \ref{sec:hepfit} and list all relevant constraints in Section \ref{sec:fit_constraints}. The results from the global fits are presented in Section \ref{sec:results}. We conclude in Section \ref{sec:conclude}. Explicit expressions of quartic couplings in terms of the physical Higgs masses are given in Appendix~\ref{app:inverse_mass}. Our results for the BFB conditions are provided in Appendices~\ref{app:pos_subspace} and \ref{app:pos_EGM}, respectively, while Appendix~\ref{app:1-loop_amp} contains the results for the one-loop scattering amplitudes. Finally, Appendix~\ref{app:2-loop_rge} includes the one-loop and two-loop renormalization group equations (RGEs), and a number of supplementary figures of our analysis are in Appendix~\ref{app:supp_figs}.
\section{Model}\label{sec:model}
We have extended the scalar sector of the SM, augmented by a real triplet $\xi$ with hypercharge $Y=0$, and a complex triplet $\chi$ with $Y=1$. With this field content, the most general $SU(2)_{L}\otimes U(1)_{Y}$ invariant scalar potential reads~\cite{Kundu:2021pcg},
\begin{align}
V&=-m_\phi^2\big(\phi^\dagger\phi\big)-m_\xi^2\big(\xi^\dagger\xi\big)-m_\chi^2\big(\chi^\dagger\chi\big)+\mu_1\big(\chi^\dagger t_a\chi\big)\xi_a+\mu_2\big(\phi^\dagger \tau_a\phi\big)\xi_a\nonumber\\
&\,\quad+\mu_3\Big[\big(\phi^T\epsilon\tau_a\phi\big)\tilde{\chi}_a+\text{h.c.}\Big]
+\lambda_\phi\big(\phi^\dagger\phi\big)^2+\lambda_\xi\big(\xi^\dagger\xi\big)^2+\lambda_\chi\big(\chi^\dagger\chi\big)^2\nonumber\\
&\,\quad+\tilde{\lambda}_\chi \big|\tilde{\chi}^\dagger\chi \big|^2+\lambda_{\phi\xi}\big(\phi^\dagger\phi\big)\big(\xi^\dagger\xi\big)
+\lambda_{\phi\chi}\big(\phi^\dagger\phi\big)\big(\chi^\dagger\chi\big)+\lambda_{\chi\xi}\big(\chi^\dagger\chi\big)\big(\xi^\dagger\xi\big)\nonumber\\
&\,\quad+\kappa_1\big|\xi^\dagger\chi \big|^2+\kappa_2\big(\phi^\dagger\tau_a\phi\big)\big(\chi^\dagger t_a\chi\big)+\kappa_3\Big[\big(\phi^T\epsilon\tau_a\phi\big)\big(\chi^\dagger t_a\xi\big)+\text{h.c.}\Big]\,,
\label{v16}
\end{align} 
where $\phi=(\phi^+ \;\phi^0)^T$ is the SM Higgs doublet with $ Y=1/2$, $\xi=(\xi^+ \;\xi^0\;-\xi^{+*})^T$, and $\chi=(\chi^{++} \;\chi^+ \;\chi^0)^T$. The charge conjugate of the complex triplet is defined as $ \tilde{\chi}=(\chi^{0*} \;-\chi^{+*} \;\chi^{++*})^T$. 
Note that $\tau_a$ and $t_a$ are the $2$-dimensional and $3$-dimensional representations of the $SU(2)$ generators, respectively, written in the spherical basis and, in general, are not hermitian. All model parameters are taken to be real to avoid explicit CP violation. We refer to Eq.~(\ref{v16}) as the scalar potential of the `generalized two-triplet model' in our paper. Following Ref.~\cite{Kundu:2021pcg}, we next discuss the conditions that give rise to $\rho = 1$ and the tree-level mass spectra.

After the EWSB, we redefine the neutral components of the fields as, 
\begin{equation}\label{eq:vev}
\phi^0=\frac{v_\phi}{\sqrt{2}}+\frac{1}{\sqrt{2}}\left(\phi'+i\phi''\right)\,,\quad \chi^0=v_\chi+\frac{1}{\sqrt{2}}\left(\chi'+i\chi''\right)\,,\quad \xi^0=v_\xi+\xi'\,,
\end{equation}
where $v_\phi$ and $v_\xi$ ($v_\chi$) are the VEVs of the SM Higgs doublet and real (complex) triplet, respectively. All VEVs are taken to be real in order to avoid spontaneous CP violation. Given the VEV structure in Eq.~(\ref{eq:vev}), the $\rho$ parameter at tree-level is
\begin{equation}
\nonumber
\rho=\frac{v_\phi^2+4\big(v_\xi^2+v_\chi^2\big)}{v_\phi^2+8 v_\chi^2}\,,\quad \text{with} \quad  v_\phi^2+4( v_\chi^2+v_\xi^2) = v^2 \approx (246 \, \text{GeV})^2\,. 
\end{equation}
We ensure $\rho=1$ remains intact at the tree-level by putting $v_{\chi}=v_{\xi}$ and imposing the following condition~\cite{Kundu:2021pcg},
\begin{equation}
\frac{\partial V}{\partial \chi}\Bigg|_{\substack{\phi^0=v_\phi, \chi^0=v_\chi,\\ \xi^0=v_\xi, v_\chi=v_\xi}} = \frac{\partial V}{\partial \xi}\Bigg|_{\substack{\phi^0=v_\phi, \chi^0=v_\chi,\\ \xi^0=v_\xi, v_\chi=v_\xi}}\,,
\label{eq:align}
\end{equation}
where the latter condition is imposed at the minimum, which ensures that the SM-like Higgs boson ($h$) is the custodial singlet state, i.e., $g_{hWW}=g_{hZZ}\times\cos^2\theta_W$.
As a consequence, four constraints are being imposed on the model parameters~\cite{Kundu:2021pcg},
\begin{equation}
m_\chi^2=2m_\xi^2\,,\quad \mu_2=\sqrt{2}\mu_3\,,\quad \lambda_{\chi\xi}=2\lambda_\chi-4\lambda_\xi\,,\quad \kappa_2=4\lambda_{\phi\xi}-2\lambda_{\phi\chi}+\sqrt{2}\kappa_3\,.
\label{eq:constrints1}
\end{equation}
These are the necessary conditions to prevent any mixing between the custodial singlet and the non-custodial singlet states in the neutral sector, while the singly charged scalars can mix with each other.
Following~\citep{Kundu:2021pcg}, we refer to this as the extended Georgi-Machacek (eGM) model. In the eGM model, the Goldstone bosons  ($G^+,G^0$) show up in the longitudinal mode of massive $W^+$ and $Z^0$ gauge bosons, and the following mass eigenstates emerge: three CP-even eigenstates ($F^0,H,h$), one CP-odd eigenstate ($A$), two singly-charged scalars ($F^{+},H^+$), and one doubly-charged scalar ($F^{++}$). Among the three CP-even eigenstates, $h$ and $H$ are the custodial singlets. The mixings among the states in the eGM model are given below,
\begingroup
\small{
\begin{equation}
\nonumber
F^{++}=\chi^{++},\quad 
\begin{bmatrix}
G^0 \\
 A\\
\end{bmatrix}
= \begin{bmatrix}
c_\beta & s_\beta \\
 -s_\beta & c_\beta\\
\end{bmatrix} 
\begin{bmatrix}
\phi'' \\
 \chi''\\
\end{bmatrix},\quad 
\begin{bmatrix}
h \\
 H\\
 F^0\\
\end{bmatrix}
=R_\alpha^{-1} R_0 
\begin{bmatrix}
\phi'\\
 \xi'\\
 \chi'\\
\end{bmatrix},\quad 
\begin{bmatrix}
G^{+} \\
 H^+\\
 F^+\\
\end{bmatrix}
=R_\delta^{-1} R_\beta R_{+} 
\begin{bmatrix}
\phi^+\\
 \xi^+\\
 \chi^+\\
\end{bmatrix}\,.
\end{equation}}
\endgroup\normalsize
The form of the rotation matrices $R_i\,(i=\alpha,\beta,\delta,0,+)$ are as follows,
\begingroup
\small{
\begin{equation}
\nonumber
R_{\alpha (\beta)}
=\begin{bmatrix}
c_{\alpha  (\beta)} & s_{\alpha  (\beta)} & 0\\
 -s_{\alpha (\beta)} & c_{\alpha (\beta)} & 0\\
 0& 0&1\\
\end{bmatrix}, \;\;
R_\delta
=\begin{bmatrix}
 1& 0&0\\
0& c_\delta & s_\delta  \\
 0& -s_\delta & c_\delta \\
\end{bmatrix}, \;\;
R_0
=\begin{bmatrix}
 1& 0&0\\
0& \frac{1}{\sqrt{3}} & \sqrt{\frac{2}{3}}  \\
 0&  -\sqrt{\frac{2}{3}} & \frac{1}{\sqrt{3}} \\
\end{bmatrix}, \;\;
R_+
=\begin{bmatrix}
 1& 0&0\\
0& \frac{1}{\sqrt{2}} & \frac{1}{\sqrt{2}}   \\
 0& -\frac{1}{\sqrt{2}}  & \frac{1}{\sqrt{2}}  \\
\end{bmatrix}\,,
\end{equation}}
\endgroup\normalsize
with  $\tan\beta = 2 \sqrt{2} v_\chi/v_\phi$\,, where $c_\theta$ and $s_\theta$ stand for $\cos\theta$ and $\sin\theta$, respectively. The mass-squared eigenvalues of the CP-even sector can be written as,
\begin{align}
\nonumber
m_{h}^2=&\left(\mathcal{M}_n^2\right)_{11} c_\alpha^2 + \left(\mathcal{M}_n^2\right)_{22}  s_\alpha^2 -2\left(\mathcal{M}_n^2\right)_{12} s_\alpha c_\alpha\,,\\
\nonumber
m_{H}^2=&\left(\mathcal{M}_n^2\right)_{11} s_\alpha^2 + \left(\mathcal{M}_n^2\right)_{22}  c_\alpha^2 +2\left(\mathcal{M}_n^2\right)_{12} s_\alpha c_\alpha\,,
\\
m_{F^0}^2=&\frac{1}{2}v^2s_\beta^2\left(4\lambda_\xi-\lambda_\chi\right)-\frac{3}{2\sqrt{2}}v^2c_\beta^2\kappa_3+\frac{1}{\sqrt{2}}c_\beta^2s_\beta^{-1}\mu_2v+\frac{1}{\sqrt{2}}s_\beta\mu_1v\,,
\label{mass_cp_even}
\end{align}
where 
\begin{equation}
\tan 2\alpha=\frac{2\left(\mathcal{M}_n^2\right)_{12}}{\left(\mathcal{M}_n^2\right)_{22}-\left(\mathcal{M}_n^2\right)_{11}}\,, \quad \text{with} \quad \alpha\in\left[-\frac{\pi}{2}\,,\frac{\pi}{2}\right]\,.
\end{equation}
$\mathcal{M}_n^2$ is the mass-squared matrix in the basis $\left\{\phi', \tfrac{\xi'+\sqrt{2}\chi'}{\sqrt{3}}\right\}$\,, whose elements are given below,
\begin{align*}
\left(\mathcal{M}_n^2\right)_{11}&=2v^2c_\beta^2\lambda_\phi\,,\\
\left(\mathcal{M}_n^2\right)_{12}&=\frac{\sqrt{3}}{2}v^2s_{\beta}c_\beta\kappa_3+\sqrt{\frac{3}{2}}v^2s_{\beta}c_\beta\lambda_{\phi\xi}-\frac{\sqrt{3}}{2}c_\beta \mu_2 v\,,\\
\left(\mathcal{M}_n^2\right)_{22}&=v^2s_\beta^2\left(\lambda_\chi-\lambda_\xi\right)-\frac{1}{2\sqrt{2}}s_\beta\mu_1 v+\frac{1}{\sqrt{2}}c_\beta^2s_\beta^{-1}\mu_2v\,.
\end{align*}
The CP-odd neutral scalar $(A)$ and the doubly-charged scalar ($F^{++}$) mass-squared eigenvalues are as follows,
\begin{align}
\nonumber
m_A^2&=\frac{1}{\sqrt{2}}s_\beta^{-1}\mu_2v-\frac{1}{2\sqrt{2}}v^2\kappa_3 \,,\\
m_{F^{++}}^2&=-\frac{1}{2\sqrt{2}}v^2c_\beta^2\left(\sqrt{2}\kappa_2+\kappa_3\right)+\frac{1}{2}v^2s_\beta^2\tilde{\lambda}_\chi+\frac{1}{\sqrt{2}}c_\beta^2s_\beta^{-1}\mu_2v+\frac{1}{\sqrt{2}}s_\beta\mu_1v\,.
\label{mass_cp_odd_doubly_charged}
\end{align}
The singly-charged mass-squared eigenvalues are given by,
\begin{align}
\nonumber
m_{H^+}^2=&\left(\mathcal{M}_c^2\right)_{11} c_\delta^2 + \left(\mathcal{M}_c^2\right)_{22}  s_\delta^2 -2\left(\mathcal{M}_c^2\right)_{12} s_\delta c_\delta\,,\\
m_{F^+}^2=&\left(\mathcal{M}_c^2\right)_{11} s_\delta^2 + \left(\mathcal{M}_c^2\right)_{22}  c_\delta^2 +2\left(\mathcal{M}_c^2\right)_{12} s_\delta c_\delta\,,
\label{mass_singly_charged}
\end{align}
where
\bea
\tan 2\delta=\frac{2\left(\mathcal{M}_c^2\right)_{12}}{\left(\mathcal{M}_c^2\right)_{22}-\left(\mathcal{M}_c^2\right)_{11}}\,, \quad \text{with} \quad \delta \in\left[-\frac{\pi}{2},\frac{\pi}{2}\right]\,.
\label{mixing_angle_delta}
\eea
$\mathcal{M}_c^2$ is the mass-squared matrix in the basis $\left\{-s_\beta\phi^++c_\beta \tfrac{\xi^+ +\chi^+}{\sqrt{2}} ,  \tfrac{-\xi^+ +\chi^+}{\sqrt{2}}\right\}$\,, with the matrix elements,
\begin{align*}
\left(\mathcal{M}_c^2\right)_{11}&=-\frac{1}{8}v^2\left(\kappa_2+\sqrt{2}\kappa_3\right)+\frac{1}{\sqrt{2}}s_\beta^{-1}\mu_2v\,,\\ 
\left(\mathcal{M}_c^2\right)_{12}&=-\frac{1}{8}v^2c_\beta\left(\kappa_2-\sqrt{2}\kappa_3\right)\,,\\
\left(\mathcal{M}_c^2\right)_{22}&=-\frac{1}{8}v^2c_\beta^2\left(\kappa_2+5\sqrt{2}\kappa_3\right)+\frac{1}{4}v^2s_\beta^2\kappa_1+\frac{1}{\sqrt{2}}c_\beta^2s_\beta^{-1}\mu_2v+\frac{1}{\sqrt{2}}s_\beta\mu_1v\,.
\end{align*}
In our analysis, we trade the mixing angle ($\delta$) between the singly charged Higgs bosons with the heavy Higgs boson mass, $m_A$, and as a result, we get the following relation, 
\begin{equation}
m_A^2=m_{H^+}^2c_\delta^2+m_{F^+}^2s_\delta^2-\big(m_{F^+}^2-m_{H^+}^2\big)c_\beta^{-1}s_\delta c_\delta \,.
\label{sum_rule_eGM}
\end{equation}
From Eq.~(\ref{sum_rule_eGM}), it can be easily seen that $m_A$ and $m_{H^+}$ are mass degenerate in the limit $\delta = 0$. 
In the eGM model, the absolute mass difference between $m_{F^{++}}$ and $m_{F^0}$, obtained from Eq.~(\ref{mass_cp_odd_doubly_charged}) and Eq.~(\ref{mass_cp_even}), is independent of the trilinear couplings, $\mu_1$ and $\mu_2$.  
All other mass differences of physical scalars depend on both the quartic and trilinear parameters of the potential.

About four decades ago, Georgi and Machacek constructed a global $SU(2)_L\otimes SU(2)_R$ symmetric potential~\cite{Georgi:1985nv} which consists of a bi-doublet ($\Phi$) and a bi-triplet ($\Delta$), where
\begin{equation}
\Phi=
\begin{pmatrix}
\phi^{0 *} & \phi^+ \\
-\phi^- & \phi^0 
\end{pmatrix}, \qquad
\Delta= 
\begin{pmatrix}
\chi^{0 *} & \xi^+ & \chi^{++} \\
-\chi^- & \xi^0 & \chi^+ \\
\chi^{--} & -\xi^- & \chi^0
\end{pmatrix}.
\label{GM_fields}
\end{equation}
Given this global symmetry, the potential consisting of a bi-doublet ($\Phi$) field and a bi-triplet ($\Delta$) field can be written, in the conventions of Ref.~\cite{Chiang:2015amq}, as
\begin{eqnarray}
\nonumber
V(\Phi, \Delta) &=& 
\frac{1}{2}m^2_1 \tr[\Phi^\dagger \Phi]+\frac{1}{2}m^2_2 \tr[\Delta^\dagger\Delta]+\lambda_1(\tr[\Phi^\dagger \Phi])^2+ \lambda_2 (\tr[\Delta^\dagger \Delta])^2\nonumber \\
&&+ \lambda_3 \tr[(\Delta^\dagger \Delta)^2]+\lambda_4 \tr[\Phi^\dagger \Phi] \tr[\Delta^\dagger \Delta]+\lambda_5 \tr[\Phi^\dagger \tau^a \Phi \tau^b] \tr[\Delta^\dagger t^a \Delta t^b] \nonumber\\
&&+\mu_1 \tr[\Phi^\dagger \tau^a \Phi \tau^b](P^\dagger \Delta P)_{ab} + \mu_2 \tr[\Delta^\dagger t^a \Delta t^b](P^\dagger \Delta P)_{ab}\,.
\label{GM_pot}
\end{eqnarray}
Here $\tau^a$'s and $t^a$'s are the $SU(2)$ generators for the doublet and triplet representations, respectively, and the matrix $P$ diagonalizes the adjoint representation of the $SU(2)$ generator~\cite{Chiang:2015amq}. All the potential parameters in this model are kept real to avoid any explicit CP-violation. This model is known as the Georgi-Machacek (GM) model. In the GM model, owing to this symmetry, $\rho=1$ at the tree-level after EWSB. Similar to the eGM model, the GM model also has one doubly charged, two singly charged, and four neutral scalars, one of them being the SM-like Higgs boson. In contrast to the eGM model, in the GM model, the physical scalars form one fiveplet, one triplet, and two singlets by their transformation properties under the custodial $SU(2)_V$ symmetry. Therefore, the scalars within each custodial multiplet are mass degenerate in the GM model. The field structures and the masses of the physical scalars of the GM model are given in Ref.~\cite{Chiang:2015amq}.
\section{Boundedness of the scalar potential}\label{sec:bfb}
A necessary requirement for absolute vacuum stability is that the scalar potential must be bounded from below in any direction of the field space. Various methodologies are extensively used in the literature to compute the bounded-from-below (BFB) conditions of the extended scalar potentials, such as copositivity~\cite{Kannike2012, Joydeep2013,Kannike2016}, geometric approaches~\cite{Abud1981,Kim1982,Abud1983,Ivanov2006,ElKaffas:2006gdt,Degee2012}, and other mathematical techniques~\cite{Klimenko1984,Murty1987,Ivanov2018}. However, it is mathematically challenging to compute the necessary and sufficient BFB conditions for the general quartic potential with multiple scalar fields. The BFB conditions, taking into account all combinations of two non-zero scalar fields, were derived in~\cite{Chiang:2012cn} for the GM model and in~\cite{Blasi:2017xmc} and  \cite{Krauss2017} for the generalized two-triplet model. Ref.~\cite{Arhrib:2011uy} provides necessary and suﬃcient conditions on the couplings of the Type II Seesaw model that are valid for all field directions. The BFB conditions considering all field directions have been studied in~\cite{Hartling:2014zca} and \cite{Moultaka2020} for the GM model, while~\cite{Moultaka2020} also computed all field BFB conditions for the generalized two-triplet model. The necessary and sufficient bounded-from-below conditions for a generic quartic scalar potential of a large number of fields cannot always be recast into a fully analytical compact form, as pointed out in~\cite{Moultaka2020}. In this section, we compute the strict copositivity criteria at the tree-level, taking into account all combinations of three non-zero scalar fields for the GM, the eGM, and the generalized two-triplet model. Following the strategy of reparameterizing the field space as given in~\cite{ElKaffas:2006gdt,Arhrib:2011uy,Moultaka2020}, we also derive all field BFB conditions for the eGM model at the tree-level.  

At large values of the fields, there should not exist any field direction that renders the potential to be unbounded from below. Strict positivity condition on the quartic part of the potential, given in Eq.~(\ref{v16}), has to be imposed to avoid such unboundedness, i.e., 
\begin{multline}
V^{(4)}\equiv\lambda_\phi(\phi^\dagger\phi)^2+\lambda_\xi(\xi^\dagger\xi)^2+\lambda_\chi(\chi^\dagger\chi)^2+\tilde{\lambda}_\chi |\tilde{\chi}^\dagger\chi |^2+\lambda_{\phi\xi}(\phi^\dagger\phi)(\xi^\dagger\xi)+\lambda_{\phi\chi}(\phi^\dagger\phi)(\chi^\dagger\chi)\\
+\lambda_{\chi\xi}(\chi^\dagger\chi)(\xi^\dagger\xi)+\kappa_1|\xi^\dagger\chi |^2+\kappa_2(\phi^\dagger\tau_a\phi)(\chi^\dagger t_a\chi)
+\kappa_3\Big[(\phi^T\epsilon\tau_a\phi)(\chi^\dagger t_a\xi)+\text{h.c.}\Big]>0\,.
\label{v16_quartic}
\end{multline}  
To compute the BFB conditions on the quartic parameters, we impose strict copositivity criteria for the potential of a specific direction in the field space. Taking into account all such directions with only two non-vanishing fields at once, we obtain the following BFB conditions for the eGM model,
\begin{align}
\ld_\phi&>0, \;&\ld_\chi&>0, \; &4\ld_{\chi}-8\ld_\xi+\kappa_1+4\sqrt{\ld_\xi\ld_\chi}&>0,\nonumber\\
\ld_{\phi\xi}+2\sqrt{\ld_\phi\ld_\xi}&>0, \;  &\ld_{\xi}&>0,\; &2\ld_{\chi}-4\ld_\xi+2\sqrt{\ld_\xi(\ld_\chi+\tilde{\ld}_\chi)}&>0,\nonumber\\
\ld_{\phi\chi}+2\sqrt{\ld_\phi(\ld_\chi+\tilde{\ld}_\chi)}&>0, \; &\ld_\chi-\ld_\xi&>0,\;&4\ld_{\phi\chi}-4\ld_{\phi\xi}-\sqrt{2}\kappa_3+4\sqrt{\ld_\phi\ld_\chi}&>0,\nonumber\\
4\ld_{\phi\xi}+\sqrt{2}\kappa_3+4\sqrt{\ld_\phi\ld_\chi}&>0, \; &\ld_\chi+\tilde{\ld}_\chi&>0,\;&2\ld_{\chi}-4\ld_\xi+\kappa_1+2\sqrt{\ld_\xi(\ld_\chi+\tilde{\ld}_\chi)}&>0\,.
\label{eq:2f_all}
\end{align}
The quartic parameters get further constrained once we consider three or more simultaneously non-vanishing field directions in the field space. For instance, if we consider the field directions where only the neutral components of the fields $\phi^0,\xi^0,$ and $\chi^0$ are non-zero, we get the following BFB conditions for the eGM model,
\begin{equation}
6\ld_{\phi\xi}+3\sqrt{2}\kappa_3+4\sqrt{3\ld_\phi(\ld_\chi-\ld_\xi)}>0,\quad 6\ld_{\phi\xi}-\sqrt{2}\kappa_3+4\sqrt{3\ld_\phi(\ld_\chi-\ld_\xi)}>0\,.
\label{eq:cust-bfb}
\end{equation}
The methodology for obtaining the above conditions is discussed in depth in Appendix~\ref{app:pos_subspace}, where we have also computed the BFB conditions taking into account all such combinations with three simultaneously non-vanishing fields for the eGM model and for the generalized two-triplet model. The BFB conditions with all non-zero field directions are given in Appendix~\ref{app:pos_EGM} for the eGM model.

We obtain the minimal set of BFB conditions considering all possible two non-vanishing field directions for the GM model,
\begin{align}
\ld_1&>0,\quad & \ld_2+\ld_3&>0,\quad &\ld_4+2\sqrt{\ld_1(\ld_2+\ld_3)}&>0,\nonumber\\
&\quad & 3\ld_2+\ld_3&>0,\quad & 4\ld_4-|\ld_5|+4\sqrt{2\ld_1(2\ld_2+\ld_3)}&>0\,.
\label{eq:2f-GM}
\end{align}
The above results agree except for the fourth condition involving $\lambda_4$ in Eq.~(3.3) of Ref.~\cite{Chiang:2012cn}. In addition, we get a more stringent BFB constraint, i.e., $3\ld_2+\ld_3>0$, from the two field directions where $\xi^0$ and $\chi^{++}/\chi^{0}$ are taken to be non-zero.  If we consider the combination where only neutral components of the fields, i.e., $\phi^0,\xi^0,$ and $\chi^0$ are non-zero, we get the following additional BFB conditions on top of the conditions listed in Eq.~(\ref{eq:2f-GM}),
\begin{equation}
\ld_4+\frac{1}{2}\ld_5+2\sqrt{\ld_1\left(\ld_2+\frac{1}{3}\ld_3\right)}>0, \quad \ld_4-\frac{1}{6}\ld_5+2\sqrt{\ld_1\left(\ld_2+\frac{1}{3}\ld_3\right)}>0\,.
\label{eq:cust-bfb-GM}
\end{equation}
In Appendix~\ref{app:pos_subspace}, we list the BFB conditions for all such combinations of three simultaneously non-vanishing fields in the GM model. In our numerical analysis in Section~\ref{sec:results}, we will compare the BFB constraints for two, three, and all thirteen non-vanishing field directions for both models.
\section{Unitarity constraints at one-loop}\label{sec:uni}
\subsection{Partial-wave analysis }\label{partial-wave analysis}
In this section, we aim to study the unitarity bounds on the quartic parameters of Eq.~(\ref{v16}). Perturbative unitarity constraints come from demanding the unitarity of the $S$-matrix, which reads as,
\bea
\left|  a_{\ell}^{2 \to 2} - \frac{1}{2}i   \right|^2 +\nsum[1.5]_{k>2} \left| a_\ell^{2 \rightarrow k} \right|^2 = \frac{1}{4}\,.
\label{eq:partial-wave analysis1}
\eea 
Here, $a_\ell^{2 \rightarrow k}$$\,$'s are the eigenvalues of the $S$-matrix consisting of $\ell$-th partial-wave amplitudes of $2 \rightarrow k$ scattering.\footnote{Note that, for $k>2$, the scattering matrix is diagonalized in the eigenbasis of the $2 \rightarrow 2$ scattering matrix.} In the limit, $s \gg \vert \lambda_i \rvert v^2$, $s \gg \vert \mu_i \rvert v$, $2 \rightarrow 3$ partial-wave amplitudes can be neglected since $\lvert a_\ell^{2 \rightarrow 3} \rvert^2 \sim \lambda_i^4 v^2/s $, where $\mu_i$ ($\lambda_i$)$\,$'s represent some linear combinations of scalar cubic (quartic) couplings of the theory. The next leading order contribution comes from $2 \rightarrow 4$ scattering amplitudes, which scale as $\lvert a_\ell^{2 \rightarrow 4} \rvert^2 \sim \lambda_i^4$. In 
the SM, the $2 \rightarrow 4$ amplitudes are significantly smaller compared to the $2 \rightarrow 2$ partial-wave amplitudes because of the smallness of the quartic coupling~\cite{PhysRevD.45.3112}. Therefore, in our analysis, only the $2 \rightarrow 2$ amplitudes will be taken into account. In the rest of the paper, we remove the superscript $2\to 2$ from partial-wave amplitudes. Under this consideration, Eq.~(\ref{eq:partial-wave analysis1})
gives an upper limit on the eigenvalues of the $S$-matrix,
\bea
\left| a_{\ell} - \frac{1}{2}i  \right|^2 \,\leq \frac{1}{4}\,.
\label{eq:partial-wave analysis2}
\eea
At the tree-level, each of the eigenvalues, $a_\ell \in \mathbb{R}$, leads to a strong bound, $\lvert\text{Re}(a_\ell) \rvert \, \leq 1/2$.  At one-loop and beyond, $a_\ell\notin \mathbb{R}$, thus the above-stated limit gets weaker when we calculate the one-loop and the higher order corrections to the $S$-matrix. Note that the unitarity bounds over the real line are employed in the tree-level analysis, and the complete unitarity circle in the complex plane is utilized once the loop corrections are incorporated into the unitarity calculations. In the limit, $s \gg \lvert \lambda_i \rvert v^2 \gg M_W^2$, $s\gg|\mu_i|v$ the most dominant contribution comes from the $\ell=0$ partial-wave at the tree-level. Therefore, we will only consider  $\ell=0$ in our analysis. To calculate the partial-wave amplitude $(a_0)$ at one-loop level, we adopt the approach of Ref.~\cite{Grinstein:2015rtl}. For a given process $i\to f$, the corresponding matrix element is 
\begin{align*}
(a_0)_{i,f}(s)=\frac{1}{16\pi s }\int_{-s}^0 dt \mathcal{M}_{i\to f}(s,t)\,,
\end{align*}
where $\mathcal{M}_{i \rightarrow f}$ represents the sum of all possible scattering amplitudes with an initial state $i$ and final state $f$. As the $SU(2)_L \otimes U(1)_Y$ symmetry is intact at high energies, the $S$-matrix can be sub-divided into smaller block diagonal forms consisting of two-particle states with their respective total charge $(Q)$ and hypercharge $(Y)$. Following this prescription, the basis states are written in Table~\ref{tab:2particle_states}. We have included $1/ \sqrt{2}$ symmetry factor in the identical initial or final states. 
However, this block diagonal structure does not hold beyond the tree level due to hypercharge interactions. At the one-loop level, in general, the off-block diagonal elements are non-zero due to the wave function renormalization terms. For a given scattering process with total charge $Q$, if the tree-level blocks have unique eigenvalues, then off-block diagonal elements do not contribute to the tree-level eigenvalues at the one-loop, while the contributions can appear at the two-loop and beyond.
\begin{table}[!h]
\begin{center}
\setlength{\tabcolsep}{0pt}
\renewcommand{\arraystretch}{1.5}
\scalebox{0.7}{
\begin{tabular}{| c| ccc | cc | cc | c | c |}
\hline
\cellcolor{lightgray!10}\diagbox{\large$\;\mathbf{Q}$}{\large$\mathbf{Y}\;$} & \multicolumn{3}{c|}{\cellcolor{lightgray!10}\large$\mathbf{0}$} & \multicolumn{2}{c|}{\cellcolor{lightgray!10}\large$\mathbf{1/2}$} & \multicolumn{2}{c|}{\cellcolor{lightgray!10}\large$\mathbf{1}$} & \cellcolor{lightgray!10}\large$\mathbf{3/2}$ & \cellcolor{lightgray!10}\large$\mathbf{2}$\\
\hline
\hline
\cellcolor{lightgray!10} &\cellcolor{isocolor0!65} \;$\phi^{0*}\phi^0$\;\;&\cellcolor{isocolor0!65} $\frac{\xi^0\xi^0}{\sqrt{2}}$ &\cellcolor{isocolor0!65} $\chi^{0*}\chi^0$ &\; \cellcolor{isocolor12m!45}$\phi^{0}\xi^0$\; \;&\cellcolor{isocolor12m!45} $\phi^{0*}\chi^0$  &\; \cellcolor{isocolor1m!37}$\frac{\phi^0\phi^0}{\sqrt{2}}$ \;\;&\cellcolor{isocolor1m!37} $\chi^0\xi^0$ & \cellcolor{isocolor32m!35}$\phi^0\chi^0$ & \cellcolor{isocolor2m!30}$\frac{\chi^0\chi^0}{\sqrt{2}}$ \\[8pt]
{\cellcolor{lightgray!10}\Large$\mathbf{0}$} &\cellcolor{isocolor0!65} $\phi^{+}\phi^-$ &\cellcolor{isocolor0!65} \;$\xi^{+}\xi^-$ &\cellcolor{isocolor0!65} $\chi^{+}\chi^-$ & \cellcolor{isocolor12m!45}$\phi^+\xi^-$ & \cellcolor{isocolor12m!45}$\chi^+\phi^-$ & \cellcolor{isocolor1m!37} &\; \cellcolor{isocolor1m!37} $\chi^+\xi^-$ \;\;&\cellcolor{isocolor32m!35} & \cellcolor{isocolor2m!30}\\[8pt]

\cellcolor{lightgray!10}& \cellcolor{isocolor0!65} &\cellcolor{isocolor0!65} & \cellcolor{isocolor0!65}\;$\chi^{++}\chi^{--}$ \;\;& \cellcolor{isocolor12m!45} &\cellcolor{isocolor12m!45}  &  \cellcolor{isocolor1m!37}& \cellcolor{isocolor1m!37}& \cellcolor{isocolor32m!35}&\cellcolor{isocolor2m!30} \\[7pt]
\hline
\hline
\cellcolor{lightgray!10}&\cellcolor{isocolor1!75}$\phi^{0*}\phi^{+}$  &\cellcolor{isocolor1!75} $\xi^{0}\xi^{+}$ &\cellcolor{isocolor1!75} $\chi^{0*}\chi^{+}$ &\cellcolor{isocolor12!99} $\phi^0\xi^+$ & \cellcolor{isocolor12!99}$\phi^{0*}\chi^+$  & \cellcolor{isocolor0!65}$\phi^{0}\phi^{+}$ &\cellcolor{isocolor0!65} $\xi^{0}\chi^{+}$ &\cellcolor{isocolor12m!45} $\phi^0\chi^+ $ &\cellcolor{isocolor1m!37} $\chi^+\chi^0 $\\[8pt]
\cellcolor{lightgray!10}\multirow{-2}{*}{\Large$\mathbf{1}$} & \cellcolor{isocolor1!75}  & \cellcolor{isocolor1!75} &\cellcolor{isocolor1!75} $\chi^{-}\chi^{++}$ &\cellcolor{isocolor12!99} $\phi^+\xi^0$ &\cellcolor{isocolor12!99} $\chi^{++}\phi^-$  &\cellcolor{isocolor0!65} $\chi^{0}\xi^{+}$ & \cellcolor{isocolor0!65}$\chi^{++}\xi^{-}$&\cellcolor{isocolor12m!45} $\phi^+\chi^0 $ & \cellcolor{isocolor1m!37}\\[7pt]
\hline
\hline
\cellcolor{lightgray!10} & \cellcolor{isocolor2!80} & \cellcolor{isocolor2!80} $\frac{\xi^+\xi^+}{\sqrt{2}}$  &\cellcolor{isocolor2!80}  $\chi^{0*}\chi^{++}$ & \cellcolor{isocolor32!85}   $\phi^+\xi^{+}$  &\;\cellcolor{isocolor32!85}  $\phi^{0*}\chi^{++}$ \;  & \cellcolor{isocolor1!75}$\frac{\phi^+\phi^+}{\sqrt{2}}$ & \cellcolor{isocolor1!75}$\xi^{+}\chi^{+}$  & \cellcolor{isocolor12!99}$\phi^+\chi^+$ & \cellcolor{isocolor0!65}$\chi^0\chi^{++} $\\[8pt]
\cellcolor{lightgray!10}\multirow{-2}{*}{\Large$\mathbf{2}$} & \cellcolor{isocolor2!80} & \cellcolor{isocolor2!80}& \cellcolor{isocolor2!80} & \cellcolor{isocolor32!85} & \cellcolor{isocolor32!85}   & \cellcolor{isocolor1!75} & \cellcolor{isocolor1!75}$\xi^0\chi^{++}$  & \cellcolor{isocolor12!99}$\phi^0\chi^{++}$ & \cellcolor{isocolor0!65}$\frac{\chi^+\chi^+}{\sqrt{2}}$\\[7pt]
\hline
\hline
\cellcolor{lightgray!10}\multirow{1.4}{*}{\Large$\mathbf{3}$} &\cellcolor{lightgray!10} &\cellcolor{lightgray!10}\multirow{1.4}{*}{\textcolor{gray}{\Large\xmark}} &\cellcolor{lightgray!10} &  \multicolumn{2}{c|}{\cellcolor{lightgray!10}\multirow{1.4}{*}{\textcolor{gray}{\Large\xmark}}}   & \cellcolor{isocolor2!80} & \cellcolor{isocolor2!80}$\xi^+\chi^{++}$&\; \cellcolor{isocolor32!85} $\phi^+\chi^{++}$\;& \cellcolor{isocolor1!75}$\chi^+\chi^{++}$\\[7pt]
\hline
\hline
\cellcolor{lightgray!10}\multirow{1.4}{*}{\Large$\mathbf{4}$} & \cellcolor{lightgray!10} & \cellcolor{lightgray!10}\multirow{1.4}{*}{\textcolor{gray}{\Large\xmark}} & \cellcolor{lightgray!10} &  \multicolumn{2}{c|}{\multirow{1.4}{*}{\cellcolor{lightgray!10}\textcolor{gray}{\Large\xmark}}}  & \multicolumn{2}{c|}{\multirow{1.4}{*}{\cellcolor{lightgray!10}\textcolor{gray}{\Large\xmark}}} &\cellcolor{lightgray!10} \multirow{1.4}{*}{\textcolor{gray}{\Large\xmark}} & \;\cellcolor{isocolor2!80}$\frac{\chi^{++}\chi^{++}}{\sqrt{2}}$\;\\[7pt]
\hline
\end{tabular}}
\caption{Two-particle basis states broken down by their total charge $Q$ and total hypercharge $Y$. We have omitted the charge conjugated states as they give the same eigenvalues. The blocks with total weak isospin $|T_3|= 0,\, 1/2,\, 1,\, 3/2, \,2$ are shown in green, lime green, olive, yellow, and brown, respectively.}
\label{tab:2particle_states}
\end{center}
\end{table}

\subsection{$2 \to 2$ scattering amplitudes}\label{sec:scattering}
Goldstone-boson equivalence theorem is a useful theoretical tool to study scattering amplitudes in high energy regime~\cite{Lee:1977eg,Chanowitz:1985hj,PhysRevD.41.264}. This theorem states that, at energies, $\sqrt{s} \gg M_W$, an amplitude involving $k$ longitudinally polarized vector bosons ($W_L^\pm,Z_L,h,...$) at the external states can be related to an amplitude with $k$ external Goldstone bosons ($w^{\pm},z,h,...$) as,
\bea
\nonumber
\mathcal{M}(W_L^{\pm},Z_L,h,...)&=&(iC)^k\mathcal{M}(w^{\pm},z,h,...)\,.
\label{eq:equivalence theorem}
\eea
In the limit, $s \gg \lvert \lambda_i \rvert v^2 \gg M_W^2$, $s\gg|\mu_i|v$, only one-particle irreducible (1PI) diagrams with two internal lines survive at the one-loop.\footnote{We have chosen this energy regime to simplify our computations. These conditions are examined uniformly throughout the energy regime of interest~\cite{PhysRevLett.62.1232,PhysRevD.40.2880}.} We choose the $\overline{\text{MS}}$ scheme to renormalize the quartic couplings in the one-loop computation to satisfy the Goldstone theorem with $C=1$~\cite{PhysRevD.41.264}.

The renormalized parameter ($\lambda$) can be defined in terms of the bare parameter ($\lambda^0$) as,
\bea
\nonumber
\lambda_i^0 = \lambda_i + \delta \lambda_i \,,\quad\, \kappa_i^0 = \kappa_i + \delta \kappa_i\,.
\label{eq:renormalized_param}
\eea
Following the convention given in \cite{Grinstein:2015rtl}, the counterterms for a given parameter $\lambda$ can be written as,
\bea
\delta \lambda = \frac{1}{16 \pi^2 \epsilon}\beta_\lambda\,,\quad \text{with} \quad \beta_\lambda = 16\pi^2 \mu^2 \frac{d \lambda}{d \mu^2}\,,
\label{eq:counterterm}
\eea
where $\mu$ is the renormalization scale. We have verified from our calculations that the counter-terms required in order
to cancel the infinities in the one-loop $S$-matrix are nothing but the beta
functions given in the literature~\cite{Blasi:2017xmc,Keeshan:2018ypw},\footnote{Note that, $\beta_{\text{Eq.~}(\ref{eq:counterterm})}=\frac{1}{2}\beta_{\textrm{Ref.~\cite{Blasi:2017xmc}}}$\,.}
\begin{align}
\beta_{\ld_\phi}&=12 \lambda _{\phi }^2+3 \lambda _{\phi \xi }^2+\frac{1}{4}\kappa _2^2+\kappa _3^2+\frac{3 }{2}\lambda _{\phi \chi }^2\,, \nonumber\\
\beta_{\ld_\xi}&=  44 \lambda _{\xi }^2+ \kappa _1 \lambda _{\chi \xi }+\frac{1}{2}\kappa _1^2+\frac{3}{2} \lambda _{\chi \xi }^2+\lambda _{\phi \xi }^2 \,,\nonumber\\
\beta_{\ld_\chi}&=  8 \tilde{\lambda }_{\chi }\left( \tilde{\lambda }_{\chi }+ \lambda _{\chi }\right)+2 \kappa _1 \lambda _{\chi \xi }+\frac{1}{2}\kappa _1^2+\frac{1}{4}\kappa _2^2+14 \lambda _{\chi }^2+3 \lambda _{\chi \xi }^2+\lambda _{\phi \chi }^2  \,,\nonumber\\
\beta_{\tilde\ld_\chi}&=   6 \tilde{\lambda }_{\chi } \left(\tilde{\lambda }_{\chi }+2 \lambda _{\chi }\right)+\frac{1}{2}\kappa _1^2-\frac{1}{4}\kappa _2^2 \,,\nonumber\\
\beta_{\kappa_1}&=  \kappa _1 \left(4 \tilde{\lambda }_{\chi }+5 \kappa _1+8 \lambda _{\xi }+2 \lambda _{\chi }+8 \lambda _{\chi \xi }\right)-\kappa _3^2  \,,\nonumber\\
\beta_{\kappa_2}&= 2 \kappa _2 \left(\lambda _{\chi }-2 \tilde{\lambda }_{\chi }+\lambda _{\phi }+2 \lambda _{\phi \chi }\right)+2\kappa _3^2   \,,\nonumber\\
\beta_{\kappa_3}&=   \kappa _3 \Big(\kappa _2-\kappa _1\Big)+2\kappa_3 \Big(\lambda _{\chi \xi }+\lambda _{\phi }+2 \lambda _{\phi \xi }+\lambda _{\phi \chi }\Big) \,,\nonumber\\
\beta_{\ld_{\phi\xi}}&=  2 \lambda _{\phi \xi } \Big(2 \lambda _{\phi \xi }+10 \lambda _{\xi }+3 \lambda _{\phi }\Big) +\lambda _{\phi \chi } \Big(\kappa _1+3 \lambda _{\chi \xi }\Big)+2 \kappa _3^2 \,,\nonumber\\
\beta_{\ld_{\phi\chi}}&=  2 \lambda _{\phi \chi } \left(\lambda _{\phi \chi} +2 \tilde{\lambda }_{\chi }+4 \lambda _{\chi }+3 \lambda _{\phi }\right)+2\lambda _{\phi \xi } \Big(\kappa _1+3 \lambda _{\chi \xi }\Big)+2\kappa _3^2+\kappa _2^2  \,,\nonumber\\
\beta_{\ld_{\chi\xi}}&=   4 \lambda _{\chi \xi } \left(\tilde{\lambda }_{\chi }+5 \lambda _{\xi }+2 \lambda _{\chi }+\lambda _{\chi \xi }\right)+2 \kappa _1 \Big(2 \lambda _{\xi }+\lambda _{\chi }\Big)+\kappa _1^2+\kappa _3^2+2 \lambda _{\phi \xi } \lambda _{\phi \chi }\,.
\label{renor_beta_func}
\end{align}
The complete set of one-loop and two-loop beta functions for the Yukawa, gauge, and quartic couplings in the theory is given in Appendix~\ref{app:2-loop_rge}. Up to symmetry factors, the one-loop amplitude for a generic 1PI diagram involving four-point vertices is given by,
\begin{equation}
\nonumber
\mathcal{M}_{\textrm{1PI}}^{2\to2}=\frac{\lambda_i\lambda_j}{16\pi^2}\left[\frac{1}{\epsilon}+2-\ln\left(\frac{-p^2-i 0_+}{\mu^2}\right)\right]\,,
\end{equation}
where the quartic couplings, $\lambda_i$ and $\lambda_j$ are evaluated at the energy scale, $\mu^2 = s$. For a given electric charge ($Q$) and hypercharge ($Y$), the $S$-matrix at the one-loop can be expressed as~\cite{Grinstein:2015rtl,Murphy:2017ojk},
\bea
256 \pi^3 \mathbf{a_0} = -16 \pi^2 \mathbf{b_0} + (i\pi - 1) \mathbf{b_0}\cdot \mathbf{b_0} + 3\mathbf{\beta_{b_0}}\,,
\label{eq:one-loop eigenvalue}
\eea
where $\mathbf{b_0}$ is the tree-level $S$-matrix and $\mathbf{\beta_{b_0}}$ represents the matrix carrying some linear combinations of the beta functions defined in Eq.~(\ref{eq:counterterm}). For example, if $b_0=a\lambda_i+b\lambda_j$, then $\beta_{b_0}=a\beta_{\lambda_i}+b\beta_{\lambda_j},\;\forall a,b\in\mathbb{R}$. Our results of the $S$-matrix elements at the one-loop are given in Appendix~\ref{app:1-loop_amp}. These results solely come from the 1PI one-loop corrections
to the tree-level processes, and wave function corrections are ignored in the energy regime of
interest~\cite{Grinstein:2015rtl,Murphy:2017ojk}. For the potential given in Eq.~(\ref{v16}), we have found 16, 15, 11, 3, and 1 unique tree-level eigenvalues for the blocks with $Q=0,1,2,3,4$, respectively. Out of these, a total of 19 eigenvalues appeared to be independent, which is in agreement with the results given in the literature~\cite{Krauss2017,Chen:2023ins},
\begin{align*}
-16\pi a_{0;1}&=2 \lambda _{\chi\xi }+\kappa _1\,,\quad & &\nonumber\\
-16\pi a_{0;2}^{\pm}&=\lambda _{\chi }-2 \tilde{\lambda}_{\chi}+\lambda _{\phi }\pm\sqrt{\kappa _2^2+\left(-\lambda _{\chi }+2 \tilde{\lambda}_{\chi}+\lambda _{\phi }\right)^2}\,,\quad &-16\pi a_{0;3}&=\lambda_{\phi\chi}+\frac{\kappa_2}{2}\,, \nonumber\\
-16\pi a_{0;4}^{\pm}&=4 \lambda _{\xi }+\lambda _{\chi }+2\tilde{\lambda}_{\chi} \pm\sqrt{2 \kappa _1^2+\left(-4 \lambda _{\xi }+\lambda _{\chi }+2 \tilde{\lambda}_{\chi}\right)^2}\,,
\quad &-16\pi a_{0;5}&=2\lambda_{\chi\xi}+4\kappa_1, \nonumber\\
-16\pi a_{0;6}^{\pm} &=\lambda _{\phi \xi }+\frac{\lambda _{\phi \chi }}{2}-\frac{\kappa _2}{4}\pm\sqrt{\kappa _3^2+\left(\frac{\kappa _2}{4}+\lambda _{\phi \xi }-\frac{\lambda _{\phi \chi }}{2}\right)^2}\,,\quad & -16\pi a_{0;7}&=2\lambda_\chi\,,\nonumber\\
-16\pi a_{0;8}^{\pm}&= \lambda _{\phi \xi }+\frac{\lambda _{\phi \chi }}{2}+\frac{\kappa _2}{2}\pm\sqrt{4\kappa _3^2+\left(\frac{\kappa _2}{2}- \lambda _{\phi \xi }+\frac{\lambda _{\phi \chi }}{2}\right)^2}\,,
\quad &-16\pi a_{0;9}&=\lambda_{\phi\chi}-\kappa_2\,,\nonumber\\
-16\pi a_{0;10}^{\pm}&=\lambda _{\chi\xi }+\lambda _{\phi }-\frac{\kappa_1}{2}\pm\sqrt{2 \kappa _3^2+\left(\frac{\kappa_1}{2}- \lambda _{\chi\xi }+ \lambda _{\phi }\right)^2}\,,  \quad&-16\pi a_{0;11}&=2\lambda_\chi+6\tilde{\lambda}_\chi\,,
\end{align*}
and $-16\pi a_{0;i} \:(i=12,13,14)$ being the eigenvalues of the following matrix,
\begin{equation}
\label{eigenvalue_matrix}
\begin{bmatrix}
 20 \lambda _{\xi } & 2 \sqrt{3} \lambda _{\phi \xi } & \sqrt{2} \left(\kappa _1+3 \lambda _{\chi \xi }\right) \\
 2 \sqrt{3} \lambda _{\phi \xi } & 6 \lambda _{\phi } & \sqrt{6} \lambda _{\phi \chi } \\
 \sqrt{2} \left(\kappa _1+3 \lambda _{\chi \xi }\right) & \sqrt{6} \lambda _{\phi \chi } & 8 \lambda _{\chi }+4\tilde{\lambda }_{\chi } \\
\end{bmatrix}.
\end{equation}
The eigenvalues of the $S$-matrix at the tree-level can be found in Refs.~\cite{Aoki:2007ah,Hartling:2014zca} for the scalar quartic couplings in the GM model.

\section{HEPfit}\label{sec:hepfit}
To constrain the parameters in the GM and the eGM models, we employ a Bayesian fit with Markov Chain Monte Carlo (MCMC) simulations. We use the open-source code \texttt{HEPfit}~\cite{DeBlas:2019ehy} to calculate various theoretical and experimental observables that depend on the model parameters and feed them into the parallelized  \texttt{BAT} library~\cite{Caldwell:2008fw} with Message Passing Interface (MPI). It offers to sample the parameter space and find out allowed regions with all sorts of constraints in a given new physics (NP) model. Approximately, each fit runs over $16$ parallel chains, generating a total of $6.4 \times 10^8$ iterations. 

In our global fits, we consider the lightest CP-even custodial singlet scalar to be the SM-like Higgs with fixed mass $m_h=125.09 $ GeV~\cite{ATLAS:2015yey}, and $v=246.22$ GeV, in both the GM and eGM models. All the other SM parameters are fixed to their best-fit values~\cite{deBlas:2016ojx}. To perform a Bayesian fit for any NP model, we need to choose a set of input parameters and their corresponding priors to sample the parameter space. Here, we use a different set of input parameters compared to~\cite{Chiang:2018cgb} for the GM model to achieve good sampling and implement the eGM model into \texttt{HEPfit} for the first time. We list the input parameters of the GM and the eGM models and their priors in Table~\ref{tab:GMprior} and~\ref{tab:eGMprior}, respectively. In the case of the eGM model, we make observables for heavy scalar masses within the range: $m_A,m_{F^{0}},m_{F^{++}}\in [130,1100]$ GeV. The triplet VEV outside of the chosen prior range is expected to be excluded from flavor observables (see~\cite{Hartling:2014aga} for the GM model). In this work, we assume that all the exotic Higgs masses are heavier than 125 GeV and focus on the parameter space that is within the reach of the LHC. For the GM model, we adopt the notation used in Ref.~\cite{Chiang:2015amq}, where $v_\Delta$ represents the triplet VEV, and $m_1,m_3,$ and $m_5$ denote the masses of the other custodial singlet, custodial triplet, and custodial fiveplet, respectively. The relations between the quartic couplings and physical masses of the GM model can be found in~\cite{Chiang:2012cn}. For the eGM model, we provide the expressions of quartic couplings in terms of the physical masses in Appendix~\ref{app:inverse_mass}.
\begin{table}[ht]
\begin{center}
\begin{tabular}{ |c|c|c| c|c|} 
 \hline
 \textbf{Parameter} & $v_\Delta$ & $\alpha$ & $\lambda_3,\lambda_5$ & $m_1,m_3,m_5$\Tstrut \\ \hline
\textbf{Range}  & \;$[0,60] $ GeV\;&\; $[-\pi/2,\pi/2]\; $ &\; $[-4\pi,4\pi] $\; & \;$[130,1100]$ GeV\;\Tstrut \\ 
 \hline
\end{tabular}
\caption{Priors on the input parameters of the GM model.}
\label{tab:GMprior}
\end{center}
\end{table}

The Bayesian statistics, in principle, do not provide a unique rule for choosing prior and posterior distributions. However, it is recommended to choose flat priors for the parameters on which the model observables depend linearly~\cite{Chowdhury:2017aav}. In order to keep our fit results to be as much prior independent as possible, we choose flat priors for all the input parameters. Following~\cite{Chowdhury:2017aav,Chiang:2018cgb}, we use both mass and mass-squared priors while performing the global fits and present the combined $95.4\%$ probability regions in this paper.
\begin{table}[ht]
\begin{center}
\begin{tabular}{ |c|c|c| c|c|c|} 
 \hline
\textbf{Parameter} & $v_\chi$ & $\alpha,\delta$ & $\lambda_\chi$ & $\tilde{\lambda}_\chi,\kappa_1,\kappa_3$ & $m_H,m_{H^+},m_{F^+}$\Tstrut\\ \hline
\textbf{Range}  &\; $[0,60] $ GeV\;&\; $[-\pi/2,\pi/2] $\; & \;$[0,4\pi] $\; &\; $[-4\pi,4\pi] $\; &\; $[130,1100]$ GeV\;\Tstrut \\  \hline
\end{tabular}
\caption{Priors on the input parameters of the eGM model.}
\label{tab:eGMprior}
\end{center}
\end{table}
\section{Global fit constraints}\label{sec:fit_constraints}
\subsection{Theoretical constraints}
From the theory perspective, we include the following constraints in our fits: 
\begin{itemize}
\item Higgs potential must satisfy the stability bounds due to all possible three simultaneously non-vanishing field directions in the field space up to the scale $1$ TeV.\footnote{In our analysis, the stability bounds refer to the bounded-from-below conditions on the scalar potential.  The metastability of the potential is not considered here and will be addressed in a future study.} 

\item Yukawa and quartic couplings of the theory are assumed to be in the perturbative regime (i.e., their magnitudes are smaller than $\sqrt{4\pi}$ and $4\pi$, respectively) up to the scale $1$ TeV.\footnote{The choice of this scale depends on the energy up to which the theory is assumed to be weakly interacting.} For the renormalization group running, we use two-loop renormalization group equations.

\item Eigenvalues of the $S$-matrix of $2\to 2$ scattering up to NLO given in Eq.~(\ref{eq:one-loop eigenvalue}) should satisfy the unitarity bounds at the scale 1 TeV.\footnote{This scale is well above the electroweak scale and we can safely use the NLO unitarity conditions given in Eq.~(\ref{eq:one-loop eigenvalue}).} We further demand that the NLO corrections to the LO eigenvalues should be smaller in magnitude. To quantify this, we define~\cite{Grinstein:2015rtl},
\bea
R_1^\prime  = \frac{\abs{a_0^{\text{NLO}}}}{\abs{a_0^{\text{LO}}}}\,,
\label{eq:R1pert}
\eea
where $a_0^{\textrm{NLO}}$ stands for NLO corrections calculated in the eigenbasis of the matrix, $\mathbf{a}_0^{\textrm{LO}}$.  
\end{itemize}
 As stated in Ref.~\cite{Grinstein:2015rtl}, a minimal requirement for perturbative expansion to hold is that the NLO corrections should be smaller in magnitude than the LO contribution, leading to $R_1'<1$. Similar criteria were used in the case of the SM~\cite{PhysRevD.45.3112} and later in the 2HDM~\cite{Grinstein:2015rtl,Cacchio:2016qyh} to analyze perturbative unitarity. However, there are certain directions in the parameter space that lead to accidental cancellations in the LO amplitudes. For example, $a_0^{\textrm{LO}}=\kappa_2-\lambda_{\phi\chi}$ is small when $\kappa_2\approx\lambda_{\phi\chi}$, while the NLO correction can be large since it depends on other quartic couplings of the theory (see Eq.~(\ref{eigenvalue_matrix})).
Therefore, while doing the $R_1^\prime$ test, we impose a cut on the tree-level eigenvalues, $|a_0^{\textrm{LO}}|>0.02$, such that the fit does not encounter any such accidental cancellations for reasonable values of the quartic couplings. Furthermore, it is worth noting that the bounds on the eigenvalues of the  $S$-matrix are obtained at the TeV energy regime.  However, the running of the VEVs destabilizes the custodial symmetric vacua~\cite{Blasi:2017xmc}. Therefore, the constraints on the model parameters given in Eq.~(\ref{eq:constrints1}) no longer hold in the high energy limit for the eGM model. The global $SU(2)_L\otimes SU(2)_R$ symmetry for the GM model also breaks down in the high energy limit. In order to make it consistent with the $S$-matrix computations, we consider the most general potential (Eq.~(\ref{v16})) containing ten quartic couplings, and we use the one-loop unitarity conditions on these couplings. The $S$-matrix elements at one-loop are given in Appendix~\ref{app:1-loop_amp}, where the quartic couplings are running couplings. The two-loop renormalization group equations (RGEs) are computed using \texttt{PyR@TE}~\cite{Sartore:2020gou}. Explicit expressions of two-loop RGEs are given in Appendix~\ref{app:2-loop_rge}. Among the Yukawa couplings, we only consider the contributions coming from the third-generation fermions. 
\begin{table}[h!]
\begin{center}
\setlength{\tabcolsep}{0pt}
\renewcommand{\arraystretch}{1.25}
\scalebox{0.8}{
\begin{tabular}{| l | c | c | cccccc | c | c |}
\hline
\;\;\;&\;\; \textbf{Signal} \;\;&\;\;\;\; \textbf{Value} \;\;\;\;& \multicolumn{6}{c|}{\textbf{Correlation matrix}} & ${\cal L}$ & \;\textbf{Source}\;\\
&\;\; \;\textbf{strength}\;\; \;&  &\multicolumn{6}{c|}{} &\;\textbf{[fb$^{-1}$]}\: & \\[3pt]
\hline
\hline
\cellcolor{sigstrcolor1} & $\mu_\text{ggF,bbh}^{\gamma \gamma}$ &$1.04\pm 0.10$  &\;1\;&$\;-0.13\;\;$&0&0&0&0 &\cellcolor{green!50}& \multirow{7}{*}{\cite{ATLAS:2022tnm}} \\[2pt]
\cellcolor{sigstrcolor1} & $\mu_\text{VBF}^{\gamma \gamma}$ & $1.20 \pm 0.26$  &\;$-0.13$\;&1&0&0&0&0 &\cellcolor{green!50}&\\[2pt]
\cellcolor{sigstrcolor1} & $\mu_\text{Wh}^{\gamma \gamma}$ & $1.5 \pm 0.55$  &0&0&1&\;$-0.37$\;\;&0&\;$-0.11$\;\; &\cellcolor{green!50}&\\[2pt]
\cellcolor{sigstrcolor1} & $\mu_\text{Zh}^{\gamma \gamma}$ & $-0.2 \pm 0.55$  &0&0&\;$-0.37$\;\;&1&0&0 &\cellcolor{green!50}&\\[2pt]
\cellcolor{sigstrcolor1} & $\mu_\text{tth}^{\gamma \gamma}$ & $0.89 \pm 0.31$  &0&0&0&0&1&$-0.44$ &\cellcolor{green!50} \multirow{-3}{*}{\parbox{23pt}{139\\ \phantom{d}}}&\\[2pt]
\cellcolor{sigstrcolor1} & $\mu_\text{th}^{\gamma \gamma}$ & $3 \pm 3.5$  &0&0&$-0.11$&0&$-0.44$&1 &\cellcolor{green!50}&\\[3pt]
\hline
\hline
\cellcolor{sigstrcolor5} & $\mu_\text{ggF}^{ZZ}$ & $0.95 \pm 0.1$  &1&$-0.22$&$-0.27$&0& & & \cellcolor{green!50} & \multirow{4.4}{*}{\cite{ATLAS:2020rej}} \\[2pt]
\cellcolor{sigstrcolor5} & $\mu_\text{VBF}^{ZZ}$ & $1.19 \pm 0.45 $  &$-0.22$&1&0&0&& &\cellcolor{green!50} &\\[2pt]
\cellcolor{sigstrcolor5} & $\mu_\text{Vh}^{ZZ}$ & $1.43 \pm 1.0$  &$-0.27$&0&1&$-0.18$ && &\cellcolor{green!50} \multirow{-1}{*}{\parbox{23pt}{139\\ \phantom{d}}}&\\[2pt]
\cellcolor{sigstrcolor5} & $\mu_\text{tth}^{ZZ}$ & $1.69 \pm 1.45$  &0&0&$-0.18$&1 && &\cellcolor{green!50} &\\[3pt]
\hline
\cellcolor{sigstrcolor5} & $\mu_\text{incl.}^{ZZ}$ & $1.0 \pm 0.1$  &&&& && &\cellcolor{green!50} 139 & \cite{ATLAS:2020rej} \\[3pt]
\hline
\hline
\cellcolor{sigstrcolor4} & $\mu_\text{ggF,bbh}^{WW}$ & $1.15 \pm 0.135$  & & & & & & & \cellcolor{green!50}&\\[2pt]
\cellcolor{sigstrcolor4} & $\mu_\text{VBF}^{WW}$ & $0.93 \pm 0.21$  & & & & & & & \cellcolor{green!50} 139 & \cite{ATLAS:2022ooq} \\[2pt]
\cellcolor{sigstrcolor4} & $\mu_\text{ggF,bbh,VBF}^{WW}$ & $1.09 \pm 0.11$  &&&&&&  &\cellcolor{green!50}&\\[3pt]
\hline
\hline
\cellcolor{sigstrcolor3} & $\mu_\text{VBF}^{\tau \tau}$ & $0.90 \pm 0.18$  &1&$-0.24$&0&0&&  &\cellcolor{green!50} & \multirow{4.4}{*}{\cite{ATLAS:2022yrq}}\\[2pt]
\cellcolor{sigstrcolor3} & $\mu_\text{ggF,bbh}^{\tau \tau}$ & $0.96 \pm 0.31$  &$-0.24$&1&$-0.29$&0&&  &\cellcolor{green!50} & \\[2pt]
\cellcolor{sigstrcolor3} & $\mu_\text{Vh}^{\tau \tau}$ & $0.98 \pm 0.60$  &0&$-0.29$&1&0&&  &\cellcolor{green!50} \multirow{-2}{*}{139}&\\[2pt]
\cellcolor{sigstrcolor3} & $\mu_\text{tth,th}^{\tau \tau}$ & $1.06 \pm 1.18$  &0&0&0&1&&  &\cellcolor{green!50} &\\[3pt]
\hline
\hline
\cellcolor{sigstrcolor2} & $\mu_\text{VBF}^{bb}$ & $0.95 \pm 0.37$  & & & & & & & \cellcolor{green!50} 126 & \cite{ATLAS:2020bhl} \\[2pt]
\cellcolor{sigstrcolor2} & $\mu_\text{Wh}^{bb}$ & $0.95 \pm 0.26$  & & & & & & & \cellcolor{green!50} 139 & \cite{ATLAS:2020fcp} \\[2pt]
\cellcolor{sigstrcolor2} & $\mu_\text{Zh}^{bb}$ & $1.08 \pm 0.24$  & & & & & & & \cellcolor{green!50} 139 & \cite{ATLAS:2020fcp} \\[2pt]
\cellcolor{sigstrcolor2} & $\mu_\text{Vh}^{bb}$ & $1.02 \pm 0.17$  & & & & & & & \cellcolor{green!50} 139 & \cite{ATLAS:2020fcp} \\[2pt]
\cellcolor{sigstrcolor2} & $\mu_\text{tth,th}^{bb}$ & $0.35 \pm 0.35$  & & & & & & & \cellcolor{green!50} 139 & \cite{ATLAS:2021qou} \\[3pt]
\hline
\hline
\cellcolor{sigstrcolor6} & $\mu_\text{pp}^{\mu\mu}$ & $1.2\pm 0.6$  & & & & & & & \cellcolor{green!50} 139 & \cite{ATLAS:2020fzp} \\[3pt]
\hline
\hline
\cellcolor{yellow} & $\mu_\text{pp}^{Z\gamma}$ & $2.0 \pm 0.95$  & & & & & & & \cellcolor{green!50} 139 & \cite{ATLAS:2020qcv} \\[3pt]
\hline
\end{tabular}}
\caption{Latest Run 2 data on $h$ signal strengths measured by ATLAS at $\sqrt{s}=13$ TeV. Correlations below $0.1$ are treated as zero. The colors in the first column represent the corresponding decay channels in Figure~\ref{fig:5}.}
\label{tab:signalstrengthsA13}
\end{center}
\end{table}
\subsection{$h$ signal strengths}
For a given process of producing the SM-like Higgs $h$ via the channel $i$ and decay to final state $f$, the signal strength for the production ($\mu_i$) and for the decay ($\mu_f$)  are given by,
\begin{equation}
\nonumber
\mu_i=\frac{\sigma_i}{(\sigma_i)_{\textrm{SM}}}\,, \quad \textrm{and}\quad \mu_f =\frac{\mathcal{B}(h\to f)}{\mathcal{B}_{\textrm{SM}}(h\to f)}\,,
\label{signal_strength_production_decay}
\end{equation}
where $\sigma_i$'s are the production cross sections via $i^{\mathrm{th}}$ channel where $i\in\{gg$F$,\, bbh,\, $VBF$,\, Wh,\,$ $Zh,\, t\bar{t}h, \,th\}$, and $ \mathcal{B}(h\to f)$'s are the decay branching fractions for the final state $f$ where $f \in\{ ZZ,\, WW,\, \gamma\gamma,\, Z\gamma,\, \mu\mu,\, bb,\,\tau\tau \}$. Signal strength for the combined process with production channel ($i$) and the decay mode ($f$) of $h$ can be defined as,
\begin{equation}
\nonumber
\mu_i^f\equiv\mu_i\cdot \mu_f= r_i\cdot \frac{r_f}{\displaystyle\sum_{\substack{f^\prime\in\{ ZZ, WW, Z\gamma,\\\gamma\gamma,  \mu\mu, bb,\tau\tau \}}}r_{f^\prime}\cdot \mathcal{B}_{\textrm{SM}}(h\to f^\prime)}\,,
\label{signal_strength_combined}
\end{equation}
where $r_i$ and $r_f$ are the ratios of the $\sigma_i$'s and the partial decay width $\Gamma^f$'s with respect to their corresponding SM values. 
\begin{table}[ht]
\begin{center}
\setlength{\tabcolsep}{0pt}
\renewcommand{\arraystretch}{1.3}
\scalebox{0.85}{
\begin{tabular}{| l | c | c | cccc | c | c |}
\hline
\;\;\;& \textbf{Signal} & \;\textbf{Value}\; & \multicolumn{4}{c|}{\;\textbf{Correlation matrix}\;\;} & \;${\cal L}$\; & \;\textbf{Source}\;\\
& \textbf{strength} &  & \multicolumn{4}{c|}{} &\;\textbf{[fb$^{-1}$]}\; & \\[3pt]
\hline
\hline
\cellcolor{sigstrcolor1} & $\mu_\text{ggh,bbh}^{\gamma \gamma}$ &\; $1.07 \pm 0.11$ \; &&&& &\cellcolor{green!50}& \multirow{5}{*}{\cite{CMS:2021kom}} \\[2pt]
\cellcolor{sigstrcolor1} & $\mu_\text{VBF}^{\gamma \gamma}$ & $1.04 \pm 0.32$  &&&& &\cellcolor{green!50}&\\[2pt]
\cellcolor{sigstrcolor1} & $\mu_\text{Vh}^{\gamma \gamma}$ & $1.34 \pm 0.34$  &&&& &\cellcolor{green!50} \multirow{1}{*}{\parbox{23pt}{137\\ \phantom{d}}}&\\[2pt]
\cellcolor{sigstrcolor1} & $\mu_\text{tth,th}^{\gamma \gamma}$ & $1.35 \pm 0.31$  &&&& &\cellcolor{green!50}&\\[3pt]
\hline
\hline
\cellcolor{sigstrcolor5} & $\;\mu_\text{ggh,bbh,tth,th}^{ZZ}$\; & $0.95 \pm 0.13$  &1&$-0.11$&&&  \cellcolor{green!50} & \multirow{2.4}{*}{\cite{CMS:2021ugl}} \\[2pt]
\cellcolor{sigstrcolor5} & $\mu_\text{VBF,Vh}^{ZZ}$ & $0.82 \pm 0.34$  &$-0.11$&1&& &\cellcolor{green!50} \multirow{1}{*}{\parbox{23pt}{137\\ \phantom{d}}}&\\[3pt]
\hline
\hline
\cellcolor{sigstrcolor4} & $\mu_\text{ggh}^{WW}$ & $0.92 \pm 0.11$  &1 &\;$-0.13$ \;\;&0 & 0 & \cellcolor{green!50}& \multirow{4.5}{*}{\cite{CMS:2022uhn}}\\[2pt]
\cellcolor{sigstrcolor4} & $\mu_\text{VBF}^{WW}$ & $0.71 \pm 0.26$  &\;$-0.13$ &1 & 0&0 &  \cellcolor{green!50}&  \\[2pt]
\cellcolor{sigstrcolor4} & $\mu_\text{Zh}^{WW}$ & $2.0\pm 0.7$  &0&0&1&0  &\cellcolor{green!50}&\\[2pt]
\cellcolor{sigstrcolor4} & $\mu_\text{Wh}^{WW}$ & $2.2\pm 0.6$  &0&0&0&1  &\cellcolor{green!50} \multirow{-3}{*}{\parbox{23pt}{138\\ \phantom{d}}}&\\[3pt]
\hline
\hline
\cellcolor{sigstrcolor3} & $\mu_\text{incl.}^{\tau \tau}$ & $0.93 \pm 0.12$  &&&&  &\cellcolor{green!50} & \multirow{4.5}{*}{\cite{CMS:2022kdi}}\\[2pt]
\cellcolor{sigstrcolor3} & $\mu_\text{ggh}^{\tau \tau}$ & $0.97 \pm 0.19$  &&&&  &\cellcolor{green!50} \cellcolor{green!50}&  \\[2pt]
\cellcolor{sigstrcolor3} & $\mu_\text{qqh}^{\tau \tau}$ & $0.68 \pm 0.23$  &&&&  &\cellcolor{green!50} &\\[2pt]
\cellcolor{sigstrcolor3} & $\mu_\text{Vh}^{\tau \tau}$ & $1.80 \pm 0.44$  &&&&  &\cellcolor{green!50}  \multirow{-3}{*}{\parbox{23pt}{138\\ \phantom{d}}}&\\[3pt]
\hline
\hline
\cellcolor{sigstrcolor2} & $\mu_\text{qqh}^{bb}$ & $1.59 \pm 0.60$  &1 &$-0.75$ & & &  \cellcolor{red!50}  & \multirow{2.4}{*}{\,\cite{CMS:2023tfj} }\\[2pt]
\cellcolor{sigstrcolor2} & $\mu_\text{ggh}^{bb}$ & $-2.7 \pm 3.89$  &$-0.75$ &1 & & &  \cellcolor{red!50} \multirow{1}{*}{\parbox{23pt}{90.8\\ \phantom{d}}} & \\[3pt]
\hline
\hline
\cellcolor{sigstrcolor6} & $\mu_\text{ggh,tth}^{\mu\mu}$ & $0.66 \pm 0.67$  & 1&$-0.24$ & & &  \cellcolor{green!50} &\multirow{2.4}{*}{\cite{CMS:2020xwi}} \\[2pt]
\cellcolor{sigstrcolor6} & $\mu_\text{VBF,Vh}^{\mu\mu}$ & $1.85 \pm 0.86$  &$-0.24$ &1 & & &  \cellcolor{green!50} \multirow{-2}{*}{137}&\\[3pt]
\hline
\hline
\cellcolor{yellow} & $\mu_\text{pp}^{Z\gamma}$ & $2.4 \pm 0.9$  & & & & &  \cellcolor{green!50} 138 & \cite{CMS:2022ahq} \\[3pt]
\hline
\end{tabular}}
\caption{Latest Run 2 data on $h$ signal strengths measured by CMS at $\sqrt{s}=13$ TeV. Correlations below $0.1$ are treated as zero. The colors in the first column represent the corresponding decay channels in Figure~\ref{fig:5}.}
\label{tab:signalstrengthsA14}
\end{center}
\end{table}
Therefore, the signal strength for a particular decay mode implicitly depends on all the other $h$-decay modes. In the $\kappa$-framework~\cite{ATLAS:2016neq}, the modifiers for SM $h$ coupling to vector bosons and fermions at the tree-level in both the GM and eGM models are given by,\footnote{Note that, the definitions of $s_\beta$ and $c_\beta$ are opposite to that given in Ref.~\cite{Chiang:2015amq} because our definition of $\tan\beta$ is inversely related to their definitions.}
\begin{equation}\label{eq:kappa}
\kappa_V=c_\alpha c_\beta -\sqrt{\frac{8}{3}}s_\alpha s_\beta\,,\quad \textrm{and} \quad \kappa_f=\frac{c_\alpha}{c_\beta}\, .
\end{equation}
At the leading order (LO), the $h$ decays into $\gamma\gamma$ and $Z\gamma$ channels are mediated via exotic charged Higgs bosons ($F^{++},F^+,H^+$) at the one-loop.  Due to the non-degenerate masses in the eGM model, both $F^{++}$ and $F^+$ contribute separately to the above loop-mediated decay modes. The general expressions of $r_{\gamma\gamma}$ and $r_{Z\gamma}$ involving the SM constituents and the exotic charged particles are given in~\cite{Gunion:1989we}. The experimental input values of $h$ signal strengths from the latest ATLAS and CMS Run 2 data at a center-of-mass energy of $13$ TeV are displayed in Table~\ref{tab:signalstrengthsA13} and \ref{tab:signalstrengthsA14}, respectively. ATLAS and CMS combination for Run 1 data on $h$ signal strengths measurements are given in Table 2 of~\cite{Chowdhury:2017aav}. 

\section{Results}\label{sec:results}
In this section, we present the allowed parameter space considering the above-mentioned theoretical constraints and the $95.4\%$ probability regions from the combined fits to the latest Run 2 LHC results on the $h$ signal strengths for the GM model. For the eGM model, we use the same sampling procedure as in the GM model. We see that, in some of the mass planes, $95.4\%$ probability limits in the eGM model appear stronger than expected, even compared to the corresponding parameter spaces of the GM model. The statistical fluctuations of the $95.4\%$ probability limits in some of the mass planes in the eGM model strongly depend on the number of iterations and the sampling procedure, while the effect on the quartic coupling planes is marginal.\footnote{It is important to mention that once we reduce the number of free parameters varied in the MCMC run, the statistical fluctuations of the $95.4\%$ probability limits in the mass planes remain marginal. For instance, fixing one of the BSM Higgs bosons near 95 GeV helps mitigate this issue, as demonstrated in Ref.~\cite{Mondal:2025tzi}.} This is because the NLO unitarity bounds quadratically depend on quartic couplings, and these couplings non-linearly depend on the mass parameters. In the eGM model, due to the presence of a large number of quartic couplings and the mass parameters that are varied in the MCMC run, small statistical fluctuations in the quartic coupling planes lead to relatively large fluctuations in some of the mass planes. Therefore, we choose a conservative approach and use both mass and mass-squared priors in the global fits with $\sim 10^9$ iterations and present the combined $99.7\%$ probability contours for the eGM model. As a result, the allowed parameter regions given in this paper become less dependent on statistical fluctuations. For the theoretical constraints, we assume flat likelihoods, and the allowed regions represent 100\% probability contours. 

\begin{figure}[!h]
     \centering
             \includegraphics[clip=true,width=0.8\columnwidth]{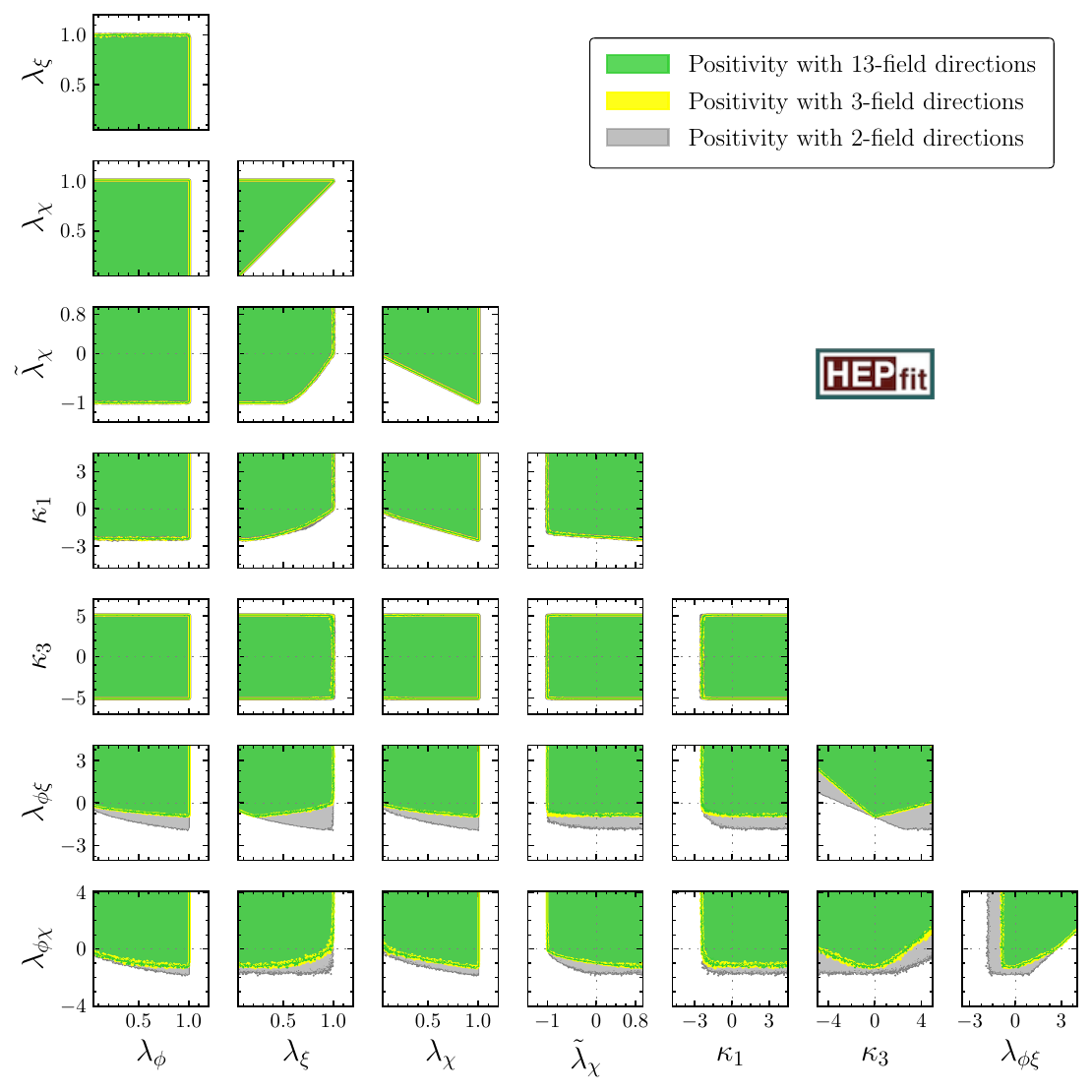}
     \caption{Allowed regions on the quartic coupling planes of the eGM model with different BFB constraints. The grey (yellow) regions show the boundedness of the potential in the 2-field (3-field) directions, while the green areas are allowed if we use BFB conditions considering all thirteen non-zero scalar field directions.}
     \label{fig:10}
\end{figure}

In Figure~\ref{fig:10}, we compare three different bounded-from-below (BFB) constraints on the quartic coupling planes of the eGM model. The grey (yellow) regions manifest allowed parameter space that satisfies the BFB conditions with all possible combinations of any two (three) non-vanishing field directions. The green areas show the allowed parameter space if we impose the BFB conditions, considering all thirteen fields have non-zero values. From Figure~\ref{fig:10}, we see that the 3-field BFB constraints largely overlap with the 13-field BFB constraints in the quartic coupling planes of the eGM model. 

\begin{figure}[!h]
     \centering
             \includegraphics[clip=true,width=0.8\columnwidth]{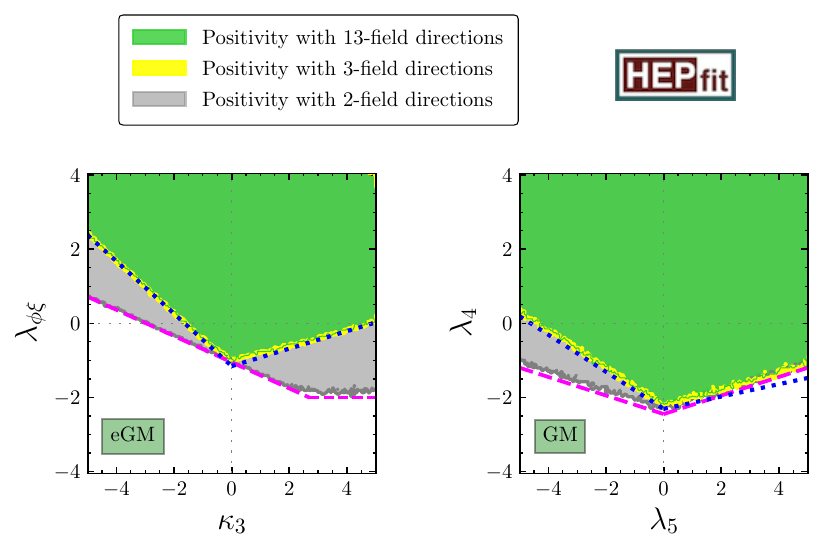}
     \caption{Comparison of different BFB conditions in the  $\kappa_3$ vs.~$\lambda_{\phi\xi}$ plane (left panel) of the eGM model. The magenta dashed boundary lines originated from the BFB conditions with all possible two non-vanishing field directions at once, while the blue dotted boundary lines originated from the BFB conditions with at least three scalar fields to be non-zero. The grey, yellow, and green colors have the same meaning as in Figure~\ref{fig:10}. The right panel shows the corresponding constraints in the $\lambda_5$ vs.~$\lambda_{4}$ plane of the GM model.}
\label{fig:1}
\end{figure}

In Figure~\ref{fig:1}, we show the boundary lines coming from 2-field and 3-field BFB conditions in the $\kappa_3$ vs.~$\lambda_{\phi\xi}$ plane (left panel) of the eGM model. The magenta dashed boundary lines depict the 2-field BFB conditions, $4\ld_{\phi\xi}+\sqrt{2}\kappa_3>-4\sqrt{\ld_\phi\ld_\chi}$ and $\ld_{\phi\xi}>-2\sqrt{\ld_\phi\ld_\xi}$. Blue dotted lines represent the boundary lines coming from one of the necessary conditions with three non-vanishing field directions given in Eq.~(\ref{eq:cust-bfb}).
 The corresponding boundary lines for the GM model are shown in the right panel of Figure~\ref{fig:1}, where the magenta dashed boundary lines represent the 2-field BFB conditions for the GM model, $ 4\ld_4-|\ld_5|>-4\sqrt{2\ld_1(2\ld_2+\ld_3)}$~\cite{Chiang:2012cn}, and the blue dotted lines represent the 3-field BFB conditions given in Eq.~(\ref{eq:cust-bfb-GM}). For completeness, we show the constraints coming from all the 2-field, 3-field, and 13-field BFB conditions in all other quartic coupling planes of the GM model in Figure~\ref{fig:14} of Appendix~\ref{app:supp_figs}. 
Note that these BFB conditions are imposed on the tree-level potential, and all the quartic couplings are interpreted at the scale of $M_Z$.\footnote{To compare the three different BFB conditions in Figures~\ref{fig:10}, \ref{fig:1}, and \ref{fig:14}, we sample the parameter space of the quartic couplings in the following ranges: (i) $\ld_\phi,\ld_\xi,\ld_\chi,\tilde{\ld}_\chi\in[-1,1]$ and $\ld_{\phi\xi},\ld_{\phi\chi},\kappa_1,\kappa_3\in[-5,5]$ for the eGM model, (ii) $\ld_1,\ld_2,\ld_3\in[-1,1]$ and $\ld_{4},\ld_{5}\in[-5,5]$ for the GM model.} 

To quantify the accuracy of the 3-field BFB approximation, we define the ratio
\begin{equation}
    f_1\equiv\frac{A_{3}-A_{13}}{A_{13}}\,,
\end{equation}
where $A_{3}$ and $A_{13}$ denote the area of the region in the quartic-coupling plane(s) allowed by the 3-field BFB conditions and all 13-field BFB conditions, respectively (see Figures~\ref{fig:10} and~\ref{fig:14}). Considering all possible quartic coupling planes, we find that the maximum discrepancy between the 3-field and all 13-field BFB conditions, quantified by the non-overlapping fraction of the parameter space, is only about $1.4\%$ in the eGM model and less than $1\%$ in the GM model. Thus, the BFB conditions with all possible 3-field directions provide a very good approximation of the full 13-field BFB conditions~\cite{Hartling:2014zca,Moultaka2020} for both models. Since the latter conditions depend on more than one field-dependent parameter (see Eq.~(\ref{eq:bfb13f-eGM}) for the eGM model), the computational cost is significantly high for this case compared to the 3-field BFB case. Therefore, in the rest of our analysis, we shall consider the BFB conditions with all possible 3-field directions to reduce computational cost. Note that these BFB conditions serve as a conservative criterion for vacuum stability, thereby providing a conservative estimate of the allowed parameter space. Inclusion of metastability analysis may reshape the parameter space allowed by the BFB conditions alone. A more accurate determination of the viable parameter space, including long-lived metastable vacua, would require a comprehensive analysis of the vacuum structure and the corresponding vacuum tunneling rates. Such an analysis is computationally challenging and lies beyond the scope of the present work.
\begin{figure}[t]
     \centering
             \includegraphics[clip=true,width=0.8\columnwidth]{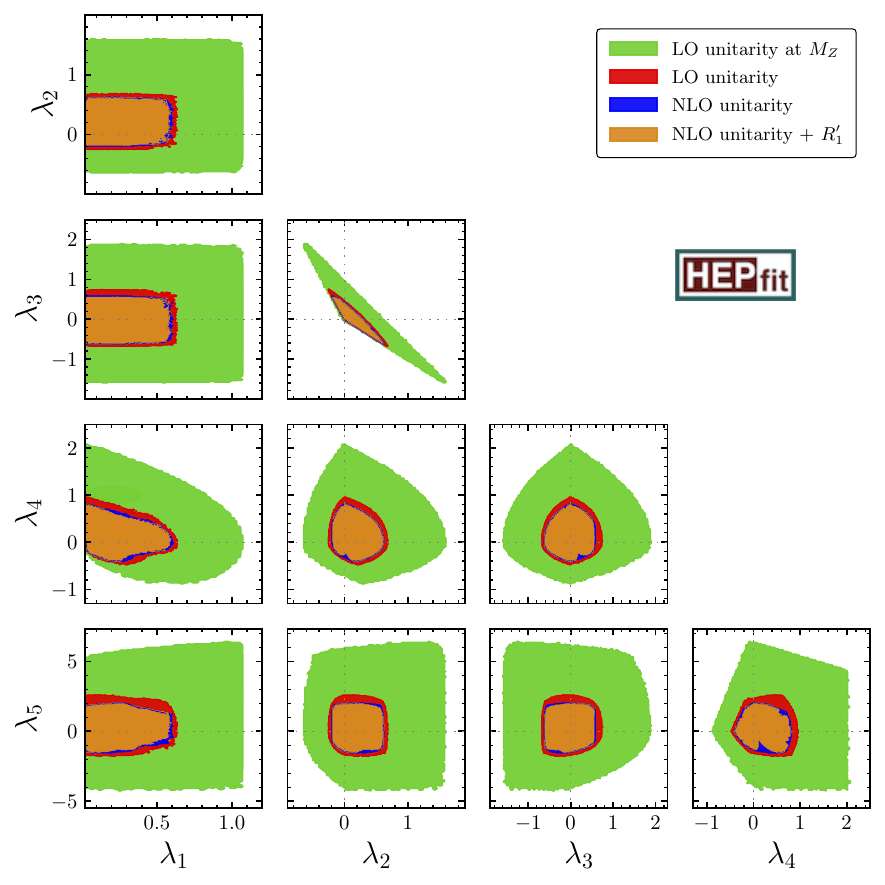}
     \caption{Allowed regions in the $\lambda_i$ vs.~$\lambda_j$ planes of the GM model.  The red (blue) colored areas show the allowed parameter space if we use  LO (NLO) unitarity bounds. The brown regions are obtained with the additional condition that the NLO value should be smaller than the LO value in magnitude (if the LO value is not accidentally small), and the green areas are allowed if we impose the unitarity conditions at the scale $M_Z$. All the $\lambda_i$ values are given at the $M_Z$ scale. Note that all these allowed regions satisfy the 3-field BFB conditions on the potential.} 
\label{fig:2}
\end{figure}

\begin{figure}[!h]
     \centering
             \includegraphics[clip=true,width=0.8\columnwidth]{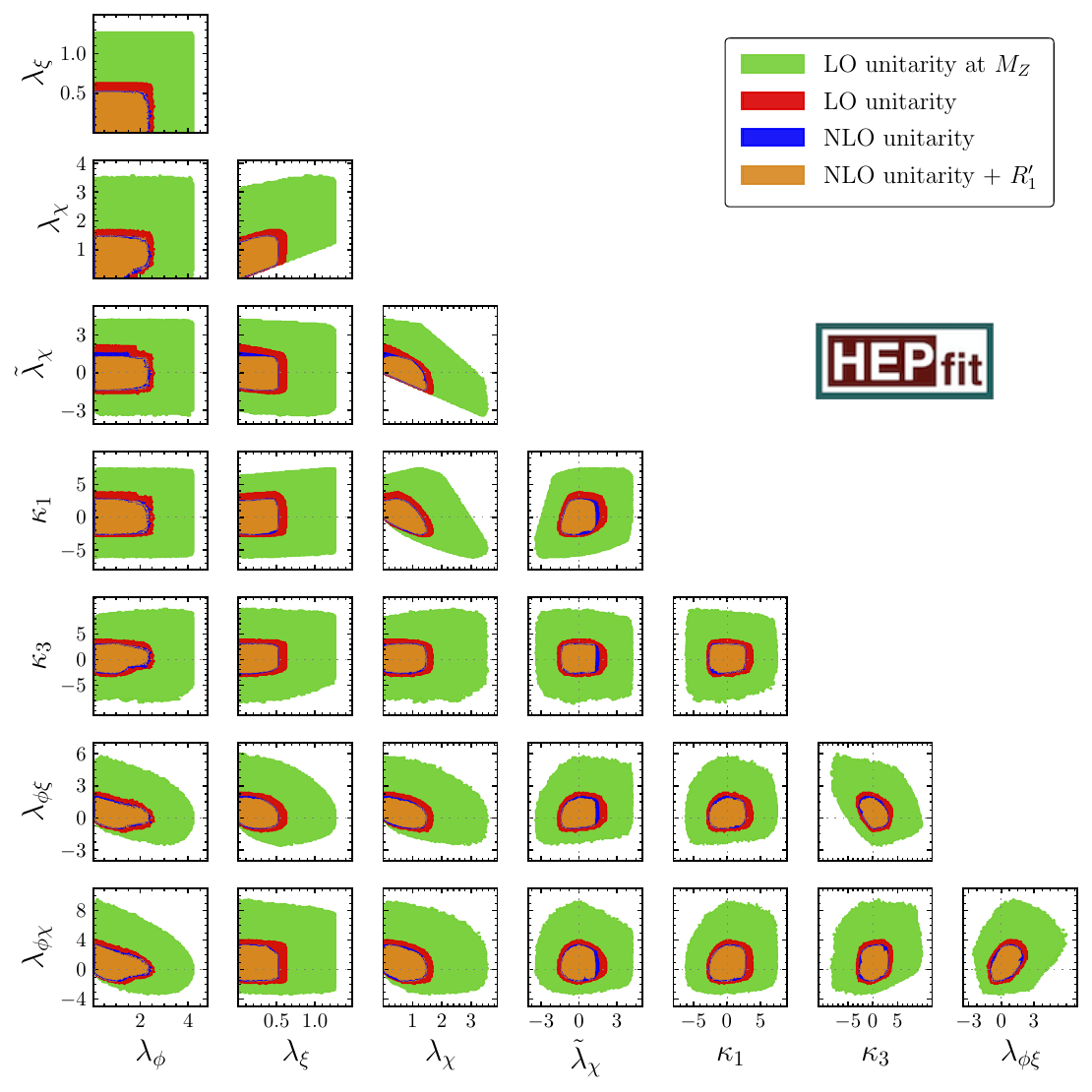}
     \caption{Allowed regions in the quartic coupling planes of the eGM model. The colors of the shaded regions have the same meaning here as in Figure~\ref{fig:2}. All the allowed regions satisfy the 3-field BFB conditions on the potential. }
     \label{fig:3}
\end{figure}

Next, we compare different unitarity constraints on the parameter spaces for the GM and eGM models. In Figure~\ref{fig:2} (Figure~\ref{fig:3}), we show the allowed regions, at the $M_Z$ scale, in the quartic coupling planes of the GM (eGM) model with LO, NLO unitarity, and NLO unitarity with $R_1^{\prime}$ (see Eq.~(\ref{eq:R1pert})), in the colors red, blue, and brown, respectively. All these regions satisfy the 3-field BFB conditions on the potential. Unless stated, the aforementioned unitarity bounds are imposed on the two-loop renormalization group improved quartic scalar couplings at the scale 1 TeV. It is distinctly evident from Figures~\ref{fig:2} and \ref{fig:3} that the NLO unitarity constraint sets substantially stringent bounds on the parameter space compared to the LO unitarity. The allowed regions get further shrunk when we include the perturbativity NLO unitarity condition ($R_1^\prime < 1$). In some of the panels in Figures~\ref{fig:2} and \ref{fig:3}, we observe sharp cuts around the origin once the perturbative unitarity is considered. An explanation of such features in the quartic coupling planes is given in Appendix~\ref{app:supp_figs}. If we use LO unitarity conditions at the $M_Z$ scale instead, the parameter space gets much more relaxed. These regions are shown in green color in the same planes of Figures~\ref{fig:2} and \ref{fig:3}. Numerically, the quartic couplings for the GM (eGM) model can have a maximum absolute value as large as  2.7, 2.2, 2.2 (4.1, 3.1, 3.1) if we apply LO unitarity, NLO unitarity, $R_1^\prime$-perturbative NLO unitarity, respectively. These values are obtained from the Figures~\ref{fig:2} and \ref{fig:3}.

\begin{figure}[t]
     \centering
             \includegraphics[clip=true,width=0.8\columnwidth]{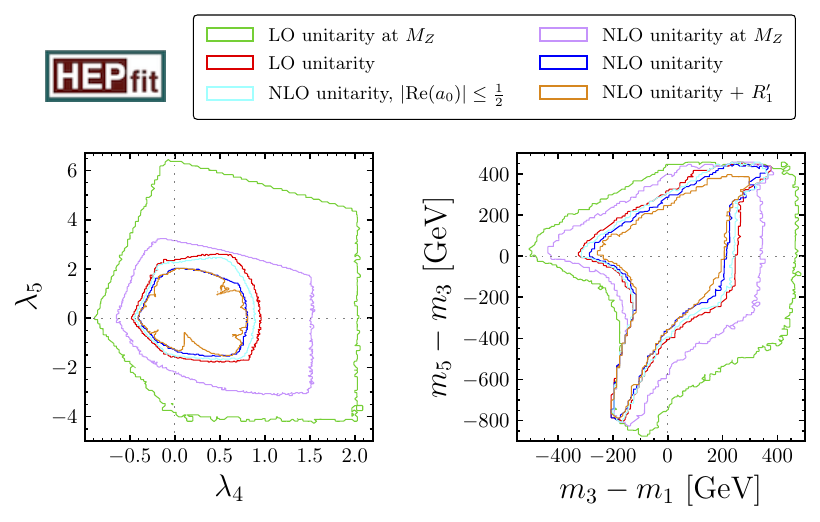}
     \caption{Comparison among different unitarity constraints in the $\lambda_4$ vs.~$\lambda_5$ plane (left) and in the $m_3-m_1$ vs.~$m_5-m_3$ plane (right) of GM model. The green, red, blue, and brown contours have the same meaning as in Figure~\ref{fig:2}. The cyan contour shows the allowed regions if we consider only real parts of the $S$-matrix eigenvalues while imposing unitarity bounds at one-loop, and the violet contour represents the allowed area if we impose NLO unitarity conditions at the scale $M_Z$. All the allowed regions satisfy the 3-field BFB conditions on the potential.}
     \label{fig:4}
\end{figure}

A comparison among different unitarity conditions for the GM model is displayed on the $\lambda_4$ vs.~$\lambda_5$ plane in the left panel of Figure~\ref{fig:4}. The green, red, blue, and brown solid lines correspond to the contours of the same color as in Figure~\ref{fig:2}.  A weaker bound comes from Eq.~(\ref{eq:partial-wave analysis2}) if we consider only the real part of the NLO unitarity eigenvalues, i.e., $\lvert\text{Re}(a_\ell) \rvert \, \leq 1/2$.  This contour is represented by a cyan line, and as expected, it is always less stringent than the blue contour. The violet contour corresponds to the allowed regions satisfying NLO unitarity conditions at the scale $M_Z$. This yields a stronger constraint than the LO unitarity evaluated at $M_{Z}$. The right panel in Figure~\ref{fig:4} shows how these constraints on the quartic couplings translate into the mass difference planes of the heavy Higgs bosons in the GM model. The different unitarity constraints follow the same ordering as in the left panel. 
\begin{figure}[h!]
     \centering
             \includegraphics[clip=true,width=0.8\columnwidth]{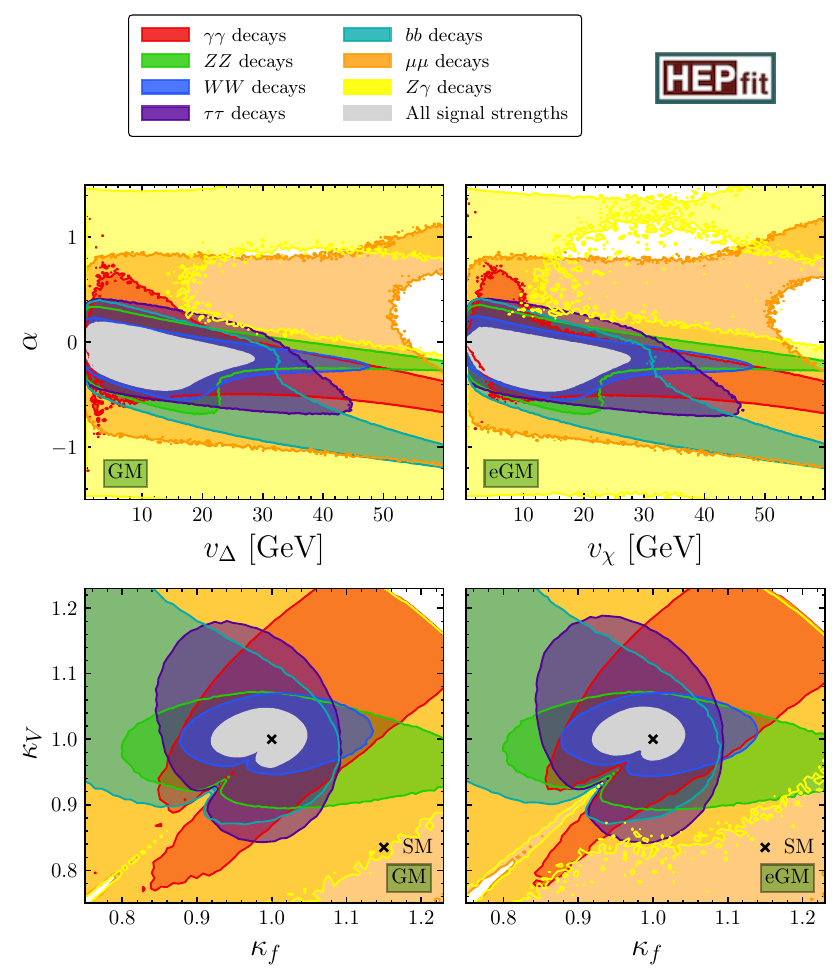}
     \caption{95.4\% probability regions show the impact of the $h$ signal strengths for different final states in the $v_\chi$ vs.~$\alpha$ planes (top) and the $\kappa_f$ vs.~$\kappa_V$ planes (bottom). The individual fits to full Run 1 and Run 2 data from $h$ decays to $\gamma\gamma,\,ZZ,\,WW,\, \tau\tau\,, \,bb,\,\mu\mu\,,$ and $Z\gamma$ in red, green, blue, purple, cyan, orange, and yellow, respectively. The results from the combined fits are shown in the grey color shaded regions, where the cross indicates the expected
value for the SM Higgs boson.}
     \label{fig:5}
\end{figure}

In the top left (top right) panel of Figure~\ref{fig:5}, we display the impacts of the individual signal strengths with different final states on the $v_\Delta$ vs.~$\alpha$ plane ($v_\chi$ vs.~$\alpha$ plane) for the GM (eGM) model. For small $v_\chi$ and the region where $\alpha$ is not close to zero, the $ZZ$, $WW$, and $\gamma\gamma$ signal strengths are most restrictive. Whereas, for large $v_\chi$ values, the most significant constraint around $\alpha=0$ region comes from $h \rightarrow b \bar{b}$ data. Amongst the loop-mediated decays, the $ h \to \gamma \gamma$ channel stringently constrains $v_\chi$ vs.~$\alpha$ plane compared to the $h \to Z \gamma$ channel. The combined fit with all signal strengths favors the negative values of $\alpha$ with $|\alpha|<0.45$, and the triplet VEV $v_\chi\lesssim 30$ GeV at a $95.4\%$ probability for both models. This result shows a significant improvement over the previous global fit reported in Ref.~\cite{Chiang:2018cgb}, where the upper limit of triplet VEV was $\approx 45$ GeV in the GM model. From Figure~\ref{fig:5}, we see that the allowed region from the $\gamma \gamma$ signal strength in the eGM model (red region) is smaller than that in the GM model. The reason is the following: $\gamma \gamma$ signal strength receives contributions from charged scalar loops, which depend on the scalar couplings in the potential. 
The relationships between the GM and eGM couplings are provided in Table 3 of Ref.~\cite{Kundu:2021pcg}. 
Since no theoretical constraints have been imposed here, certain allowed coupling values in the GM model are not accessible in the eGM model, given the chosen prior ranges listed in Tables~\ref{tab:GMprior} and \ref{tab:eGMprior}. Consequently, the parameter space of the eGM model is more constrained than the GM model. This also reflects on the allowed parameter space once all the signal strength channels are combined (grey region). Once the theoretical constraints are considered, the allowed couplings in both models remain well below the chosen prior ranges, eliminating this apparent discrepancy.

In the decoupling limit, $\alpha,v_\chi \to 0$~\cite{Hartling:2014zca}, all additional Higgs bosons become heavy, and the couplings of the SM-like Higgs approach towards its SM values, which translate into $\kappa_V =\kappa_f = 1$ (see Eq.~(\ref{eq:kappa})). Away from the decoupling limit, $\kappa_V$ and $\kappa_f$ are no longer unity at the tree-level. In the bottom left (right) panel of Figure~\ref{fig:5}, we show the allowed ranges in $\kappa_f$ vs.~$\kappa_V$ plane for all different final states for the GM (eGM) model. The SM predictions of $\kappa_f=1$ and $\kappa_V=1$ lie within the allowed regions (grey shaded area) of all signal strengths. In the combined fit, we find that the $h$ signal strength data disfavor the region where $\kappa_V > 1.05$, $\kappa_V < 0.95$, and $\kappa_f>1.05$, $\kappa_f<0.92$ at a $95.4\%$ probability for both models. At leading order, the SM-like Higgs boson $h$ couplings to $\gamma\gamma$ and $Z\gamma$ are modified by $\kappa_f$, $\kappa_V$, and contributions from charged scalars loops~\cite{Gunion:1989we}. To quantify new physics effects on these couplings, we define~\cite{Gunion:1989we}, $r_{VV}\equiv \Gamma(h\to VV)/\Gamma_{\textrm{SM}} (h\to VV)$ with $VV = \{\gamma\gamma, Z\gamma, gg\}$, where $\Gamma(h\to VV)$ denotes the partial decay width of the $h\to VV$ channel. The $95.4\%$ probability contours in the $r_{V\gamma}$ vs.~$r_{gg}$ ($V = \gamma, Z$) planes are shown in Figure~\ref{fig:15} of Appendix~\ref{app:supp_figs}. The maximal deviations of $r_{\gamma\gamma}$ and $r_{Z\gamma}$ from their SM values are approximately 25\% and 10\%, respectively, in both models. 
\begin{figure}[t]
     \centering
             \includegraphics[clip=true,width=0.8\columnwidth]{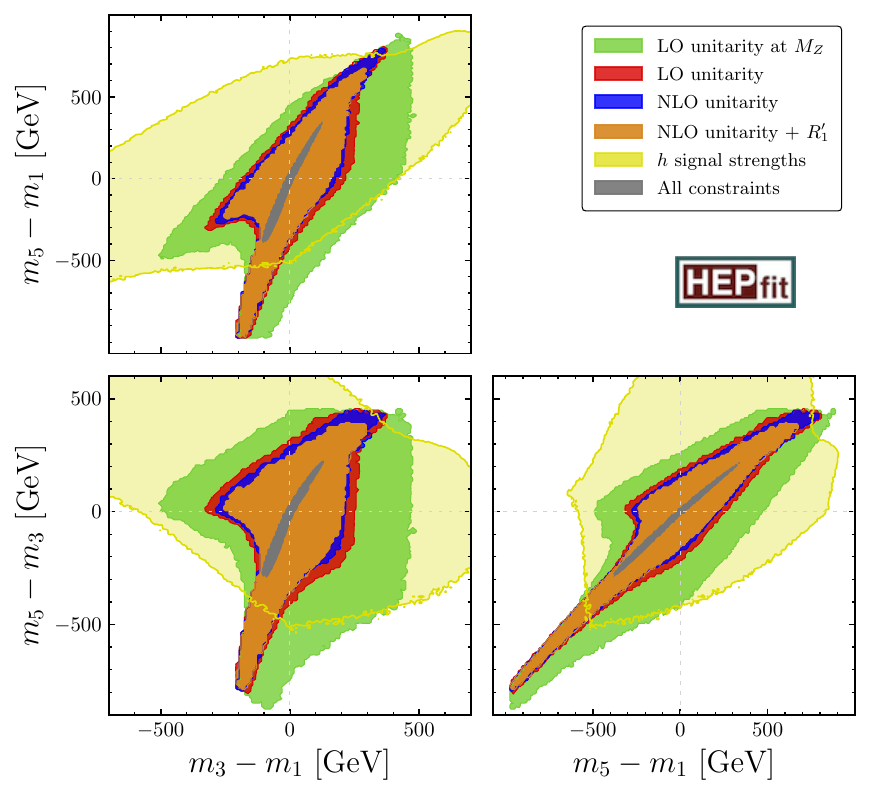}
     \caption{Allowed regions in the mass difference planes of the GM model. The green, red, blue, and brown areas have the same meaning as the regions of the same color in Figure~\ref{fig:2}. The yellow regions are allowed from the $h$ signal strengths measurements with a $95.4\%$ probability. Finally, the grey areas are allowed from the combined fits to the $h$ signal strengths data with the theoretical constraints.}
     \label{fig:6}
\end{figure}
\begin{figure}[t]
     \centering
             \includegraphics[clip=true,width=0.8\columnwidth]{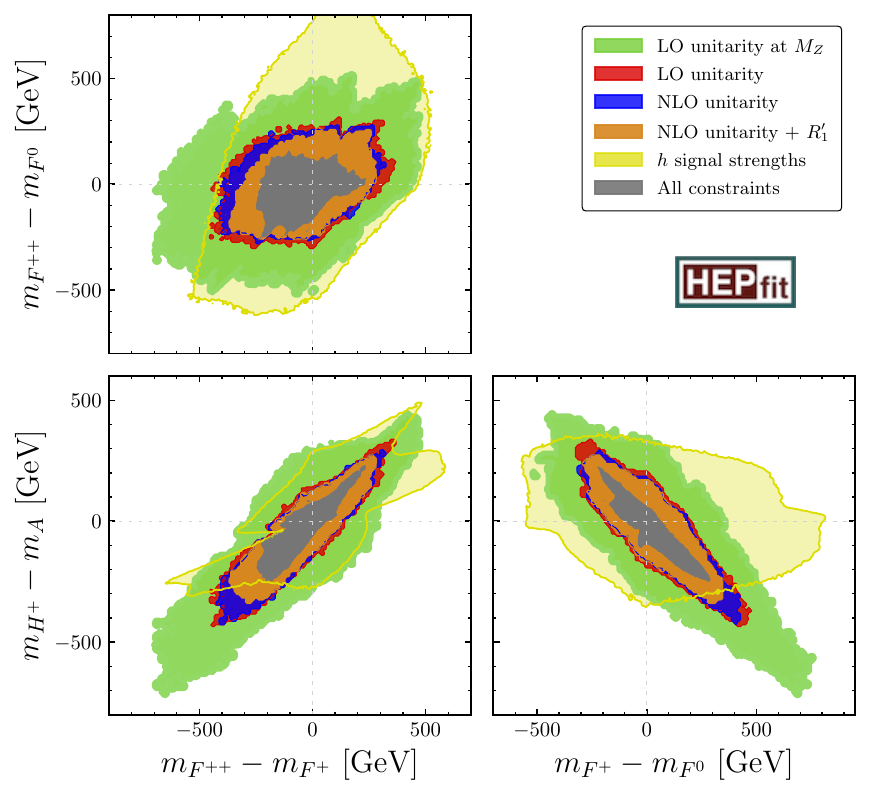}
     \caption{Effects of different constraints on the mass splittings in the eGM model. The shaded regions are allowed from various constraints, and the colors have the same meaning as in Figure~\ref{fig:6}.}
     \label{fig:7}
\end{figure}

In Figure~\ref{fig:6}, we present the effects of different theoretical constraints and the latest $h$ signal strength data on the mass differences of the heavy Higgs bosons in the GM model. The green, red, blue, and brown areas correspond to the same regions as in Figure~\ref{fig:2}. The yellow shaded areas show the allowed region from the $h$ signal strength data at a 95.4\% probability, without imposing any theoretical bounds on the quartic couplings. From Figure~\ref{fig:6}, we find that the latest $h$ signal strength data reduce the allowed mass differences by approximately 100 GeV in the GM model compared to the previous global fit~\citep{Chiang:2018cgb}. Additionally, imposing theoretical constraints on the fit to the $h$ signal strength data significantly constrains the resulting parameter space (grey regions). We see that the combined fit excludes most of the regions in parameter space that are individually allowed by theoretical and experimental constraints. To understand this feature, we perform a fit to the $h$ signal strength data with a naive perturbativity bound on the quartic couplings, i.e., $|\lambda_i| < 4\pi$ for $i=1,...,5$, and we observe that the constraints on $m_3-m_1$ and its correlation with other mass differences severely reduce the allowed regions in the parameter space (see Figure~\ref{fig:16} of Appendix~\ref{app:supp_figs}). These allowed regions are further reduced when we include the perturbative unitarity bounds on the quartic couplings into the fit. Importantly, we find that the NLO unitarity bounds constrain the parameter space more tightly than the LO unitarity bounds, and the mass splitting among three multiplets ($m_1$, $m_3$, and $m_5$) remains within the range of $\Delta m_{ij}=|m_i-m_j|\approx 400$ GeV for $i,j=1,3,5$. Note that the maximal mass splitting among these multiplets is also reduced by approximately 100 GeV compared to the previous global fits~\citep{Chiang:2018cgb,Chen:2022zsh}, and the constraint on $\Delta m_{13}$ is dominated by the theoretical bounds.

One of the distinctive features of the eGM model is the mixing among the custodial triplet and fiveplet scalars. Hence, the scalar mass eigenstates are not necessarily degenerate at the tree-level. In Figure~\ref{fig:7}, we show the effect of all the above-mentioned constraints on the corresponding mass splittings. We find that these mass differences can reach up to 300 GeV in the eGM model. From Figure~\ref{fig:7}, we see that the mass splitting, $m_{F^{++}}-m_{F^0}$, receives the strongest constraint from theory bounds. This is because the mass difference between $m_{F^{++}}$ and $m_{F^0}$ is parameterized by the quartic couplings and VEVs only (see Eq.~(\ref{mass_cp_even}) and Eq.~(\ref{mass_cp_odd_doubly_charged})). However, the mass differences among the other Higgs bosons also depend on the trilinear couplings, $\mu_1$ and $\mu_2$. The effects of different constraints on the other mass differences in the eGM model are shown in Figure~\ref{fig:13} of Appendix~\ref{app:supp_figs}. As in the GM model, $m_A-m_H$ is also strongly constrained from theory bounds in the eGM model and is restricted to be smaller than $140$ GeV in magnitude. 

\begin{figure}[!h]
     \centering
             \includegraphics[clip=true,width=0.9\columnwidth]{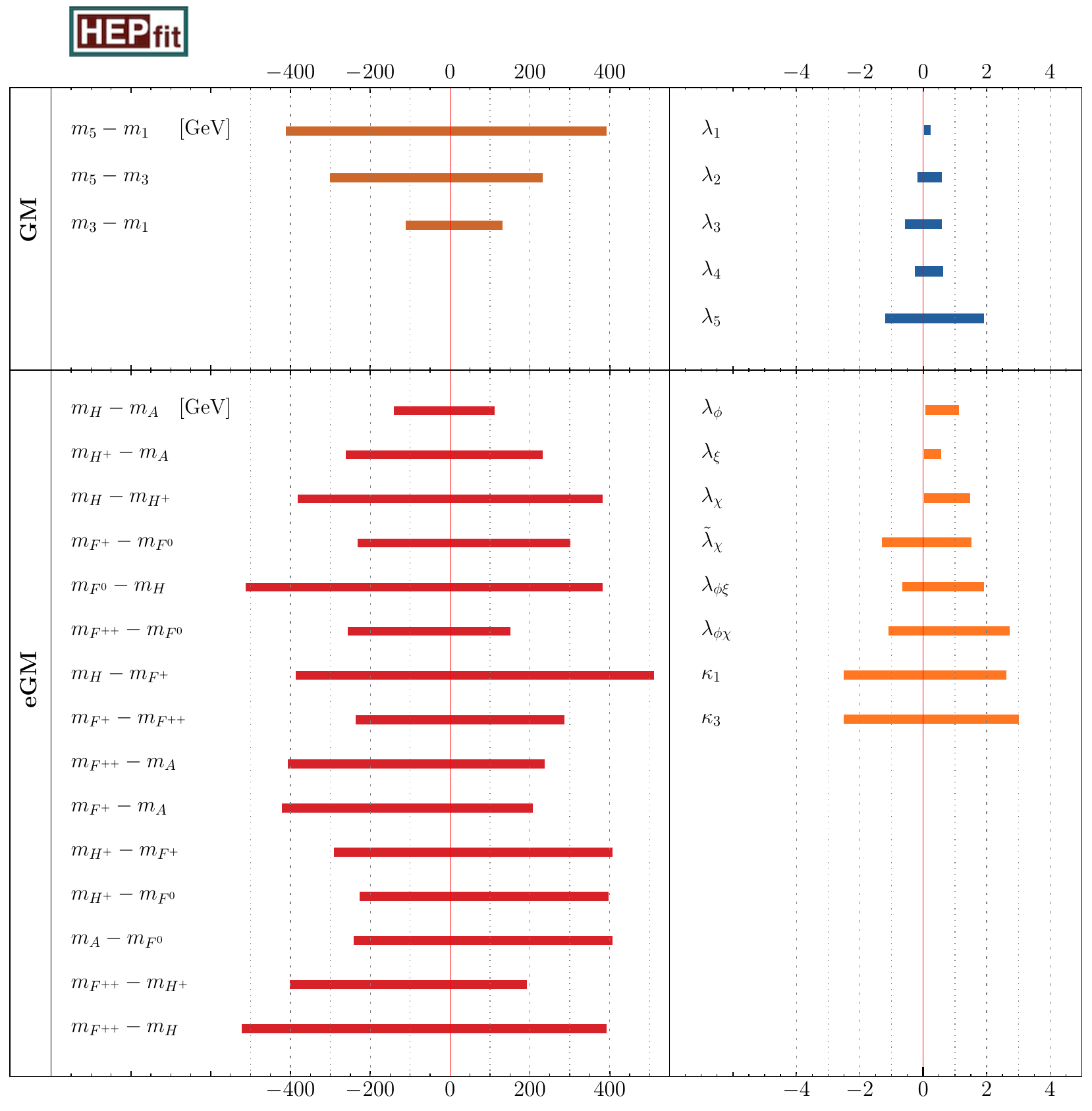}
     \caption{Limits on the mass differences of the heavy Higgs bosons and quartic couplings from the global fit in both the GM and eGM models. The limits on these mass parameters are obtained from the Figures~\ref{fig:6},~\ref{fig:7},~\ref{fig:12} and \ref{fig:13}.}
     \label{fig:8}
\end{figure}

In summary, Figure~\ref{fig:8} shows the limits on the quartic couplings and heavy Higgs mass differences once all constraints are taken into account. The mass differences of the heavy Higgs bosons are restricted to be smaller than 410 GeV and 520 GeV for GM and eGM models, respectively. The maximum magnitude of $\kappa_3$ ($\lambda_5$) can reach $3.0$ ($1.91$) without violating the aforementioned theoretical bounds in the eGM (GM) model.
\section{Conclusions}\label{sec:conclude}
We present the global fits of the GM and eGM models. The global fits combine the latest $h$ signal strength measurements from the $13$ TeV LHC and the improved theoretical constraints, including NLO unitarity and 3-field BFB conditions.

We first derive the tree-level BFB conditions for the generalized two-triplet model by considering all combinations of three non-zero scalar fields, and subsequently obtain the corresponding conditions for the GM and eGM models. We observe that the BFB conditions with all possible 3-field directions are a very good approximation of all 13-field BFB conditions in both the GM and eGM models and significantly reduce computational cost. 

We compute one-loop corrections to the tree-level eigenvalues of $S$-matrix for all $2\to2$ bosonic scatterings in the generalized two-triplet model, and obtain the corresponding results for the GM and eGM models. We find that the NLO unitarity bounds have a significant impact on the allowed parameter space of these models, especially when the stability and unitarity bounds are imposed on the two-loop renormalization group improved quartic couplings. Considering the NLO unitarity and 3-field BFB conditions, we have observed that the allowed values of the quartic couplings are less than 2.2 and 3.1 in magnitude in the GM and eGM models, respectively.

We present the constraints derived for the 125 GeV $h$ boson couplings in the $\kappa$-framework. From the fit results, we see that the $h$ signal strength data disfavor the regions where $\kappa_V > 1.05$, $\kappa_V < 0.95$, and $\kappa_f>1.05$, $\kappa_f<0.92$ at a 95.4\% probability in both the GM and eGM models. In terms of model parameters, we find an upper limit on the triplet VEV and on the mixing angle, $|\alpha|,$ around 30 GeV and 0.45, respectively. At one-loop, the maximal deviation of $r_{\gamma\gamma}$ and $r_{Z\gamma}$ from their SM values is roughly 25\% and 10\% in both models, respectively.

Combining all theoretical constraints and the latest $h$ signal strength data, we find the following allowed ranges of the model parameters:  (i) values of the quartic couplings to be less than 1.91 (3.0) in magnitude in the GM (eGM) model and (ii) the mass differences between the heavy Higgs bosons are restricted to be smaller than 410 GeV and 520 GeV in the GM and eGM models, respectively.

We find that the improved theoretical bounds reduce the maximal allowed mass difference among the heavy Higgs bosons by approximately 100 GeV relative to earlier results for the GM model~\citep{Chiang:2018cgb,Chen:2022zsh}. One of the striking features of the eGM model is that the scalar mass eigenstates are not necessarily degenerate at the
tree-level. We find that the global fit permits these mass splittings to be less than 300 GeV. Our results would impact search strategies for heavy Higgs bosons at the LHC and other future colliders; however, a detailed collider analysis is beyond the scope of this paper. In a subsequent paper~\cite{Mondal:2025tzi}, we have presented the global fits to the most important flavor observables, along with constraints from the latest direct searches for exotic Higgs bosons in these models. 

Additionally, the EWPO can be used to further constrain these models. Among the EWPO, the Peskin-Takeuchi $T$ parameter is ultraviolet divergent~\cite{Gunion:1990dt,Drauksas:2023isd,Chowdhury:2026rus}, whereas the $S$ and $U$ parameters are ultraviolet finite in both models~\cite{Hartling:2014aga,Drauksas:2023isd}. However, as pointed out in Refs.~\cite{Gunion:1990dt,Drauksas:2023isd,Chowdhury:2026rus}, a global electroweak fit with four input parameters is required to derive constraints on the model parameters from these observables. Also, $S$ and $U$ are correlated with $T$ within the existing three-input-parameter global electroweak fit~\cite{deBlas:2016ojx, ParticleDataGroup:2022pth}. Hence, a  comprehensive global analysis of these models, including all the electroweak precision observables, lies beyond the scope of the present work and will be carried out in a subsequent publication.

\begin{acknowledgments}
We thank Joydeep~Chakrabortty, Sanmay~Ganguly, Anirban~Kundu, Swagata~Mukherjee, Christopher~W.~Murphy, Ulrich Nierste, and Sudhir~K.~Vempati  for useful discussions. We are indebted to Gilbert~Moultaka and Palash~B.~Pal for their valuable comments on the manuscript. We are grateful to Ayan~Paul and Luca~Silvestrini for numerous discussions regarding the \texttt{HEPfit} code. This work is supported by the SERB (Govt.~of India) under grant SERB/CRG/2021/007579.  
SS acknowledges funding from the MHRD (Govt.~of India) under the Prime Minister’s Research Fellows (PMRF) Scheme, 2023. PM and DC acknowledge funding from the SERB (Govt.~of India) under grant SERB/CRG/2021/007579. DC also acknowledges funding from an initiation grant
IITK/PHY/2019413 at IIT Kanpur.
\end{acknowledgments}

\appendix
\section{Quartic couplings from physical masses}\label{app:inverse_mass}
In this appendix, we provide the expressions of the quartic couplings ($\lambda_i,\kappa_i$) in terms of the physical masses for the eGM model.
\begin{align}
\ld_\phi&=\frac{1}{2v^2c_\beta^2}\Big(m_h^2c_\alpha^2+m_H^2 s_\alpha^2\Big),\notag\\
\ld_\xi&=\frac{1}{3v^2s_\beta^2}\Big(m_h^2s_\alpha^2+m_H^2c_\alpha^2+2m_{F^{0}}^2+3\Big(M_2^2-2m_A^2\Big)c_\beta^2-\frac{3}{2}M_1^2\Big),\notag\\
\ld_\chi&=\frac{2}{3v^2s_\beta^2}\Big(2m_h^2s_\alpha^2+2m_H^2c_\alpha^2+m_{F^0}^2-3m_A^2c_{\beta}^2\Big),\notag\\
\tilde{\ld}_\chi&=\frac{2}{v^2s_\beta^{2}}\left[m_{F^{++}}^2-M_1^2+c_\beta^2\Big(2M_2^2-3m_A^2\Big)\right]\notag\\
&\quad\qquad \qquad \qquad  +\frac{2c_\beta^2}{v^2s_\beta^{2}(c_{2\delta}-c_\beta s_{2\delta})}\left[2m_{F^+}^2-2m_{H^+}^2+c_{2\delta} \Big(4m_A^2-2m_{F^+}^2-2m_{H^+}^2\Big)\right],\notag\\
\ld_{\phi\xi}&=\frac{1}{v^2s_\beta c_\beta}\Big(s_\beta c_\beta \Big(2m_A^2-M_2^2\Big)+\sqrt{\frac{2}{3}}\Big(M_H^2-m_h^2\Big)c_\alpha s_\alpha\Big),\notag\\
\ld_{\phi\chi}&=\frac{2}{v^2}\left[m_{A}^2+\sqrt{\frac{2}{3}}\frac{s_{2\alpha}}{s_{2\beta}}\Big(m_H^2-m_h^2\Big)\right]-\frac{2}{v^2(c_{2\delta}-c_\beta s_{2\delta})}\Big[ 2s_\delta^2m_{F^+}^2-2c_\delta^2m_{H^+}^2\notag\\
&\quad\qquad \qquad \qquad  +\Big(M_2^2+m_A^2\Big)c_{2\delta}-\Big(M_2^2-m_A^2\Big)c_\beta s_{2\delta}\Big],\notag\\
\kappa_1 &= \frac{4}{v^2s_\beta^2(c_{2\delta}-c_\beta s_{2\delta})}\left[-M_1^2c_{2\delta}+\Big(c_{\delta}^2m_{F^+}^2-s_\delta^2m_{H^+}^2\Big)s_\beta^2+2\Big(M_2^2-m_A^2\Big)c_\beta^2 c_{2\delta}\Big) \right.\notag\\
&\qquad \left.+\Big(m_{F^+}^2-m_{H^+}^2\Big)c_\beta^2 +c_\beta s_{2\delta}\Big(M_1^2+m_A^2-m_{F^+}^2-m_{H^+}^2-\Big(2M_2^2-3m_A^2\Big)c_\beta^2\Big)\right],\notag\\
\kappa_3&=  \frac{2\sqrt{2}}{v^2}\Big(M_2^2-m_A^2\Big),
\end{align}
where we have defined,
\begin{equation}
M_1^2=\frac{\mu_1s_\beta v}{\sqrt{2}}\,,\quad M_2^2= \frac{\mu_2v}{\sqrt{2}s_\beta}\,.
\end{equation}
These eight quartic couplings can be reduced to five $\lambda_i\,(i=1,2,3,4,5)$\,, in the limit $\delta=0,\;m_A=m_{H^+},$ and $m_{F^0}=m_{F^+}=m_{F^{++}}$, which are consistent  with the relations given in~\cite{Chiang:2012cn}. 
\section{Bounded-from-below constraints in the field subspace}\label{app:pos_subspace}
Here, we consider the subspaces of the potential where only two or three fields are non-vanishing at once. At large field values, only quartic terms in the potential dominate as shown in Eq.~(\ref{v16_quartic}). 
In the field space, first we consider a field direction where only electrically neutral components, $\chi^0$ and $\xi^0$ are non-vanishing. The corresponding potential reads,
\eqn{2f_0}{V^{(4)}=\ld_\chi|\chi^0|^4+\ld_{\chi\xi}|\chi^0|^2(\xi^0)^2+\ld_\xi(\xi^0)^4\,.}
Due to the bi-quadratic form of the potential in Eq.~(\ref{eq:2f_0}), the necessary and sufficient BFB conditions in this direction are nothing but the criteria for strict copositivity of the potential~\cite{Hadeler1983}, given by, 
\eqn{2f_0_con}{\ld_\chi>0\;\land\;\ld_\xi>0\;\land\;\ld_{\chi\xi}+2\sqrt{\ld_\chi\ld_\xi}>0\,.}
Similarly, if we consider the field direction where only $\phi^0, \xi^0,$ and, $\chi^+$ are non-vanishing, 
\begin{align}
V^{(4)}&=\ld_\phi |\phi^0|^4 + \ld_\xi (\xi^0)^4 +(\ld_\chi+\tilde{\ld}_\chi) |\chi^+|^4 + \ld_{\phi\xi}|\phi^0|^2(\xi^0)^2\nonumber\\
&\mathrel{\phantom{=}} +\ld_{\phi\chi} |\phi^0|^2 |\chi^+|^2 +(\ld_{\chi\xi}+\kappa_1)|\chi^+|^2(\xi^0)^2\,.
\end{align}
The above equation can be written in a matrix form $x^T\Lambda x$, in the basis $x^T=\{ |\phi^0|^2,(\xi^0)^2,\\
 |\chi^+|^2\}$, where $\Lambda$ is given by,
\begin{equation}
\label{cop_matrix_3f1}
\Lambda=\begin{bmatrix}
 \lambda _{\phi } & \frac{\ld_{\phi\xi}}{2} & \frac{\lambda _{\phi\chi }}{2} \\[3pt]
\frac{\ld_{\phi\xi}}{2} & \lambda _{\xi } &  \frac{\kappa _1+ \lambda _{\chi \xi }}{2} \\[3pt]
\frac{\lambda _{\phi\chi }}{2} & \frac{\kappa _1+ \lambda _{\chi \xi }}{2} &  \lambda _{\chi }+\tilde{\lambda }_{\chi } \\
\end{bmatrix}.
\end{equation}
The symmetric matrix $\Lambda$ is strictly copositive (i.e., $x^T\Lambda x>0$) if and only if~\cite{Hadeler1983,Chang1994},
\begin{align}
  \quad\quad \ld_{\phi} &> 0, & \ld_{\chi}+\tilde{\ld}_\chi &> 0, &  a_{1} = \ld_{\phi\xi} + 2\sqrt{ \ld_{\phi} \ld_{\xi} }&> 0,
  \notag
  \\
  \ld_{\xi}  &> 0,
  &
  b_{1} = \ld_{\phi\chi} + 2\sqrt{ \ld_{\phi}( \ld_{\chi}+\tilde{\ld}_\chi) } &> 0,
  &
  c_{1} = \ld_{\chi\xi} +\kappa_1+ 2\sqrt{ \ld_{\xi}( \ld_{\chi}+\tilde{\ld}_\chi) } &>0,\notag
  \\
  \span \span \span \span \ld_{\phi\xi}\sqrt{\ld_\chi+\tilde{\ld}_\chi}+\ld_{\phi\chi}\sqrt{\ld_\xi}+(\ld_{\chi\xi}+\kappa_1)\sqrt{\ld_\phi}+2\sqrt{\ld_\phi\ld_\xi(\ld_\chi+\tilde{\ld}_\chi)}+\sqrt{a_1b_1c_1} &> 0\,.\label{eq:3f1-GGM}
\end{align}
Among these 3-field directions, we write down the part of the potential where only neutral components of the fields  $\phi^0,\xi^0,$ and $\chi^0$ are non-vanishing,
\begin{multline}\label{eq:3f_0}
V^{(4)}=\ld_\phi |\phi^0|^4 +\ld_\xi (\xi^0)^4+\ld_\chi |\chi^0|^4+\ld_{\phi\xi}|\phi^0|^2(\xi^0)^2 +\ld_{\chi\xi}|\chi^0|^2 (\xi^0)^2\\
+\left(\ld_{\phi\chi}+\frac{\kappa_2}{2}\right)|\phi^0|^2|\chi^0|^2+\frac{\kappa_3}{\sqrt{2}}\left[(\phi^0)^2\xi^0\chi^{0*}+\textrm{h.c.}\right].
\end{multline}
Following the methodology given in~\cite{ElKaffas:2006gdt,Arhrib:2011uy}, we define,
\eqn{def_ind}{x=\sqrt{\frac{|\chi^0|^2}{(\xi^0)^2}}\;\in \mathbb{R}_+\,,\quad X=\frac{(\xi^0)^2}{|\phi^0|^2}\;\in \mathbb{R}_+\,,\quad \zeta=\frac{1}{2}\frac{\left[(\phi^0)^2\xi^0\chi^{0*}+\text{h.c.}\right]}{|\phi^0|^2\sqrt{|\chi^0|^2(\xi^0)^2}}\,,}
where $x,X,$ and $\zeta$ are three independent variables.\footnote{Here, the parameter, $\zeta$ can be seen as a cosine of angles in the field space.} The quartic potential in Eq.~(\ref{eq:3f_0}) can be written as, 
\begin{equation}\label{eq:3f_0_short}
\frac{V^{(4)}}{|\phi^0|^4}=\ld_\phi+\left(\ld_{\phi\xi} +\sqrt{2}\kappa_3 \zeta x+\left(\ld_{\phi\chi}+\frac{\kappa_2}{2}\right)x^2\right)X+ \Big(\ld_\xi+\ld_{\chi\xi}x^2+\ld_\chi x^4\Big)X^2\,,
\end{equation}
where $\zeta$ takes the values $\zeta=0,\pm1$.\footnote{It is worth to note that the parameter, $\zeta$ varies in the bounded domain, $\zeta\in[-1,1]$, if we consider four fields at once.} The potential in Eq.~(\ref{eq:3f_0}) is strictly copositive  if and only if,
\begin{align}
 \quad\quad\quad \ld_{\phi} &> 0, & \ld_{\chi} &> 0, &  a_{2} = \ld_{\phi\xi} + 2\sqrt{ \ld_{\phi} \ld_{\xi} }&> 0,
  \notag
  \\
  \ld_{\xi}  &> 0,
  &
  b_{2} = \ld_{\chi\xi} +2\sqrt{\ld_\chi\ld_\xi} &> 0,
  &
  c_{2} = 2 \ld_{\phi\chi} +\kappa_2 +4\sqrt{\ld_\phi\ld_\chi}  &>0,\notag
  \\
   \span \span \span \span  2\ld_{\phi\xi}+(2\ld_{\phi\chi}+\kappa_2)x^2-2\sqrt{2}|\kappa_3|x+4\sqrt{\ld_\phi(\ld_\xi+\ld_{\chi\xi}x^2+\ld_\chi x^4)} &>0,\notag\\
   \span \span \span \span 2\ld_{\phi\xi}\sqrt{\ld_\chi}+2\ld_{\chi\xi}\sqrt{\ld_\phi}+(2\ld_{\phi\chi}+\kappa_2)\sqrt{\ld_\xi}+4\sqrt{\ld_\phi\ld_\xi\ld_\chi}+\sqrt{2a_2b_2c_2} &> 0\,.
   \label{eq:3f0-GGM}
\end{align}
The condition given in the third line of Eq.~(\ref{eq:3f0-GGM}) contains a field dependent parameter, $x$. We check this condition numerically by scanning over the range of $x \in \mathbb{R}_+$. Considering all such directions with three simultaneously non-vanishing fields, we get a minimal set of necessary and sufficient 3-field BFB conditions:
\begin{align}
 \quad\quad\quad \ld_{\phi} &> 0, & \Lambda_{1} = \ld_{\phi\xi} + 2\sqrt{ \ld_{\phi} \ld_{\xi} }&> 0,\kern-2em &\!  \Lambda_{2}  = 2\ld_{\chi\xi} +\kappa_1+ 4\sqrt{ \ld_{\xi} \ld_{\chi}}  &> 0,
  \notag
  \\[5pt]
  \ld_{\xi}  &> 0,
  &
  \Lambda_{3}  = \ld_{\chi\xi} +2\sqrt{\ld_\chi\ld_\xi} &> 0,\kern-2em
  &\!
  \Lambda_{4}  = 2 \ld_{\phi\chi} +\kappa_2 +4\sqrt{\ld_\phi\ld_\chi}  &>0,\notag
  \\[5pt]
   \ld_{\chi} &> 0,
  &
  \Lambda_{5}  = \ld_{\chi} +2\sqrt{\ld_\chi(\ld_{\chi}+\tilde{\ld}_{\chi})}  &> 0,\kern-2em
  &\!
    \Lambda_{6} = 2 \ld_{\phi\chi} -\kappa_2 +4\sqrt{\ld_\phi\ld_\chi}  &>0,\notag
  \\[5pt]
  \ld_{\chi}+\tilde{\ld}_{\chi} &> 0,
  &
   \Lambda_{7} =\ld_{\chi\xi} +2\sqrt{\ld_\xi(\ld_{\chi}+\tilde{\ld}_{\chi})} &> 0,\kern-2em
  &\!
   \Lambda_{8} =  \ld_{\phi\chi} +2\sqrt{\ld_\phi(\ld_{\chi}+\tilde{\ld}_{\chi})} &>0,\nonumber\\[5pt]
   & & & &\Lambda_{9} = \ld_{\chi\xi} +\kappa_1+ 2\sqrt{ \ld_{\xi}( \ld_{\chi}+\tilde{\ld}_\chi) } &>0,\notag\\[5pt]
   \span \span \span \span 2\ld_{\phi\chi}\sqrt{\ld_\phi}+4\ld_{\phi}\sqrt{\ld_\chi}+\sqrt{\ld_\phi\Lambda_4\Lambda_6}&>0,\nonumber\\[5pt]
    \span \span \span \span 8\ld_{\xi}\sqrt{\ld_\chi}+(4\ld_{\chi\xi}+\kappa_1)\sqrt{\ld_\xi}+2\sqrt{2\ld_\xi\Lambda_2\Lambda_3}&>0,\nonumber\\[3pt]
   \span \span \span \span  4\Big(\ld_{\chi}+\tilde{\ld}_\chi\Big)\sqrt{\ld_\xi}+(2\ld_{\chi\xi}+\kappa_1)\sqrt{\ld_\chi}+\Lambda_2\sqrt{\ld_{\chi}+\tilde{\ld}_\chi}&>0,\nonumber\\[5pt]
     \span \span \span \span \ld_{\chi\xi}\sqrt{\ld_\chi}+ 2\Big(\ld_{\chi}+\tilde{\ld}_\chi\Big)\sqrt{\ld_\xi}+\Lambda_3\sqrt{\ld_{\chi}+\tilde{\ld}_\chi}&>0,\nonumber
       \end{align}
 \begin{align}
   \span \span \span \span 4\ld_{\xi}\sqrt{\ld_\chi+\tilde{\ld}_\chi}+(2\ld_{\chi\xi}+\kappa_1)\sqrt{\ld_\xi}+2\sqrt{\ld_\xi\Lambda_7\Lambda_9}&>0,\nonumber\\[5pt]
   \span \span \span \span  \ld_{\phi\xi}\sqrt{\ld_\chi+\tilde{\ld}_\chi}+\ld_{\phi\chi}\sqrt{\ld_\xi}+\Lambda_9\sqrt{\ld_\phi}+\sqrt{\Lambda_1\Lambda_8\Lambda_9}&>0,\nonumber\\[3pt]
   \span \span \span \span  \ld_{\phi\xi}\sqrt{\ld_\chi+\tilde{\ld}_\chi}+\ld_{\phi\chi}\sqrt{\ld_\xi}+\Lambda_7\sqrt{\ld_\phi}+\sqrt{\Lambda_1\Lambda_7\Lambda_8}&>0,\nonumber\\[3pt]
   \span \span \span \span   \ld_{\chi\xi}\sqrt{\ld_\chi+\tilde{\ld}_\chi}+2\ld_\chi\sqrt{\ld_\xi}
+\Lambda_9\sqrt{\ld_\chi}+\sqrt{2\Lambda_3\Lambda_5\Lambda_9}&>0,\nonumber\\[3pt]
   \span \span \span \span 2\ld_{\phi\chi}\sqrt{\ld_\chi}+4\Big(\ld_{\chi}+\tilde{\ld}_\chi\Big)\sqrt{\ld_\phi}+\sqrt{\Lambda_4\Lambda_6(\ld_{\chi}+\tilde{\ld}_\chi)}&>0,\nonumber\\[3pt]
   \span \span \span \span 2\ld_{\chi\xi}\sqrt{\ld_\chi}+4\ld_\chi\sqrt{\ld_\xi}+
\Lambda_2\sqrt{\ld_\chi+\tilde{\ld}_\chi}+2\sqrt{\Lambda_2\Lambda_5\Lambda_7}&>0,\nonumber\\[3pt]
    \span \span \span \span 2\ld_{\phi\chi}\sqrt{\ld_\chi}+4\ld_\chi\sqrt{\ld_\phi}
+\Lambda_4\sqrt{\ld_\chi+\tilde{\ld}_\chi}+2\sqrt{\Lambda_4\Lambda_5\Lambda_8}&>0,\nonumber\\[3pt]
    \span \span \span \span 2\ld_{\phi\chi}\sqrt{\ld_\chi}+4\ld_\chi\sqrt{\ld_\phi}
+\Lambda_6\sqrt{\ld_\chi+\tilde{\ld}_\chi}+2\sqrt{\Lambda_5\Lambda_6\Lambda_8}&>0,\nonumber\\[3pt]
   \span \span \span \span  2\ld_{\phi\xi}\sqrt{\ld_\chi}+(2\ld_{\phi\chi}-\kappa_2)\sqrt{\ld_\xi}
+\Lambda_2\sqrt{\ld_\phi}+\sqrt{\Lambda_1\Lambda_2\Lambda_6}&>0,\notag\\[3pt]
    \span \span \span \span 2\ld_{\phi\xi}\sqrt{\ld_\chi}+(2\ld_{\phi\chi}+\kappa_2)\sqrt{\ld_\xi}
+\Lambda_2\sqrt{\ld_\phi}+\sqrt{\Lambda_1\Lambda_2\Lambda_4}&>0,\notag\\[3pt]
     \span \span \span \span   2\ld_{\phi\xi}\sqrt{\ld_\chi}+(2\ld_{\phi\chi}-\kappa_2)\sqrt{\ld_\xi}+2\Lambda_3\sqrt{\ld_\phi}+\sqrt{2\Lambda_1\Lambda_3\Lambda_6} &> 0\,,\notag\\[3pt]
    \span \span \span \span    2\ld_{\phi\xi}\sqrt{\ld_\chi}+(2\ld_{\phi\chi}+\kappa_2)\sqrt{\ld_\xi}+2\Lambda_3\sqrt{\ld_\phi}+\sqrt{2\Lambda_1\Lambda_3\Lambda_4} &> 0\,,\notag\\[3pt]
\span \span \span \span \ld_{\phi\xi}+\ld_{\phi\chi}x^2-|\kappa_3|x+2\sqrt{\ld_\phi(\ld_\xi+\ld_{\chi\xi}x^2+(\ld_\chi+\tilde{\ld}_\chi) x^4)} &>0,\notag\\[3pt]
   \span \span \span \span 2\ld_{\phi\xi}+(2\ld_{\phi\chi}+\kappa_2)x^2-2\sqrt{2}|\kappa_3|x+4\sqrt{\ld_\phi(\ld_\xi+\ld_{\chi\xi}x^2+\ld_\chi x^4)} &>0,\notag\\[3pt]
   \span \span \span \span 2\ld_{\chi\xi}+\kappa_1+(2\ld_{\chi\xi}+\kappa_1)x^2-2|\kappa_1|x+4\sqrt{\ld_\xi(\ld_\chi+(2\ld_{\chi}+4\tilde{\ld}_\chi)x^2+\ld_\chi x^4)} &>0,
   \label{eq:3fall2-GGM}
\end{align}

For the eGM model, the minimal set of necessary and sufficient 3-field BFB conditions reads as,
\begin{align}
 \quad\quad\quad \ld_{\phi} &> 0, & \Sigma_{1} = \ld_{\phi\xi} + 2\sqrt{ \ld_{\phi} \ld_{\xi} }&> 0,\kern-2em &\!  \Sigma_{2}  = 4\ld_\chi-8\ld_\xi +\kappa_1+ 4\sqrt{ \ld_{\xi} \ld_{\chi}}  &> 0,
  \notag
  \\[3pt]
  \ld_{\xi}  &> 0,
  &
  \ld_{\chi}+\tilde{\ld}_{\chi} &> 0,\kern-2em
  &\!
   \Sigma_{3}  = 2\ld_\chi-4\ld_\xi +2\sqrt{\ld_\chi\ld_\xi} &>0,\notag
  \\[3pt]
   \ld_{\chi} &> 0,
  &
 \Sigma_{5}  = \ld_{\chi} +2\sqrt{\ld_\chi(\ld_{\chi}+\tilde{\ld}_{\chi})}    &> 0,\kern-2em
  &\!
  \Sigma_{4}  = \sqrt{2} \kappa _3+4 \left(\sqrt{\lambda _{\chi } \lambda _{\phi }}+\lambda _{\phi \xi }\right)   &>0,\notag
  \\[3pt]
   & & & &  \Sigma_{6} = 4 \left(\sqrt{\lambda _{\chi } \lambda _{\phi }}-\lambda _{\phi \xi }+\lambda _{\phi \chi }\right)-\sqrt{2} \kappa _3 &>0,\notag\\
   & & & & \Sigma_{7} =2\ld_\chi-4\ld_\xi +2\sqrt{\ld_\xi(\ld_{\chi}+\tilde{\ld}_{\chi})} &>0,\notag\\[3pt]
   & & & & \Sigma_{8} =  \ld_{\phi\chi} +2\sqrt{\ld_\phi(\ld_{\chi}+\tilde{\ld}_{\chi})} &> 0,\notag\\[3pt]
   & & & &\Sigma_{9} = 2\ld_\chi-4\ld_\xi +\kappa_1+ 2\sqrt{ \ld_{\xi}( \ld_{\chi}+\tilde{\ld}_\chi) } &>0,\notag
    \end{align}
   \begin{align}
   \span \span \span \span 2\ld_{\phi\chi}\sqrt{\ld_\phi}+4\ld_{\phi}\sqrt{\ld_\chi}+\sqrt{\ld_\phi\Sigma_4\Sigma_6}&>0,\nonumber\\[3pt]
    \span \span \span \span 8\ld_{\xi}\sqrt{\ld_\chi}+(8\ld_\chi-16\ld_\xi)+\kappa_1)\sqrt{\ld_\xi}+2\sqrt{2\ld_\xi\Sigma_2\Sigma_3}&>0,\nonumber\\[3pt]
   \span \span \span \span  4\Big(\ld_{\chi}+\tilde{\ld}_\chi\Big)\sqrt{\ld_\xi}+(4\ld_\chi-8\ld_\xi+\kappa_1)\sqrt{\ld_\chi}+\Sigma_2\sqrt{\ld_{\chi}+\tilde{\ld}_\chi}&>0,\nonumber\\[5pt]
     \span \span \span \span (2\ld_\chi-4\ld_\xi)\sqrt{\ld_\chi}+ 2\Big(\ld_{\chi}+\tilde{\ld}_\chi\Big)\sqrt{\ld_\xi}+\Sigma_3\sqrt{\ld_{\chi}+\tilde{\ld}_\chi}&>0,\nonumber\\[5pt]
   \span \span \span \span 4\ld_{\xi}\sqrt{\ld_\chi+\tilde{\ld}_\chi}+(4\ld_\chi-8\ld_\xi+\kappa_1)\sqrt{\ld_\xi}+2\sqrt{\ld_\xi\Sigma_7\Sigma_9}&>0,\nonumber\\[5pt]
   \span \span \span \span  \ld_{\phi\xi}\sqrt{\ld_\chi+\tilde{\ld}_\chi}+\ld_{\phi\chi}\sqrt{\ld_\xi}+\Sigma_9\sqrt{\ld_\phi}+\sqrt{\Sigma_1\Sigma_8\Sigma_9}&>0,\nonumber\\[5pt]
   \span \span \span \span  \ld_{\phi\xi}\sqrt{\ld_\chi+\tilde{\ld}_\chi}+\ld_{\phi\chi}\sqrt{\ld_\xi}+\Sigma_7\sqrt{\ld_\phi}+\sqrt{\Sigma_1\Sigma_7\Sigma_8}&>0,\nonumber\\[5pt]
   \span \span \span \span   (2\ld_\chi-4\ld_\xi)\sqrt{\ld_\chi+\tilde{\ld}_\chi}+2\ld_\chi\sqrt{\ld_\xi}
+\Sigma_9\sqrt{\ld_\chi}+\sqrt{2\Sigma_3\Sigma_5\Sigma_9}&>0,\nonumber\\[5pt]
   \span \span \span \span 2\ld_{\phi\chi}\sqrt{\ld_\chi}+4\Big(\ld_{\chi}+\tilde{\ld}_\chi\Big)\sqrt{\ld_\phi}+\sqrt{\Sigma_4\Sigma_6(\ld_{\chi}+\tilde{\ld}_\chi)}&>0,\nonumber\\[5pt]
   \span \span \span \span (4\ld_\chi-8\ld_\xi)\sqrt{\ld_\chi}+4\ld_\chi\sqrt{\ld_\xi}+
\Sigma_2\sqrt{\ld_\chi+\tilde{\ld}_\chi}+2\sqrt{\Sigma_2\Sigma_5\Sigma_7}&>0,\nonumber\\[5pt]
    \span \span \span \span 2\ld_{\phi\chi}\sqrt{\ld_\chi}+4\ld_\chi\sqrt{\ld_\phi}
+\Sigma_4\sqrt{\ld_\chi+\tilde{\ld}_\chi}+2\sqrt{\Sigma_4\Sigma_5\Sigma_8}&>0,\nonumber\\[5pt]
    \span \span \span \span 2\ld_{\phi\chi}\sqrt{\ld_\chi}+4\ld_\chi\sqrt{\ld_\phi}
+\Sigma_6\sqrt{\ld_\chi+\tilde{\ld}_\chi}+2\sqrt{\Sigma_5\Sigma_6\Sigma_8}&>0,\notag\\[5pt]
   \span \span \span \span  2\ld_{\phi\xi}\sqrt{\ld_\chi}+(4 \lambda _{\phi \chi }-\sqrt{2} \kappa _3-4 \lambda _{\phi \xi })\sqrt{\ld_\xi}
+\Sigma_2\sqrt{\ld_\phi}+\sqrt{\Sigma_1\Sigma_2\Sigma_6}&>0,\notag\\[5pt]
    \span \span \span \span 2\ld_{\phi\xi}\sqrt{\ld_\chi}+(\sqrt{2} \kappa _3+4 \lambda _{\phi \xi })\sqrt{\ld_\xi}
+\Sigma_2\sqrt{\ld_\phi}+\sqrt{\Sigma_1\Sigma_2\Sigma_4}&>0,\notag\\[5pt]
     \span \span \span \span   2\ld_{\phi\xi}\sqrt{\ld_\chi}+(4 \lambda _{\phi \chi }-\sqrt{2} \kappa _3-4 \lambda _{\phi \xi })\sqrt{\ld_\xi}+2\Sigma_3\sqrt{\ld_\phi}+\sqrt{2\Sigma_1\Sigma_3\Sigma_6} &> 0\,,\notag\\[5pt]
    \span \span \span \span    2\ld_{\phi\xi}\sqrt{\ld_\chi}+(\sqrt{2} \kappa _3+4 \lambda _{\phi \xi })\sqrt{\ld_\xi}+2\Sigma_3\sqrt{\ld_\phi}+\sqrt{2\Sigma_1\Sigma_3\Sigma_4} &> 0\,,\notag\\[5pt]
\span \span \span \span \ld_{\phi\xi}+\ld_{\phi\chi}x^2-|\kappa_3|x+2\sqrt{\ld_\phi(\ld_\xi+(2\ld_\chi-4\ld_\xi)x^2+(\ld_\chi+\tilde{\ld}_\chi) x^4)} &>0,\notag\\[5pt]
   \span \span \span \span 2\ld_{\phi\xi}+(\sqrt{2} \kappa _3+4 \lambda _{\phi \xi })x^2-2\sqrt{2}|\kappa_3|x+4\sqrt{\ld_\phi(\ld_\xi+(2\ld_\chi-4\ld_\xi)x^2+\ld_\chi x^4)} &>0,\notag\\[5pt]
   \span \span \span \span 4\ld_\chi-8\ld_\xi+\kappa_1+(4\ld_\chi-8\ld_\xi+\kappa_1)x^2-2|\kappa_1|x+4\sqrt{\ld_\xi(\ld_\chi+(2\ld_{\chi}+4\tilde{\ld}_\chi)x^2+\ld_\chi x^4)} &>0,
   \label{eq:3fall2-EGM}
\end{align}

In the case of the GM model, we get the following 3-field BFB conditions: 

\begin{align}
 \quad\quad\quad \ld_{1} &> 0, & \Omega_{1} = 2 \sqrt{\lambda _1 \left(\lambda _2+\lambda _3\right)}+\lambda _4&> 0,\kern-2em &\!  \Omega_{4}  = 4 \sqrt{\lambda _1 \left(4 \lambda _2+2 \lambda _3\right)}+4 \lambda _4+\lambda _5  &> 0,
  \notag
  \\[5pt]
  \ld_{2}+\ld_3  &> 0,
  &
  2\ld_{2}+\ld_3 &> 0,\kern-2em
    &\!
  \Omega_{6} = 4 \sqrt{\lambda _1 \left(4 \lambda _2+2 \lambda _3\right)}+4 \lambda _4-\lambda _5   &>0,\notag
  \\[5pt]
   & & & & \Omega_{2}  =  \sqrt{\left(\lambda _2+\lambda _3\right) \left(4 \lambda _2+2\lambda _3\right)}+2 \lambda _2+\lambda _3 &>0,\notag\\[5pt]
 & & & & \Omega_{3}  = \sqrt{\left(\lambda _2+\lambda _3\right) \left(4 \lambda _2+2 \lambda _3\right)}+2 \lambda _2&>0,\notag\\[5pt]
   \span \span \span \span 16 \sqrt{4 \lambda _2+2 \lambda _3} \lambda _1+16 \sqrt{\lambda _1} \lambda _4+4 \sqrt{\lambda _1 \Omega_{4} \Omega_{6}}&>0,\nonumber
\end{align}
\begin{align}
    \span \span \span \span 8 \sqrt{4 \lambda _2+2 \lambda _3} \left(\lambda _2+\lambda _3\right)+4 \sqrt{\lambda _2+\lambda _3} \left(4 \lambda _2+\lambda _3\right)+8 \sqrt{\left(\lambda _2+\lambda _3\right) \Omega_2 \Omega_3}&>0,\nonumber\\[5pt]
   \span \span \span \span  4 \left(\lambda _2+\lambda _3\right){}^{3/2}+\sqrt{2} \left(2 \lambda _2+\lambda _3\right){}^{3/2}+2 \sqrt{\lambda _2+\lambda _3} \Omega_2&>0,\nonumber\\[5pt]
     \span \span \span \span 8 \left(\lambda _2+\lambda _3\right){}^{3/2}+4 \sqrt{4 \lambda _2+2 \lambda _3} \lambda _2+4 \sqrt{\lambda _2+\lambda _3} \Omega_3&>0,\nonumber\\[5pt]
   \span \span \span \span 16 \sqrt{\lambda _2+\lambda _3} \lambda _2+12 \sqrt{\lambda _2+\lambda _3} \lambda _3+8 \sqrt{2} \sqrt{2 \lambda _2+\lambda _3} \left(\lambda _2+\lambda _3\right) &>0,\nonumber\\[5pt]
   \span \span \span \span 8 \lambda _4 \sqrt{\lambda _2+\lambda _3}+2 \sqrt{\lambda _1} \left(\lambda _2+\lambda _3\right)+\sqrt{\lambda _2+\lambda _3} \Omega_1 &>0,\nonumber\\[5pt]
   \span \span \span \span 8 \sqrt{\lambda _1} \left(2 \lambda _2+\lambda _3\right)+8 \sqrt{\lambda _2+\lambda _3} \lambda _4+4 \sqrt{4 \lambda _2+2 \lambda _3} \Omega_1 &>0,\nonumber\\[5pt]
   \span \span \span \span   4 \sqrt{\lambda _2+\lambda _3} (4\lambda _2+\ld_3)+8 \sqrt{4 \lambda _2+2 \lambda _3} \left(\lambda _2+\lambda _3\right)+8 \sqrt{\left(\lambda _2+\lambda _3\right) \Omega_2 \Omega_3}&>0,\nonumber\\[5pt]
   \span \span \span \span 8 \sqrt{\lambda _1} \left(\lambda _2+\lambda _3\right)+2 \sqrt{4 \lambda _2+2 \lambda _3} \lambda _4+\sqrt{\left(\lambda _2+\lambda _3\right) \Omega_4 \Omega_6}&>0,\nonumber\\[5pt]
   \span \span \span \span  \left(2 \lambda _2+\lambda _3\right) \sqrt{\lambda _2+\lambda _3}+ \sqrt{4 \lambda _2+2 \lambda _3} \lambda _2+ \sqrt{\lambda _2+\lambda _3} \Omega_2+ \sqrt{4 \lambda _2+2 \lambda _3} \Omega_2 &>0,\nonumber\\[5pt]
    \span \span \span \span 16 \sqrt{\lambda _1} \left(2 \lambda _2+\lambda _3\right)+8 \sqrt{4 \lambda _2+2 \lambda _3} \lambda _4+8 \sqrt{\Omega_1 \Omega_2 \Omega_4}+4 \sqrt{\lambda _2+\lambda _3} \Omega_4&>0,\nonumber\\[5pt]
    \span \span \span \span 16 \sqrt{\lambda _1} \left(2 \lambda _2+\lambda _3\right)+8 \sqrt{4 \lambda _2+2 \lambda _3} \lambda _4+8 \sqrt{\Omega_1 \Omega_2 \Omega_6}+4 \sqrt{\lambda _2+\lambda _3} \Omega_6&>0,\notag\\[5pt]
   \span \span \span \span  4 \sqrt{4 \lambda _2+2 \lambda _3} \lambda _4+\sqrt{\lambda _2+\lambda _3} \left(8 \lambda _4-2 \lambda _5\right)+4 \sqrt{\Omega_1 \Omega_2 \Omega_6}+8 \sqrt{\lambda _1} \Omega_2&>0,\notag\\[5pt]
    \span \span \span \span 4 \sqrt{4 \lambda _2+2 \lambda _3} \lambda _4+2 \sqrt{\lambda _2+\lambda _3} \left(4 \lambda _4+\lambda _5\right)+4 \sqrt{\Omega_1 \Omega_2 \Omega_4}+8 \sqrt{\lambda _1} \Omega_2&>0,\notag\\[5pt]
     \span \span \span \span   4 \sqrt{4 \lambda _2+2 \lambda _3} \lambda _4+\sqrt{\lambda _2+\lambda _3} \left(8 \lambda _4-2 \lambda _5\right)+4 \sqrt{\Omega_1 \Omega_3 \Omega_6}+8 \sqrt{\lambda _1} \Omega_3 &> 0\,,\notag\\[5pt]
    \span \span \span \span 4 \sqrt{4 \lambda _2+2 \lambda _3} \lambda _4+2 \sqrt{\lambda _2+\lambda _3} \left(4 \lambda _4+\lambda _5\right)+4 \sqrt{\Omega_1 \Omega_3 \Omega_4}+8 \sqrt{\lambda _1} \Omega_3 &> 0\,,\notag\\[5pt]
\span \span \span \span \lambda _4 \left(4 x^2+2\right)+4 \sqrt{\lambda _1 \left(\lambda _3 \left(4 x^4+1\right)+\lambda _2 \left(2 x^2+1\right)^2\right)}-\sqrt{2} |\lambda _5| x&>0,\notag\\[5pt]
   \span \span \span \span 4 \lambda _4+2 \left(4 \lambda _4+\lambda _5\right) x^2+8 \sqrt{\lambda _1 \left(\lambda _3 \left(2 x^4+1\right)+\lambda _2 \left(2 x^2+1\right)^2\right)}- 4 |\ld _{5} |x&>0,\notag\\[5pt]
   \span \span \span \span (2 \lambda _2 + \lambda _3) \left(x^2+1\right)+ \sqrt{2\left(\lambda _2+\lambda _3\right) \left(2 \lambda _2 \left(x^2+1\right)^2+\lambda _3 \left(x^4+6 x^2+1\right)\right)}-2 |\ld_3| x &>0,
   \label{eq:3fall2-GM}
\end{align}
\section{Necessary and sufficient bounded-from-below conditions}\label{app:pos_EGM}
\begin{figure}[!h]
     \centering
            \includegraphics[clip=true,width=0.85\columnwidth]{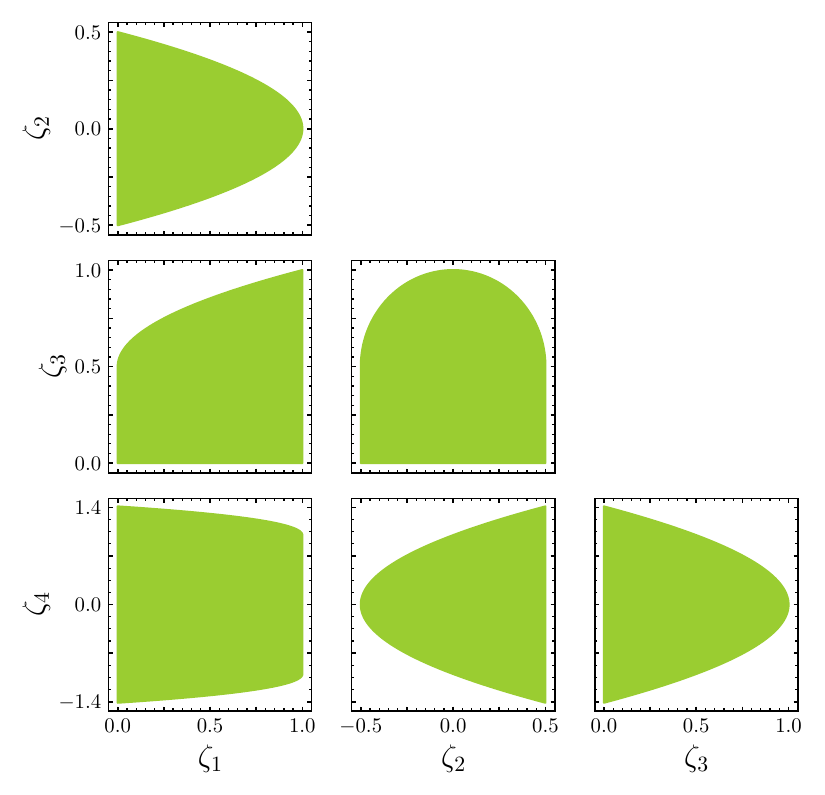}
     \caption{The green region manifests allowed $\zeta$-parameter space onto the projected planes.}
    \label{fig:9}
\end{figure}

In general, the above BFB conditions for specific field directions are necessary but not sufficient to ensure boundedness of the scalar potential of the GM and the eGM models. To obtain the necessary and sufficient BFB conditions, one needs to consider all the field directions simultaneously in the field space. The necessary and sufficient BFB conditions considering all 13-field directions for the GM model are given in Refs.~\cite{Hartling:2014zca,Moultaka2020}, while Ref.~\cite{Moultaka2020} also provides the necessary and sufficient BFB conditions considering all 13-field directions for the most general two-triplet model as given in Eq.~(\ref{v16_quartic}). To obtain all 13-field BFB conditions for the eGM model, we follow the approach of Ref.~\cite{Moultaka2020}. We parameterize the scalar quartic potential in Eq.~(\ref{v16_quartic}) by considering a number of dimensionless ratio of invariants, 
\begin{align}
\zeta_1&=\frac{|\tilde{\chi}^\dagger\chi |^2}{(\chi^\dagger\chi)^2}\,,\quad &\zeta_2&=\frac{(\phi^\dagger\tau_a\phi)(\chi^\dagger t_a\chi)}{(\phi^\dagger\phi)(\chi^\dagger\chi)},\nonumber\\  \zeta_3&=\frac{|\xi^\dagger\chi |^2}{(\chi^\dagger\chi)(\xi^\dagger\xi)}\,,\quad 
&\zeta_4&=\frac{[(\phi^T\epsilon\tau_a\phi)(\chi^\dagger t_a\xi)+\text{h.c.}]}{(\phi^\dagger\phi)\sqrt{(\chi^\dagger\chi)(\xi^\dagger\xi)}}\,,\label{eq:zeta} 
\end{align}
where
\begin{equation}
\nonumber
\zeta_1\in[0,1]\,,\quad\zeta_2\in[-1/2,1/2]\,,\quad\zeta_3\in[0,1]\,,\quad\zeta_4\in[-\sqrt{2},\sqrt{2}]\,.
\end{equation}
Following Ref.~\cite{Moultaka2020}, we obtain the necessary and sufficient 13-field BFB conditions for the eGM model, 
\begin{multline}
\lambda_\phi>0, \quad \lambda_\xi>0, \quad \lambda_\chi>0, \quad \lambda_\chi+\tilde{\lambda}_\chi>0,\quad \zeta_3\kappa_1+2\lambda_{\chi}-4\ld_{\xi}+2\sqrt{\lambda_\xi\left(\zeta_1\tilde{\lambda}_\chi+\lambda_\chi\right)}>0,\\
\lambda _{\phi \xi }+x\kappa _3 \left(\sqrt{2} \zeta_{2} x+\zeta_{4}\right)+4 x^2\zeta_{2}  \lambda _{\phi \xi }+\Big(1-2 \zeta_{2}\Big) x^2 \lambda _{\phi \chi }\\
+2 \sqrt{\lambda _{\phi } \left(\lambda _{\xi }+x^4 \zeta_{1}  \tilde{\lambda}_{\chi}+x^2(\zeta_{3} \kappa _1-4 \lambda_{\xi })+\left(x^4+2x^2\right) \lambda _{\chi }\right)}>0\,,
\label{eq:bfb13f-eGM}
\end{multline}
where $x\:(\equiv\sqrt{\chi^\dagger\chi/\xi^\dagger\xi})\in \mathbb{R}_+$. Note that, the parameters $\zeta_i \;(i=1,..,4)$ are correlated. Despite considering the entire four-dimensional domain of the $\zeta$-parameter space, we show the six possible projected planes. The boundary curves in the $\zeta_i$ vs.~$\zeta_j$ planes are listed in Table~\ref{tab:zeta-planes}, and their allowed domains are displayed in Figure~\ref{fig:9}. In numerical analysis, while considering the 13-field BFB conditions, we scan over the allowed parameter spaces of $\zeta_i$'s and $x$ for given model parameters. 
\begin{table}[ht]
\begin{center}
\setlength{\tabcolsep}{8pt}
\renewcommand{\arraystretch}{1.6}
\scalebox{1.}{
\begin{tabular}{|c|l|} 
 \hline
Projected planes & Boundary curves \\
\hline\hline
$\zeta_1$ vs. $\zeta_2$  & (i) $0\leq \zeta_1\leq 1-4\zeta_2^2,\quad \forall\, \zeta_2\in\left[-\frac{1}{2},\frac{1}{2}\right].$\\[3pt]
  \hline
 $\zeta_1$ vs. $\zeta_3$ & (i) $\zeta_3=0,\;\;\forall \,\zeta_1\in[0,1].$\\
 &(ii) $\zeta_1=1,\;\;\forall\, \zeta_3\in[0,1].$\\
 & (iii) $\zeta_1=0,\;\;\forall \,\zeta_3\in\left[0,\frac{1}{2}\right].$\\
 & (iv) $\zeta_1=1-4\zeta_3(1-\zeta_3),\;\;\forall\, \zeta_3\in\left[\frac{1}{2},1\right].$\\[3pt]
  \hline
  $\zeta_1$ vs. $\zeta_4$ & (i) $\zeta_1=0\,,\;\;\forall\, \zeta_4\in[-\sqrt{2},\sqrt{2}].$\\
 &(ii) $\zeta_1=1,\;\;\forall \,\zeta_4\in[-1,1].$\\
 & (iii) $\zeta_1=2\zeta_4^2-\zeta_4^4,\;\;\forall \,\zeta_4\in[-\sqrt{2},-1)\cup(1,\sqrt{2}].$\\[3pt]
 \hline
 $\zeta_2$ vs. $\zeta_3$ & (i) $\zeta_2=-\frac{1}{2},\;\;\forall \,\zeta_3\in\left[0,\frac{1}{2}\right].$\\
 &(ii) $\zeta_2=\frac{1}{2},\;\;\forall \,\zeta_3\in\left[0,\frac{1}{2}\right].$\\
 & (iii) $\zeta_3=0, \;\;\forall \,\zeta_2\in\left[-\frac{1}{2},\frac{1}{2}\right].$\\
 & (iv) $\zeta_3=\frac{1}{2}+\sqrt{\frac{1}{4}-\zeta_2^2}\:,\;\forall\,\zeta_2\in\left[-\frac{1}{2},\frac{1}{2}\right].$\\[3pt]
 \hline
 $\zeta_2$ vs. $\zeta_4$  & (i) $\zeta_2=\frac{1}{2}\,.$\\
 & (ii) $\zeta_2=\frac{1}{2}(\zeta_4^2-1), \quad\forall\, \zeta_4\in[-\sqrt{2},\sqrt{2}]$\\[3pt]
 \hline
 $\zeta_3$ vs. $\zeta_4$  & (i) $0\leq \zeta_3\leq 1-\frac{1}{2}\zeta_4^2\,,\quad \forall \,\zeta_4\in[-\sqrt{2},\sqrt{2}].$\\[3pt]
 \hline
\end{tabular}}
\caption{Boundary curves in the six projected planes of the $\zeta$-parameter space~\cite{Moultaka2020}. }
\label{tab:zeta-planes}
\end{center}
\end{table}
%
\section{Results for one-loop scattering amplitudes}\label{app:1-loop_amp}
For a given process, $i \to f$, the amplitudes given in this Appendix correspond to $256\pi^3(a_0)_{i,f}$. In these amplitudes, the contributions from wave function renormalization are not considered. Therefore, all of the scattering amplitudes form a block diagonal structure in the $S$-matrix of $2\to 2$ scattering. We have cross-checked our results by generating amplitudes using \texttt{FeynRules-2.3}~\cite{Christensen:2008py} and \texttt{FeynArts-3.11}~\cite{Hahn:2000kx}.
\begin{itemize}
\item[] {$\bullet$ \,\boldmath $Q=0, Y=0$ :}
\begin{align}
\phi^{0*}\phi^0\to\phi^{0*}\phi^0 =& -64\pi^2\lambda_\phi+12\beta_{\lambda_\phi}+(i\pi-1)\Big[
20\lambda_\phi^2+6\lambda_{\phi \xi}^2+3\lambda_{\phi\chi}^2+\frac{1}{2}\kappa_2^2\Big], \notag \\ 
\phi^{0*}\phi^0\to\chi^{0*}\chi^0 =&-16\pi^2\left(\lambda_{\phi\chi}+\frac{1}{2}\kappa_2\right)+3\left(\beta_{\lambda_{\phi\chi}}+\frac{1}{2}\beta_{\kappa_2}\right)+(i\pi-1)\Big[2\lambda_{\phi\xi}\left(\kappa_1+3\lambda_{\chi\xi}\right)\notag\\
&+\kappa_2\left(\lambda_\phi+\lambda_\chi-2\tilde{\lambda}_\chi\right)
+2\lambda_{\phi\chi}\left(3\lambda_\phi+4\lambda_\chi+2\tilde{\lambda}_\chi\right)\Big],\notag
\end{align}
\begin{align}
\phi^{0*}\phi^0\to\phi^+\phi^-=&-32\pi^2\lambda_\phi+6\beta_{\lambda_{\phi}}+(i\pi-1)\Big[
16\lambda_\phi^2+6\lambda_{\phi\xi}^2+3\lambda_{\phi\chi}^2-\frac{1}{2}\kappa_2^2\Big],\notag\\[5pt]
\phi^{0*}\phi^0\to\xi^+\xi^-=&-32\pi^2\lambda_{\phi\xi}+6\beta_{\lambda_{\phi\xi}}\notag\\
&+2(i\pi-1)\Big[
20\lambda_\xi\lambda_{\phi\xi}+6\lambda_{\phi}\lambda_{\phi\xi}+\lambda_{\phi\chi}\left(\kappa_1+3\lambda_{\chi\xi}\right)\Big],\notag\\[5pt]
\phi^{0*}\phi^0\to\chi^+\chi^-=&-16\pi ^2 \lambda _{\phi \chi }+3 \beta _{\lambda _{\phi \chi }}\notag\\
&+(i\pi-1)\Big[2 \lambda _{\phi \chi }\left(2 \tilde{\lambda }_{\chi }+4 \lambda _{\chi}+3 \lambda _{\phi }\right)+2 \lambda _{\phi \xi }\left(\kappa _1+3 \lambda _{\chi \xi}\right)\Big],\notag
\end{align}
\begin{align*}
\chi^{0*}\chi^0\to\phi^{0*}\phi^0&=\phi^{0*}\phi^0\to\chi^{0*}\chi^0,\\[2pt]
\phi^+\phi^-\to\phi^{0*}\phi^0&=\phi^{0*}\phi^0\to\phi^+\phi^-,\\[2pt]
\xi^+\xi^-\to\phi^{0*}\phi^0&=\phi^{0*}\phi^0\to\xi^+\xi^-,\\[2pt]
\chi^+\chi^-\to\phi^{0*}\phi^0&=\phi^{0*}\phi^0\to\chi^+\chi^-,
\end{align*}
\begin{align}
\phi^{0*}\phi^0\to\frac{1}{\sqrt{2}}(\xi^0\xi^0) =&-16\sqrt{2}\pi^2\lambda_{\phi\xi}+3\sqrt{2}\beta_{\lambda_{\phi\xi}}\notag\\
&+\sqrt{2}(i\pi-1)\Big[6\lambda_\phi\lambda_{\phi\xi}+\lambda_{\phi\chi}\left(\kappa_1+3\lambda_{\chi\xi}\right)+20\lambda_{\phi\xi}\lambda_\xi\Big],  \notag\\
\phi^{0*}\phi^0\to\chi^{++}\chi^{--} =&-16 \pi ^2\left(\lambda _{\phi \chi }-\frac{1}{2}\kappa_2\right)+3 \left(\beta _{\lambda_{\phi \chi}}-\frac{1}{2}\beta _{\kappa _2}\right)+(i\pi-1)\Big[2 \lambda _{\phi \xi }\left(\kappa _1\right.
\notag\\
&+\left.3 \lambda _{\chi \xi}\right)+\kappa _2 \left(2\tilde{\lambda }_{\chi }-\lambda _{\chi }-\lambda_{\phi }\right)+2 \lambda _{\phi \chi } \left(2\tilde{\lambda }_{\chi }+4 \lambda _{\chi }+3\lambda _{\phi }\right)
\Big], \notag
\end{align}
\begin{align*}
\frac{1}{\sqrt{2}}(\xi^0\xi^0)\to\phi^{0*}\phi^0&=\phi^{0*}\phi^0\to\frac{1}{\sqrt{2}}(\xi^0\xi^0),\\
\chi^{++}\chi^{--}\to\phi^{0*}\phi^0&=\phi^{0*}\phi^0\to\chi^{++}\chi^{--},
\end{align*}
\begin{align*}
\frac{1}{\sqrt{2}}(\xi^0\xi^0)\to\frac{1}{\sqrt{2}}(\xi^0\xi^0) =& -192 \pi ^2 \lambda _{\xi }+36\beta_{\lambda _{\xi }}+2(i\pi-1)\Big[2 \kappa _1 \lambda _{\chi \xi }\\
&+\kappa_1^2+88 \lambda _{\xi }^2+3 \lambda _{\chi \xi}^2+2 \lambda _{\phi \xi }^2\Big],\\
\frac{1}{\sqrt{2}}(\xi^0\xi^0)\to\chi^{0*}\chi^0 =&-16 \sqrt{2} \pi ^2 \lambda _{\chi \xi }+3\sqrt{2} \beta _{\lambda _{\chi \xi}} +2 \sqrt{2}(i\pi-1)\Big[
2\lambda _{\chi\xi}\left(\tilde{\lambda }_{\chi}\right.\\
&+5 \lambda_{\xi }+2 \lambda _{\chi }\Big)+\kappa _1\left(2 \lambda _{\xi }+\lambda _{\chi}\right)+\lambda _{\phi \xi } \lambda _{\phi \chi}\Big],\\
\frac{1}{\sqrt{2}}(\xi^0\xi^0)\to\chi^{++}\chi^{--}=&-16 \sqrt{2}\pi ^2 \lambda _{\chi \xi }+3 \sqrt{2} \beta _{\lambda _{\chi \xi }}+2\sqrt{2}(i\pi-1)\Big[ \lambda _{\phi \xi } \lambda _{\phi \chi}\\
&+2 \lambda _{\chi \xi }\left(\tilde{\lambda }_{\chi}+5 \lambda _{\xi }+2 \lambda_{\chi }\right)+\kappa _1 \left(2 \lambda _{\xi }+\lambda_{\chi }\right)\Big],
\end{align*}
\begin{align*}
\chi^{0*}\chi^0\to\frac{1}{\sqrt{2}}(\xi^0\xi^0)&=\frac{1}{\sqrt{2}}(\xi^0\xi^0)\to\chi^{0*}\chi^0,\\
\chi^{++}\chi^{--}\to\frac{1}{\sqrt{2}}(\xi^0\xi^0)&=\frac{1}{\sqrt{2}}(\xi^0\xi^0)\to\chi^{++}\chi^{--},
\end{align*}
\begin{align*}
\frac{1}{\sqrt{2}}(\xi^0\xi^0)\to \phi^+\phi^- =& -16 \sqrt{2}\pi ^2 \lambda _{\phi \xi }+3 \sqrt{2} \beta _{\lambda _{\phi \xi }}\\
&+\sqrt{2}(i\pi-1)\Big[\lambda _{\phi \chi }\left(\kappa _1+3 \lambda _{\chi \xi }\right)+20\lambda _{\xi } \lambda _{\phi \xi }+6 \lambda_{\phi } \lambda _{\phi \xi }\Big],\\
\frac{1}{\sqrt{2}}(\xi^0\xi^0)\to\xi^+\xi^- =&-64 \sqrt{2} \pi ^2 \lambda _{\xi }+12 \sqrt{2} \beta _{\lambda _{\xi }}\\
&+2 \sqrt{2}(i\pi-1)\Big[
\lambda _{\chi \xi }\left(2 \kappa _1+3 \lambda _{\chi \xi }\right)+56\lambda _{\xi }^2+2 \lambda _{\phi \xi}^2\Big],\\
\frac{1}{\sqrt{2}}(\xi^0\xi^0)\to\chi^+\chi^-=& -16 \sqrt{2} \pi ^2 \left(\kappa _1+\lambda _{\chi
\xi }\right)+3 \sqrt{2}\left(\beta _{\kappa_1}+\beta _{\lambda _{\chi \xi}}\right)+\sqrt{2}(i\pi-1)\Big[2\lambda _{\phi \xi } \lambda _{\phi \chi }\\
&+4 \kappa _1\left(\tilde{\lambda }_{\chi }+3 \lambda _{\xi }+\lambda_{\chi }\right)+4\lambda _{\chi \xi } \left(\tilde{\lambda}_{\chi }+5 \lambda _{\xi }+2 \lambda _{\chi }\right)\Big],\\
\end{align*}
\begin{align*}
\phi^+\phi^-\to\frac{1}{\sqrt{2}}(\xi^0\xi^0)&=\frac{1}{\sqrt{2}}(\xi^0\xi^0)\to \phi^+\phi^-,\\
\xi^+\xi^-\to\frac{1}{\sqrt{2}}(\xi^0\xi^0)&=\frac{1}{\sqrt{2}}(\xi^0\xi^0)\to\xi^+\xi^-,\\
\chi^+\chi^-\to\frac{1}{\sqrt{2}}(\xi^0\xi^0)&=\frac{1}{\sqrt{2}}(\xi^0\xi^0)\to\chi^+\chi^-,
\end{align*}
\begin{align*}
\chi^{0*}\chi^0\to\chi^{0*}\chi^0=&-64 \pi ^2\lambda_\chi+12 \beta _{\lambda _{\chi }}+(i\pi-1)\Big[4 \kappa _1 \lambda _{\chi \xi }\\
&+2 \left(8 \tilde{\lambda }_{\chi }^2+8 \lambda _{\chi }\tilde{\lambda }_{\chi }+12 \lambda _{\chi }^2+3 \lambda _{\chi \xi}^2+\lambda _{\phi \chi }^2\right)+\kappa_1^2+\frac{1}{2}\kappa _2^2\Big],\\
\chi^{0*}\chi^0\to\phi^+\phi^-=&-16 \pi ^2 \left(\lambda_{\phi \chi }-\frac{1}{2}\kappa _2\right)+3 \left(\beta _{\lambda _{\phi \chi
   }}-\frac{1}{2}\beta _{\kappa _2}\right)+(i\pi-1)\Big[2 \lambda _{\phi \xi } \left(\kappa _1\right.\\
   &\left.+3 \lambda_{\chi \xi }\right)+\kappa _2 \left(2 \tilde{\lambda }_{\chi}-\lambda _{\chi }-\lambda _{\phi }\right)+2 \lambda _{\phi \chi }\left(2 \tilde{\lambda }_{\chi }+4 \lambda _{\chi }+3 \lambda_{\phi }\right)\Big],\\
\chi^{0*}\chi^0\to\xi^+\xi^- =&-16 \pi ^2 \left(\kappa _1+2 \lambda _{\chi \xi }\right)+3\left(\beta _{\kappa_1}+2 \beta _{\lambda _{\chi \xi}}\right)+(i\pi-1)\Big[4 \lambda _{\phi \xi } \lambda _{\phi \chi }\\
&+2 \kappa _1 \left(2 \tilde{\lambda }_{\chi}+8 \lambda _{\xi }+3 \lambda _{\chi }\right)+8 \lambda _{\chi \xi} \left(\tilde{\lambda }_{\chi }+5 \lambda _{\xi }+2 \lambda _{\chi}\right)\Big],\\
\chi^{0*}\chi^0\to\chi^+\chi^- =&-32 \pi ^2 \lambda _{\chi }+6\beta _{\lambda _{\chi }}+2(i\pi-1)\Big[8 \lambda _{\chi } \tilde{\lambda }_{\chi}\\
&+\lambda _{\chi \xi } \left(2 \kappa _1+3 \lambda _{\chi \xi}\right)+10 \lambda _{\chi }^2+\lambda _{\phi \chi }^2\Big],\\
\chi^{0*}\chi^0\to\chi^{++}\chi^{--} =&-32 \pi ^2 \left(2 \tilde{\lambda }_{\chi}+\lambda _{\chi }\right)+6 \left(2 \beta _{\tilde{\lambda}_{\chi }}+ \beta _{\lambda_{\chi }}\right)+(i\pi-1)\Big[4 \kappa _1 \lambda_{\chi \xi }\\
&+4 \lambda _{\chi } \left(8\tilde{\lambda }_{\chi }+5 \lambda _{\chi }\right)+\kappa _1^2-\frac{\kappa _2^2}{2}+6 \lambda _{\chi \xi }^2+2\lambda _{\phi \chi }^2\Big],
\end{align*}
\begin{align*}
\phi^+\phi^-\to\chi^{0*}\chi^0&=\chi^{0*}\chi^0\to\phi^+\phi^-,\\
\xi^+\xi^-\to\chi^{0*}\chi^0&=\chi^{0*}\chi^0\to\xi^+\xi^-,\\
\chi^+\chi^-\to\chi^{0*}\chi^0&=\chi^{0*}\chi^0\to\chi^+\chi^- ,\\
\chi^{++}\chi^{--}\to\chi^{0*}\chi^0&=\chi^{0*}\chi^0\to\chi^{++}\chi^{--},
\end{align*}
\begin{align*}
\phi^+\phi^-\to\phi^+\phi^-=&-64 \pi ^2 \lambda _{\phi }+12 \beta _{\lambda _{\phi }}+(i\pi-1)\Big[20 \lambda _{\phi }^2+6 \lambda _{\phi \xi }^2+3 \lambda _{\phi\chi }^2+\frac{1}{2}\kappa_2^2\Big],\\
\phi^+\phi^-\to\xi^+\xi^- =&-32\pi ^2 \lambda _{\phi \xi }+6 \beta _{\lambda _{\phi \xi }}\\
&+2(i\pi-1)\Big[\lambda _{\phi\chi } \left(\kappa _1+3 \lambda _{\chi \xi }\right)+20 \lambda _{\xi }\lambda _{\phi \xi }+6 \lambda _{\phi } \lambda _{\phi \xi }\Big],\\
\phi^+\phi^-\to\chi^+\chi^- =&-16 \pi ^2 \lambda _{\phi \chi }+3 \beta_{\lambda _{\phi \chi }}\\
&+(i\pi-1)\Big[2 \lambda _{\phi \chi } \left(2 \tilde{\lambda}_{\chi }+4 \lambda _{\chi }+3 \lambda _{\phi }\right)+2 \lambda _{\phi\xi } \left(\kappa _1+3 \lambda _{\chi \xi }\right)\Big],\\
\phi^+\phi^-\to\chi^{++}\chi^{--} =& -16\pi ^2 \left(\frac{1}{2}\kappa _2+\lambda _{\phi \chi }\right)+3\left(\frac{1}{2}\beta _{\kappa _2}+\beta _{\lambda _{\phi \chi }}\right)+(i\pi-1)\Big[2\lambda _{\phi \xi } \left(\kappa _1\right.\\
&\left.+3 \lambda _{\chi \xi }\right)+\kappa _2 \left(-2 \tilde{\lambda }_{\chi }+\lambda_{\chi }+\lambda _{\phi }\right)+2 \lambda _{\phi \chi } \left(2\tilde{\lambda }_{\chi }+4 \lambda _{\chi }+3 \lambda _{\phi }\right)\Big],
\end{align*}
\begin{align*}
\xi^+\xi^-\to\phi^+\phi^-&=\phi^+\phi^-\to\xi^+\xi^-,\\
\chi^+\chi^- \to \phi^+\phi^-&=\phi^+\phi^-\to\chi^+\chi^- ,\\
\chi^{++}\chi^{--} \to \phi^+\phi^-&=\phi^+\phi^-\to\chi^{++}\chi^{--} ,
\end{align*}
\begin{align*}
\xi^+\xi^-\to\xi^+\xi^- =&-256 \pi ^2 \lambda _{\xi }+48 \beta _{\lambda _{\xi }}+(i\pi-1)\Big[2 \left(\kappa _1+2\lambda _{\chi \xi }\right)^2\\
&+288 \lambda _{\xi }^2+4 \lambda _{\chi \xi}^2+8 \lambda _{\phi \xi }^2\Big],\\
\xi^+\xi^-\to\chi^+\chi^- =&-32 \pi ^2 \lambda _{\chi \xi }+6 \beta _{\lambda _{\chi \xi }}\\
&+4(i\pi-1)\Big[2 \lambda _{\chi \xi } \left(\tilde{\lambda }_{\chi}+5 \lambda _{\xi }+2 \lambda _{\chi }\right)+\kappa _1 \left(2 \lambda_{\xi }+\lambda _{\chi }\right)+\lambda _{\phi \xi } \lambda _{\phi \chi}\Big],\\
\xi^+\xi^-\to\chi^{++}\chi^{--} =&-16 \pi ^2\left(\kappa _1+2 \lambda _{\chi \xi }\right)+3 \left(\beta_{\kappa_1}+2 \beta _{\lambda _{\chi \xi }}\right)+(i\pi-1)\Big[4 \lambda _{\phi \xi } \lambda _{\phi \chi }\\
&+2 \kappa _1 \left(2 \tilde{\lambda }_{\chi }+8\lambda _{\xi }+3 \lambda _{\chi }\right)+8 \lambda _{\chi \xi }\left(\tilde{\lambda }_{\chi }+5 \lambda _{\xi }+2 \lambda _{\chi}\right)\Big],
\end{align*}
\begin{align*}
\chi^+\chi^-\to\xi^+\xi^-&=\xi^+\xi^-\to\chi^+\chi^-,\\
\chi^{++}\chi^{--} \to \xi^+\xi^-&=\xi^+\xi^-\to\chi^{++}\chi^{--}  ,
\end{align*}
\begin{align*}
\chi^+\chi^-\to\chi^+\chi^- =&-64 \pi ^2 \left(\tilde{\lambda} _{\chi }+\lambda _{\chi}\right)+12 \left( \beta _{\tilde{\lambda}_{\chi }}+ \beta _{\lambda_{\chi}}\right)+2(i\pi-1)\Big[8 \left(\tilde{\lambda }_{\chi}+\lambda _{\chi }\right)^2\\
&+\left(\kappa _1+\lambda _{\chi \xi}\right){}^2+4 \lambda _{\chi }^2+2 \lambda _{\chi \xi }^2+\lambda _{\phi\chi }^2\Big],\\
\chi^+\chi^-\to\chi^{++}\chi^{--} =&-32 \pi ^2 \lambda _{\chi }+6 \beta_{\lambda _{\chi }}+2(i\pi-1)\Big[8 \lambda _{\chi } \tilde{\lambda }_{\chi}+10 \lambda _{\chi }^2\\
&+\lambda _{\chi \xi } \left(2 \kappa _1+3 \lambda _{\chi \xi}\right)+\lambda _{\phi \chi }^2\Big],
\end{align*}
\begin{align*}
\chi^{++}\chi^{--} \to \chi^+\chi^-&=\chi^+\chi^-\to\chi^{++}\chi^{--},
\end{align*}
\begin{align}
\chi^{++}\chi^{--}\to\chi^{++}\chi^{--} =&-64 \pi ^2 \lambda _{\chi }+12 \beta_{\lambda _{\chi }}+(i\pi-1)\Big[4 \kappa _1 \lambda_{\chi \xi }\notag\\
&+2 \left(8 \tilde{\lambda }_{\chi }^2+8 \lambda_{\chi } \tilde{\lambda }_{\chi }+12 \lambda _{\chi }^2+3 \lambda_{\chi \xi }^2+\lambda _{\phi \chi }^2\right)+\kappa _1^2+\frac{1}{2}\kappa _2^2\Big].
\end{align}
\item[] {$\bullet$ \,\boldmath $Q=0, Y=1/2$ : }
\begin{align*}
\phi^{+}\xi^-\to\phi^{+}\xi^- =& -32\pi^2\lambda_{\phi\xi}+6 \beta _{\lambda _{\phi \xi }}+(i\pi-1)\Big[3 \kappa _3^2+4 \lambda _{\phi \xi}^2\Big],\\
\phi^{+}\xi^-\to\phi^{0}\xi^0 =& \sqrt{2}(i\pi-1)\kappa _3^2,\\
\phi^{+}\xi^-\to\chi^{+}\phi^- =& -16\sqrt{2}\pi^2\kappa_3+3 \sqrt{2} \beta _{\kappa _3}+ \frac{1}{\sqrt{2}}(i\pi-1)\Big[\kappa _3 \left(\kappa _2+4 \lambda _{\phi \xi }+2\lambda _{\phi \chi }\right)\Big],\\
\phi^{+}\xi^-\to\phi^{0*}\chi^0 =& -16\pi^2\kappa_3+3 \beta _{\kappa _3}+ (i\pi-1)\Big[\frac{1}{2}\kappa _3 \left(3 \kappa _2+4 \lambda _{\phi \xi}+2 \lambda _{\phi \chi }\right)\Big],
\end{align*}
\begin{align*}
\phi^{0}\xi^0\to\phi^{+}\xi^-&=\phi^{+}\xi^-\to\phi^{0}\xi^0,\\
\chi^{+}\phi^-\to \phi^{+}\xi^-&=\phi^{+}\xi^-\to\chi^{+}\phi^-,\\
\phi^{0*}\chi^0\to\phi^{+}\xi^-&=\phi^{+}\xi^-\to\phi^{0*}\chi^0,
\end{align*}
\begin{align*}
\phi^{0}\xi^0\to\phi^{0}\xi^0 =& -32\pi^2\lambda_{\phi\xi}+6 \beta _{\lambda _{\phi \xi }}+2(i\pi-1)\Big[\kappa _3^2+2 \lambda _{\phi \xi}^2\Big],\\
\phi^{0}\xi^0\to\chi^{+}\phi^- =& (i\pi-1)\kappa _2 \kappa _3,\\
\phi^{0}\xi^0\to\phi^{0*}\chi^0 =& -16\sqrt{2}\pi^2\kappa_3+3 \sqrt{2} \beta _{\kappa _3}+\frac{1}{\sqrt{2}}(i\pi-1)\Big[\kappa _3 \left(\kappa _2+4 \lambda _{\phi \xi }+2\lambda _{\phi \chi }\right)\Big],
\end{align*}
\begin{align*}
\chi^{+}\phi^-\to\phi^{0}\xi^0&=\phi^{0}\xi^0\to\chi^{+}\phi^-,\\
\phi^{0*}\chi^0\to\phi^{0}\xi^0&=\phi^{0}\xi^0\to\phi^{0*}\chi^0,
\end{align*}
\begin{align*}
\chi^{+}\phi^-\to\chi^{+}\phi^- =&  -16\pi^2\lambda_{\phi\chi}+3 \beta _{\lambda _{\phi \chi }}+(i\pi-1)\Big[\frac{\kappa _2^2}{2}+2 \kappa _3^2+\lambda_{\phi \chi }^2\Big],\\
\chi^{+}\phi^-\to\phi^{0*}\chi^0 =& -8\sqrt{2}\pi^2\kappa_2+\frac{3}{\sqrt{2}}\beta _{\kappa_2}+\frac{1}{2\sqrt{2}}(i\pi-1)\Big[4\kappa _2 \lambda _{\phi \chi }+\kappa _2^2+4\kappa _3^2\Big],
\end{align*}
\begin{align*}
\phi^{0*}\chi^0\to\chi^{+}\phi^-&=\chi^{+}\phi^-\to\phi^{0*}\chi^0,
\end{align*}
\begin{align}
\phi^{0*}\chi^0\to\phi^{0*}\chi^0 =& -16\pi^2\left(\lambda_{\phi\chi}+\frac{1}{2}\kappa_2\right)+3\left(\beta _{\lambda_{\phi \chi }}+\frac{1}{2}\beta _{\kappa _2}\right)\notag\\
&+(i\pi-1)\Big[\kappa _2\lambda _{\phi \chi }+\frac{3 \kappa _2^2}{4}+3\kappa _3^2+\lambda _{\phi \chi }^2\Big].
\end{align}
\item[] {$\bullet$ \,\boldmath $Q=0, Y=1$ :}
\begin{align*}
\chi^{+}\xi^-\to\chi^{+}\xi^- &= -32 \pi ^2 \lambda _{\chi \xi }+6 \beta _{\lambda _{\chi \xi }}+(i\pi-1)\Big[\kappa _1^2+\kappa _3^2+4 \lambda _{\chi \xi}^2\Big],\\
\chi^{+}\xi^-\to\chi^0\xi^0 &= 16 \pi ^2 \kappa _1-3 \beta _{\kappa _1}+(i\pi-1)\Big[\kappa _3^2-4 \kappa _1 \lambda _{\chi \xi}\Big],\\
\chi^{+}\xi^-\to\frac{1}{\sqrt{2}}(\phi^{0}\phi^0) &=-16 \pi^2 \kappa _3+3 \beta _{\kappa _3}+(i\pi-1)\Big[\kappa _3 \left(2 \left(\lambda _{\chi \xi}+\lambda _{\phi }\right)-\kappa _1\right)\Big],
\end{align*}
\begin{align*}
\chi^0\xi^0\to\chi^{+}\xi^-&=\chi^{+}\xi^-\to\chi^0\xi^0,\\
\frac{1}{\sqrt{2}}(\phi^{0}\phi^0)\to\chi^{+}\xi^-&=\chi^{+}\xi^-\to\frac{1}{\sqrt{2}}(\phi^{0}\phi^0),
\end{align*}
\begin{align*}
\chi^{0}\xi^0\to\chi^{0}\xi^0 &= -32 \pi ^2 \lambda _{\chi \xi }+6 \beta _{\lambda _{\chi \xi }}+(i\pi-1)\Big[\kappa _1^2+\kappa _3^2+4 \lambda _{\chi \xi}^2\Big],\\
\chi^{0}\xi^0\to\frac{1}{\sqrt{2}}(\phi^{0}\phi^0) &=-16 \pi^2 \kappa _3+3 \beta _{\kappa _3} +(i\pi-1)\Big[\kappa _3 \left(2 \left(\lambda _{\chi \xi}+\lambda _{\phi }\right)-\kappa _1\right)\Big],
\end{align*}
\begin{align*}
\frac{1}{\sqrt{2}}(\phi^{0}\phi^0)\to\chi^{0}\xi^0&=\chi^{0}\xi^0\to\frac{1}{\sqrt{2}}(\phi^{0}\phi^0),
\end{align*}
\begin{align}
\frac{1}{\sqrt{2}}(\phi^{0}\phi^0) \to\frac{1}{\sqrt{2}}(\phi^{0}\phi^0) &= -32\pi ^2 \lambda _{\phi }+6 \beta _{\lambda _{\phi }}+2(i\pi-1)\Big[\kappa _3^2+2 \lambda _{\phi }^2\Big].
\label{eq:z33}
\end{align}
\item[] {$\bullet$ \,\boldmath $Q=0, Y=\:3/2 $ :}
\begin{align}
\phi^{0}\chi^{0}\to\phi^{0}\chi^{0} =&-16 \pi ^2\left(\frac{1}{2}\kappa _2+\lambda _{\phi \chi }\right)+3\left(\beta _{\lambda _{\phi \chi}}+\frac{1}{2} \beta _{\kappa _2}\right)\notag\\
&+(i\pi-1)\Big[\kappa _2 \lambda _{\phi \chi
   }+\frac{\kappa _2^2}{4}+\lambda _{\phi \chi }^2\Big].
\label{eq:s2}
\end{align}
\item[] {$\bullet$ \,\boldmath $Q=0, Y=\:2 $ :}
\begin{align}
\frac{1}{\sqrt{2}}(\chi^{0}\chi^0)\to\frac{1}{\sqrt{2}}(\chi^{0}\chi^0) &=-32 \pi ^2 \lambda _{\chi }+6 \beta _{\lambda _{\chi }}+4(i\pi-1)\lambda _{\chi }^2.
\label{eq:s3}
\end{align}
\item[] {$\bullet$ \,\boldmath $Q=1, Y=0$ :}
\begin{align*}
\chi^{++}\chi^{-}\to\chi^{++}\chi^{-} &= -32\pi^2\lambda_\chi+6\beta_{\lambda_\chi}+(i\pi-1)\Big[
16\tilde{\lambda}_\chi^2+\kappa_1^2+\frac{\kappa_2^2}{2}+4\lambda_\chi^2\Big],\\
\chi^{++}\chi^{-}\to\chi^{+}\chi^{0*} &= 64\pi^2\tilde{\lambda}_\chi-12\beta_{\tilde{\lambda}_\chi}+\frac{1}{2}(i\pi-1)\Big[
-2\kappa_1^2+\kappa_2^2-32\lambda_\chi\tilde{\lambda}_\chi\Big],\\
\chi^{++}\chi^{-}\to\xi^{+}\xi^{0} &=-16\pi^2\kappa_1+3\beta_{\kappa_1}+2(i\pi-1)\Big[
\kappa_1\left(4\lambda_\xi+\lambda_\chi+2\tilde{\lambda}_\chi\right)\Big],\\
\chi^{++}\chi^{-}\to\phi^{+}\phi^{0*} &= -8\sqrt{2}\kappa_2+\frac{3}{\sqrt{2}}\beta_{\kappa_2}+\sqrt{2}(i\pi-1)\Big[
\kappa_2\left(\lambda_\phi+\lambda_\chi-2\tilde{\lambda}_\chi\right)\Big],
\end{align*}
\begin{align*}
\chi^{+}\chi^{0*}\to\chi^{++}\chi^{-}&=\chi^{++}\chi^{-}\to\chi^{+}\chi^{0*},\\
\xi^{+}\xi^{0}\to\chi^{++}\chi^{-}&=\chi^{++}\chi^{-}\to\xi^{+}\xi^{0},\\
\phi^{+}\phi^{0*}\to\chi^{++}\chi^{-}&=\chi^{++}\chi^{-}\to\phi^{+}\phi^{0*},
\end{align*}
\begin{align*}
\chi^{+}\chi^{0*}\to\chi^{+}\chi^{0*} &= -32\pi^2\lambda_\chi+6\beta_{\lambda_\chi}
+(i\pi-1)\Big[
\kappa_1^2+\frac{\kappa_2^2}{2}+4\lambda_\chi^2+16\tilde{\lambda}_\chi^2\Big],\\
\chi^{+}\chi^{0*}\to\xi^{+}\xi^{0} &= 16\pi^2\kappa_1-3\beta_{\kappa_1}-2(i\pi-1)\Big[
\kappa_1\left(4\lambda_\xi+\lambda_\chi+2{\tilde{\lambda}}_\chi\right)\Big],\\
\chi^{+}\chi^{0*}\to\phi^{+}\phi^{0*} &= -8\sqrt{2}\pi^2\kappa_2+\frac{3}{\sqrt{2}}\beta_{\kappa_2}+\sqrt{2}(i\pi-1)\Big[
\kappa_2\left(\lambda_\phi+\lambda_\chi-2\tilde{\lambda}_\chi\right)\Big],
\end{align*}
\begin{align*}
\xi^{+}\xi^{0}\to\chi^{+}\chi^{0*}&=\chi^{+}\chi^{0*}\to\xi^{+}\xi^{0},\\
\phi^{+}\phi^{0*}\to\chi^{+}\chi^{0*}&=\chi^{+}\chi^{0*}\to\phi^{+}\phi^{0*},
\end{align*}
\begin{align}
\xi^{+}\xi^{0}\to\xi^{+}\xi^{0} &=-128\pi^2\lambda_\xi+24\beta_{\lambda_\xi}+2(i\pi-1)\Big[
\kappa_1^2+32\lambda_\xi^2\Big],\notag\\
\xi^{+}\xi^{0}\to\phi^{+}\phi^{0*} &=0=\phi^{+}\phi^{0*}\to\xi^{+}\xi^{0},\notag\\
\phi^{+}\phi^{0*}\to\phi^{+}\phi^{0*} &= -32\pi^2\lambda_\phi+6\beta_{\lambda_\phi}+(i\pi-1)\Big[
\kappa_2^2+4\lambda_\phi^2\Big].
\end{align}
\item[] {$\bullet$ \,\boldmath $Q=1, Y=1/2$ :}
\begin{align*}
\phi^{0}\xi^{+}\to\phi^{0}\xi^{+} &=-32\pi^2\lambda_{\phi\xi}+6\beta_{\lambda_{\phi\xi}}+ (i\pi-1)\Big[ 3\kappa_3^2+4\lambda_{\phi\xi}^2\Big],\\
\phi^{0}\xi^{+}\to\phi^{+}\xi^{0} &=-\sqrt{2}(i\pi-1)\kappa_3^2,\\
\phi^{0}\xi^{+}\to\chi^{+}\phi^{0*} &=-16\sqrt{2}\pi^2\kappa_3+3\sqrt{2}\beta_{\kappa_3}+\frac{(i\pi-1)}{\sqrt{2}}\Big[
\kappa_3\left(\kappa_2+4\lambda_{\phi\xi}+2\lambda_{\phi\chi}\right),\\
\phi^{0}\xi^{+}\to\phi^{-}\chi^{++} &=-16\pi^2\kappa_3+3\beta_{\kappa_3}+(i\pi-1)\Big[
\frac{1}{2}\kappa_3\left(3\kappa_2+4\lambda_{\phi\xi}+2\lambda_{\phi\chi}\right)\Big],
\end{align*}
\begin{align*}
\phi^{+}\xi^{0}\to\phi^{0}\xi^{+}&=\phi^{0}\xi^{+}\to\phi^{+}\xi^{0},\\
\chi^{+}\phi^{0*}\to\phi^{0}\xi^{+}&=\phi^{0}\xi^{+}\to\chi^{+}\phi^{0*} ,\\
\phi^{-}\chi^{++}\to\phi^{0}\xi^{+}&=\phi^{0}\xi^{+}\to\phi^{-}\chi^{++} ,
\end{align*}
\begin{align*}
\phi^{+}\xi^{0}\to\phi^{+}\xi^{0} &=-32\pi^2\lambda_{\phi\xi}+6\beta_{\lambda_{\phi\xi}}+ 2(i\pi-1)\Big[
\kappa_3^2+2\lambda_{\phi\xi}^2\Big],\\
\phi^{+}\xi^{0}\to\chi^{+}\phi^{0*} &=-(i\pi-1)\kappa_2\kappa_3,\\
\phi^{+}\xi^{0}\to\phi^{-}\chi^{++} &=16\sqrt{2}\pi^2\kappa_3-3\sqrt{2}\beta_{\kappa_3}-\frac{1}{\sqrt{2}}(i\pi-1)\Big[
\kappa_3\left(\kappa_2+4\lambda_{\phi\xi}+2\lambda_{\phi\chi}\right)\Big],
\end{align*}
\begin{align*}
\chi^{+}\phi^{0*}\to\phi^{+}\xi^{0}&=\phi^{+}\xi^{0}\to\chi^{+}\phi^{0*},\\
\phi^{-}\chi^{++}\to\phi^{+}\xi^{0}&=\phi^{+}\xi^{0}\to\phi^{-}\chi^{++} ,
\end{align*}
\begin{align*}
\phi^{-}\chi^{++}\to\phi^{-}\chi^{++} =&-16\pi^2\left(\lambda_{\phi\chi}+\frac{\kappa_2}{2}\right)+3\left(\beta_{\lambda_{\phi\chi}}+\frac{1}{2}\beta_{\kappa_2}\right)\\
&+(i\pi-1)\Big[\frac{3}{4}\kappa_2^2+3\kappa_3^2+\kappa_2\lambda_{\phi\chi}+\lambda_{\phi\chi}^2\Big],\\
\phi^{-}\chi^{++}\to\chi^{+}\phi^{0*} =&-8\sqrt{2}\kappa_2+\frac{3}{\sqrt{2}}\beta_{\kappa_2}+\frac{1}{2\sqrt{2}}(i\pi-1)\Big[\kappa_2^2+4\kappa_3^2+4\kappa_2\lambda_{\phi\chi}\Big],
\end{align*}
\begin{align*}
\chi^{+}\phi^{0*}\to\phi^{-}\chi^{++}&=\phi^{-}\chi^{++}\to\chi^{+}\phi^{0*} ,
\end{align*}
\begin{align}
\chi^{+}\phi^{0*}\to\chi^{+}\phi^{0*} &=-16\pi^2\lambda_{\phi\chi}+3\beta_{\lambda_{\phi\chi}}+(i\pi-1)\Big[
\frac{\kappa_2^2}{2}+2\kappa_3^2+\lambda_{\phi\chi}^2\Big].
\end{align}
\item[] {$\bullet$ \,\boldmath $Q=1, Y=1$ :}
\begin{align*}
\phi^{+}\phi^{0}\to\phi^{+}\phi^{0} &= -32\pi^2\lambda_\phi+6\beta_{\lambda_\phi}+2(i\pi-1)\Big[
\kappa_3^2+2\lambda_\phi^2\Big],\\
 \phi^{+}\phi^{0}\to\chi^{+}\xi^{0} &= 0=\chi^{+}\xi^{0}\to \phi^{+}\phi^{0},\\
 \phi^{+}\phi^{0}\to\chi^{0}\xi^{+} &= -16\pi^2\kappa_3+3\beta_{\kappa_3}+(i\pi-1)\Big[\kappa_3\left(-\kappa_1+2(\lambda_\phi+\lambda_{\chi\xi})\right)\Big],\\
 \phi^{+}\phi^{0}\to\chi^{++}\xi^{-} &=  -16\pi^2\kappa_3+3\beta_{\kappa_3}+(i\pi-1)\Big[\kappa_3\left(-\kappa_1+2(\lambda_\phi+\lambda_{\chi\xi})\right)\Big],
\end{align*}
\begin{align*}
\chi^{0}\xi^{+}\to\phi^{+}\phi^{0}&=\phi^{+}\phi^{0}\to\chi^{0}\xi^{+} ,\\
\chi^{++}\xi^{-}\to\phi^{+}\phi^{0}&= \phi^{+}\phi^{0}\to\chi^{++}\xi^{-} ,
\end{align*}
\begin{align*}
 \chi^{+}\xi^{0}\to\chi^{+}\xi^{0} &=-32\pi^2\left(\lambda_{\chi\xi}+\kappa_1\right) +6(\beta_{\lambda_{\chi\xi}}+\beta_{\kappa_1})+(i\pi-1)\Big[ 2\kappa_1^2+4(\kappa_1+\lambda_{\chi\xi})^2\Big],\\
 \chi^{+}\xi^{0}\to\chi^{++}\xi^{-} &= -16\pi^2\kappa_1+3\beta_{\kappa_1}+(i\pi-1)\Big[
\kappa_1\left(5\kappa_1+4\lambda_{\chi\xi}\right)\Big],\\
   \chi^{+}\xi^{0}\to\chi^{0}\xi^{+} &= 16\pi^2\kappa_1-3\beta_{\kappa_1}-(i\pi-1)\Big[
\kappa_1\left(5\kappa_1+4\lambda_{\chi\xi}\right)\Big],
\end{align*}
\begin{align*}
\chi^{++}\xi^{-}\to\chi^{+}\xi^{0}&= \chi^{+}\xi^{0}\to\chi^{++}\xi^{-} ,\\
\chi^{0}\xi^{+}\to  \chi^{+}\xi^{0}&=    \chi^{+}\xi^{0}\to\chi^{0}\xi^{+}  ,
\end{align*}
\begin{align*}
  \chi^{++}\xi^{-}\to\chi^{++}\xi^{-} =& -16\pi^2(\kappa_1+2\lambda_{\chi\xi})+3\left(\beta_{\kappa_1}+2\beta_{\lambda_{\chi\xi}}\right)\\
 & +(i\pi-1)\Big[
5\kappa_1^2+\kappa_3^2+(\kappa_1+2\lambda_{\chi\xi})^2\Big],\\
\chi^{++}\xi^{-}\to\chi^{0}\xi^{+} &=32\pi^2\kappa_1-6\beta_{\kappa_1}+ (i\pi-1)\Big[
-5\kappa_1^2+\kappa_3^2-8\kappa_1\lambda_{\chi\xi}\Big],
\end{align*}
\begin{align*}
\chi^{0}\xi^{+}\to  \chi^{++}\xi^{-}&=\chi^{++}\xi^{-}\to\chi^{0}\xi^{+},
\end{align*}
\begin{align}
 \chi^{0}\xi^{+}\to\chi^{0}\xi^{+} =&-16\pi^2\left(2\lambda_{\chi\xi}+\kappa_1\right)+3(2\beta_{\chi\xi}+\beta_{\kappa_1})\notag\\
& +(i\pi-1)\Big[
5\kappa_1^2+\kappa_3^2+(\kappa_1+2\lambda_{\chi\xi})^2\Big].
 \label{eq:Q44}
 \end{align}
\item[] {$\bullet$ \,\boldmath $Q=1, Y=\:3/2 $ : }
\begin{align*}
\phi^{+}\chi^{0}\to\phi^{0}\chi^{+} =& -8 \sqrt{2} \pi ^2 \kappa_2+\frac{3}{\sqrt{2}}\beta_{\kappa_2}-\frac{1}{2\sqrt{2}}(i\pi-1)\Big[\kappa _2 \left(\kappa _2-4\lambda _{\phi \chi }\right)\Big],\\
\phi^{0}\chi^{+}\to\phi^{0}\chi^{+} =& -16 \pi ^2 \lambda_{\phi \chi }+3 \beta _{\lambda _{\phi \chi}}+(i\pi-1)\Big[\frac{\kappa_2^2}{2}+\lambda _{\phi \chi}^2\Big],
\end{align*}
\begin{align*}
\phi^{0}\chi^{+}\to\phi^{+}\chi^{0}&=\phi^{+}\chi^{0}\to\phi^{0}\chi^{+},
\end{align*}
\begin{align}
\phi^{+}\chi^{0}\to\phi^{+}\chi^{0} =&-16 \pi ^2 \left(\lambda_{\phi \chi }-\frac{1}{2}\kappa_2\right)+3 \left(\beta _{\lambda _{\phi \chi}}-\frac{1}{2}\beta _{\kappa_2}\right)\notag\\
&+(i\pi-1)\Big[-\kappa _2 \lambda _{\phi\chi }+\frac{3 }{4}\kappa_2^2+\lambda _{\phi \chi}^2\Big].
\end{align}
\item[] {$\bullet$ \,\boldmath $Q=1, Y= 2$ : }
\begin{align}
\chi^{+}\chi^{0}\to\chi^{+}\chi^{0} &= -32\pi^2\lambda_\chi+6\beta_{\lambda_\chi}+4(i\pi-1)\lambda_\chi^2.
\label{eq:s1}
 \end{align}
\item[] {$\bullet$ \,\boldmath $Q=2, Y=0 $ : }
\begin{align*}
\chi^{++}\chi^{0*}\to\chi^{++}\chi^{0*} =&  -32\pi^2(\lambda_\chi + 2 \tilde{\lambda}_\chi)+6\left(\beta_{\ld_\chi}+2\beta_{\tilde{\ld}_{\chi}}\right)\\
&+ (i\pi-1)\Big[2\kappa_1^2 +4\left(\lambda_\chi + 2 \tilde{\lambda}_\chi\right)^{2} \Big],\\
\chi^{++}\chi^{0*}\to\frac{1}{\sqrt{2}}(\xi^{+}\xi^+) =&  16\sqrt{2}\pi^2\kappa_1-3\sqrt{2}\beta_{\kappa_{1}} \\
&+\frac{1}{\sqrt{2} }(i\pi-1)\Big[-4\kappa_1\left(4\lambda_{\xi}+\lambda_{\chi}+2\tilde{\lambda}_{\chi}\right)\Big],
\end{align*}
\begin{align*}
\frac{1}{\sqrt{2}}(\xi^{+}\xi^+)\to\chi^{++}\chi^{0*}&=\chi^{++}\chi^{0*}\to\frac{1}{\sqrt{2}}(\xi^{+}\xi^+),
\end{align*}
\begin{align}
 \frac{1}{\sqrt{2}}(\xi^{+}\xi^+)\to\frac{1}{\sqrt{2}}(\xi^{+}\xi^+) &=  -128\pi^2\lambda_{\xi}+24\beta_{\ld_{\xi}}+2(i\pi-1)\Big[\left(\kappa_{1}^{2}+32{\lambda_{\xi}}^{2}\right)\Big]. 
\end{align}
\item[] {$\bullet$ \,\boldmath $Q=2, Y=1/2$ : }
\begin{align*}
 \phi^{+}\xi^+\to\phi^{+}\xi^+ &= -32\pi^2 \lambda_{\phi \xi}+6\beta_{\ld_{\phi\xi}}+(i\pi-1)\Big[\left(\kappa_{3}^{2}+4 \lambda_{\phi \xi}^{2}\right)\Big],\\
 \phi^{+}\xi^+\to\chi^{++}\phi^{0*} &=-16\pi^2\kappa_{3}+3\beta_{\kappa_{3}}+(i\pi-1)\Big[
 \kappa_{3}\left(-\frac{\kappa_2}{2}+2\lambda_{\phi \xi}+\lambda_{\phi\chi}\right)\Big],
\end{align*}
\begin{align*}
\chi^{++}\phi^{0*}\to\phi^{+}\xi^+&= \phi^{+}\xi^+\to\chi^{++}\phi^{0*},
\end{align*}
\begin{align}
\chi^{++}\phi^{0*}\to\chi^{++}\phi^{0*} =&-16\pi^2\left(\lambda_{\phi\chi}-\frac{1}{2}\kappa_{2}\right)+3\left(\beta_{\lambda_{\phi\chi}}-\frac{1}{2}\beta_{\kappa_{2}}\right)\notag\\
&+(i\pi-1)\Big[ \kappa_{3}^{2}+\left(-\frac{\kappa_{2}}{2}+\lambda_{\phi\chi}\right)^{2} \Big].  
\end{align}
\item[] {$\bullet$ \,\boldmath $Q=2, Y=3/2$ :}
\begin{align*}
\phi^{+}\chi^{+}\to\phi^{+}\chi^{+} &=  -16\pi^2 \lambda_{\phi\chi}+ 3\beta_{\lambda_{\phi\chi}} + (i\pi-1)\Big[ \frac{1}{2} \kappa_{2}^{2}+ \lambda_{\phi \chi}^{2}\Big] ,\\
\phi^{+}\chi^{+}\to\chi^{++}\phi^{0} &=  -8\sqrt{2}\pi^2 \kappa_{2} + \frac{3}{\sqrt{2}}\beta_{\kappa_{2}}-\frac{ 1}{2\sqrt{2}}(i\pi-1)\Big[\kappa_{2}\left(\kappa_{2}-4\lambda_{\phi\chi}\right)\Big] ,
\end{align*}
\begin{align*}
\chi^{++}\phi^{0}\to\phi^{+}\chi^{+}&=\phi^{+}\chi^{+}\to\chi^{++}\phi^{0},
\end{align*}
\begin{align}
 \chi^{++}\phi^{0}\to\chi^{++}\phi^{0} =& -16\pi^2\left(\lambda_{\phi\chi}-\frac{\kappa_{2}}{2}\right)+3\left(\beta_{\lambda_{\phi\chi}}-\frac{1}{2}\beta_{\kappa_{2}}\right)\notag\\
 &+(i\pi-1)\Big[\frac{3\kappa_{2}^{2}}{4}-\kappa_{2}\lambda_{\phi\chi}+\lambda_{\phi\chi}^{2}\Big].
\end{align}
\item[] {$\bullet$ \,\boldmath $Q=2, Y= 2$ : }
\begin{align*}
\chi^{++}\chi^{0}\to\chi^{++}\chi^{0} =& -32\pi^2\left(\lambda_\chi+2\tilde{\lambda}_\chi\right) +6\left(\beta_{\lambda_\chi}+2\beta_{\tilde{\lambda}_\chi}\right)\\
&+(i\pi-1)\Big[ 8{\tilde{\lambda}_\chi}^2+4\left(\lambda_\chi+2\tilde{\lambda}_\chi\right)^2\Big],\\
\chi^{++}\chi^{0}\to\frac{1}{\sqrt{2}}(\chi^{+}\chi^+) =& 32\sqrt{2}\pi^2\tilde{\lambda}_{\chi}-6\sqrt{2}\beta_{\tilde{\lambda}_\chi}+(i\pi-1)\Big[-4\sqrt{2}\tilde{\lambda}_\chi\left(2 \lambda_\chi+3\tilde{\lambda}_\chi\right)\Big],
\end{align*}
\begin{align*}
\frac{1}{\sqrt{2}}(\chi^{+}\chi^+)\to\chi^{++}\chi^{0}&=\chi^{++}\chi^{0}\to\frac{1}{\sqrt{2}}(\chi^{+}\chi^+),
\end{align*}
\begin{align}
\frac{1}{\sqrt{2}}(\chi^{+}\chi^+)\to\frac{1}{\sqrt{2}}(\chi^{+}\chi^+) =&-32\pi^2\left(\lambda_\chi+\tilde{\lambda}_{\chi}\right)+6\left(\beta_{\lambda_\chi}+\beta_{\tilde{\lambda}_\chi}\right)\notag\\
&+4(i\pi-1)\Big[2{\tilde{\lambda}_\chi}^2+\left(\tilde{\lambda}_\chi+\lambda_\chi\right)^2\Big].
\end{align}
\item[] {$\bullet$ \,\boldmath $Q=2, Y= 1$ : }
 \begin{align*}
\frac{1}{\sqrt{2}}\left(\phi^{+}\phi^+\right)\to\chi^{++}\xi^0 =&16\pi^2\kappa_3-3\beta_{\kappa_3}+(i\pi-1)\Big[\kappa_3\left(\kappa_1-2\left(\lambda_\phi+\lambda_{\chi \xi}\right)\right)\Big],\\
 \frac{1}{\sqrt{2}}\left(\phi^{+}\phi^+\right)\to\chi^{+}\xi^+ =&-16\pi^2\kappa_{3}+3\beta_{\kappa_{3}}+(i\pi-1)\Big[ \kappa_3\left(-\kappa_1+2\left(\lambda_\phi+\lambda_{\chi \xi}\right)\right)\Big],\\
 \frac{1}{\sqrt{2}}\left(\phi^{+}\phi^+\right)\to\frac{1}{\sqrt{2}}\left(\phi^{+}\phi^+\right) =&-32\pi^2 \lambda_\phi+6\beta_{\lambda_\phi}+2(i\pi-1)\Big[ \kappa_3^2+2\lambda_\phi^2 \Big],
\end{align*}
 \begin{align*}
\chi^{++}\xi^0\to\frac{1}{\sqrt{2}}\left(\phi^{+}\phi^+\right)&= \frac{1}{\sqrt{2}}\left(\phi^{+}\phi^+\right)\to\chi^{++}\xi^0,\\
 \chi^{+}\xi^+\to \frac{1}{\sqrt{2}}\left(\phi^{+}\phi^+\right)&= \frac{1}{\sqrt{2}}\left(\phi^{+}\phi^+\right)\to\chi^{+}\xi^+,
 \end{align*}
\begin{align*}
\chi^{++}\xi^0\to\chi^{++}\xi^0 & = -32 \pi^2 \lambda_{\chi\xi}+6\beta_{\lambda_{\chi\xi}}+(i\pi-1)\Big[\kappa_1^2+\kappa_3^2+4\lambda_{\chi\xi}^2\Big],\\
\chi^{++}\xi^0\to\chi^{+}\xi^+ & = -16\pi^2\kappa_1+3\beta_{\kappa_1}+(i\pi-1)\Big[-\kappa_3^2+4\kappa_1\lambda_{\chi\xi}\Big],
 \end{align*}
\begin{align*}
\chi^{+}\xi^+\to\chi^{++}\xi^0&=\chi^{++}\xi^0\to\chi^{+}\xi^+,
 \end{align*}
\begin{align}
\chi^{+}\xi^+\to\chi^{+}\xi^+ & =-32\pi^2\lambda_{\chi\xi}+6\beta_{\lambda_{\chi\xi}}+(i\pi-1)\Big[\Big.
 \begin{aligned}[t]
 \kappa_1^2+\kappa_3^2+4\lambda_{\chi\xi}^2\Big].
 \end{aligned}
 \label{eq:E33}
 \end{align}
\item[] {$\bullet$ \,\boldmath $Q=3, Y=1$ :}
\begin{align}
\chi^{++}\xi^{+}\to\chi^{++}\xi^{+} =&-16\pi^2\left( 2\lambda_{\chi\xi}+\kappa_1\right)+3\left(2\beta_{\lambda_{\chi\xi}}+\beta_{\kappa_1}\right)\notag\\
&+(i\pi-1)\Big[4 \kappa _1 \lambda _{\chi \xi}+\kappa _1^2+4 \lambda _{\chi \xi}^2\Big].
\label{eq:1.9.1}
\end{align}
\item[] {$\bullet$ \,\boldmath $Q=3, Y=3/2$ :}
\begin{align}
\chi^{++}\phi^{+}\to\chi^{++}\phi^{+} =&-16\pi^2\left( \lambda_{\phi\chi}+\frac{\kappa_2}{2}\right)+3\left(\beta_{\lambda_{\phi\chi}}+\frac{1}{2}\beta_{\kappa_2}\right)\notag\\
&+(i\pi-1)\Big[\kappa _2 \lambda _{\phi \chi}+\frac{\kappa _2^2}{4}+\lambda_{\phi \chi }^2\Big].
\label{eq:1.9.2}
\end{align}
\item[] {$\bullet$ \,\boldmath $Q=3,Y=2$ :}
\begin{align}
\chi^{++}\chi^{+}\to\chi^{++}\chi^{+} &=-32\pi^2\lambda_\chi+6\beta_{\lambda_\chi}+4(i\pi-1)\lambda_\chi^2.
\label{eq:1.9.3}
\end{align}
\item[] {$\bullet$ \,\boldmath $Q=4, Y=2$ :}
\begin{align}
\frac{1}{\sqrt{2}}\left(\chi^{++}\chi^{++}\right)\to\frac{1}{\sqrt{2}}\left(\chi^{++}\chi^{++}\right) &=-32\pi^2\lambda_\chi+6\beta_{\lambda_\chi}+4(i\pi-1)\lambda_\chi^2\,.
\label{eq:1.10}
\end{align}
\end{itemize}
\section{Two-loop renormalization group equations}\label{app:2-loop_rge}
In this appendix, we list the renormalization group equations (RGEs) for the quartic couplings of the most general potential given in Eq.~(\ref{v16}), obtained using the public code PyR@TE~\cite{Sartore:2020gou}. For any coupling $A$, the complete $\beta$ function up to two-loop order can be sub-divided into leading and next-to-leading order contributions separately as,
\begin{equation}
\beta_A\equiv \frac{dA}{d\ln\mu}=\frac{1}{16\pi^2}\beta_A^{(0)}+\frac{1}{(16\pi^2)^2}\beta_A^{(1)}\,.
\end{equation}
The two-loop RGEs for the Yukawa couplings are the same as in the SM~\cite{Luo:2002ey} and the type-I 2HDM~\cite{Chowdhury:2015yja}, since the contributions of Higgs triplets to the Yukawa Lagrangian are not considered in this analysis~\cite{Chiang:2012cn}.  The RGEs of the gauge couplings up to two-loop order are given by, 
\begin{align*}
\beta_{g_1}^{(0)}&=\frac{47}{6}g_1^3,\quad &\beta_{g_1}^{(1)}&=\frac{1}{18}g_1^3 \Big(415 g_1^2 + 513 g_2^2 + 264 g_3^2 -15 y_b^2 -51 y_t^2 -45 y_\tau^2\Big),\\[10pt]
\beta_{g_2}^{(0)}&=-\frac{13}{6}g_2^3,\quad &\beta_{g_2}^{(1)}&=\frac{1}{6}g_2^3 \Big(57 g_1^2 + 203 g_2^2 + 72 g_3^2 -9 y_b^2 -9 y_t^2 -3 y_\tau^2\Big),\\[10pt]
\beta_{g_3}^{(0)}&=-7g_3^3,\quad &\beta_{g_2}^{(1)}&=\frac{1}{6}g_3^3 \Big(11 g_1^2 + 27 g_2^2 - 156 g_3^2 -12 y_b^2 -12 y_t^2 \Big),\\[10pt]
\end{align*}

The running of the quartic couplings is given by the following:
\begin{align*}
\beta_{\lambda_\phi}^{(0)} &= +24\lambda _{\phi }^2+\frac{3}{8} \Big(g_1^4+2 g_1^2 g_2^2+3 g_2^4\Big)  - 6 y_b^4 - 6 y_t^4 - 2 y_{\tau }^4  \\
& \quad+\lambda _{\phi } \Big(12y_b^2+12y_t^2+4y_{\tau }^2-3 g_1^2-9 g_2^2\Big)+6 \lambda _{\phi \xi }^2+3
\lambda _{\phi \chi }^2+ \frac{1}{2}\kappa _2^2+2 \kappa _3^2\,,\\[25pt]
\beta_{\lambda_\phi}^{(1)} &= -312 \lambda _{\phi }^3+\frac{1}{48} g_1^4 \Big(60 y_b^2-643 g_2^2-228
   y_t^2-300 y_{\tau }^2\Big)+\frac{1}{48} g_1^2
   \Big(24 g_2^2 \Big(9 y_b^2+21 y_t^2+11 y_{\tau
   }^2\Big)\\
   &\quad +64 \Big(y_b^4-2 y_t^4 -3 y_{\tau}^4\Big)-373 g_2^4\Big) -\frac{3}{4} g_2^4
   \Big(3y_b^2+3y_t^2+y_{\tau}^2\Big)+\frac{25}{6} g_1^2 y_b^2 \lambda _{\phi}+\frac{45}{2} g_2^2 y_b^2 \lambda _{\phi }\\
   &\quad +80g_3^2 y_b^2 \lambda _{\phi }-32 g_3^2 y_b^4 
   -42 y_b^2 \lambda _{\phi } y_t^2-6 y_b^2
   y_t^4-6 y_b^4 y_t^2-144 y_b^2 \lambda _{\phi }^2-3
   y_b^4 \lambda _{\phi }+30 y_b^6\\
   &\quad +2 \kappa _3^2
   \Big(g_1^2+10 g_2^2-14 \lambda _{\phi }-20 \lambda_{\phi \xi } -10 \lambda _{\phi \chi }\Big)+\kappa_2^2 \Big(4 g_1^2+5 g_2^2 -7 \lambda _{\phi }-10
   \lambda _{\phi \chi }\Big)\\
   &\quad +2 \kappa _2 \Big(5g_1^2 g_2^2-3 \kappa _3^2\Big)+6 \lambda _{\phi\chi }^2 \Big(4 g_1^2+8 g_2^2-5 \lambda _{\phi}\Big)+\frac{761}{24} g_1^4 \lambda_{\phi }+36
   g_1^2 \lambda _{\phi }^2+\frac{39}{4} g_2^2 g_1^2
   \lambda _{\phi }\\
   &\quad +108 g_2^2 \lambda _{\phi}^2+\frac{59}{8} g_2^4 \lambda _{\phi }+96 g_2^2
   \lambda _{\phi \xi }^2+30 g_2^4 \lambda _{\phi \xi}+15 \Big(g_1^4+2 g_2^4\Big) \lambda _{\phi \chi} +\frac{85}{6} g_1^2 \lambda _{\phi }
   y_t^2\\
   &\quad+\frac{45}{2} g_2^2 \lambda _{\phi }y_t^2+\frac{25}{2} g_1^2 \lambda _{\phi } y_{\tau}^2+\frac{15}{2} g_2^2 \lambda _{\phi } y_{\tau}^2-\frac{463 g_1^6}{48}+\frac{221 g_2^6}{16}+80g_3^2 \lambda _{\phi } y_t^2-32 g_3^2y_t^4\\
   &\quad -60 \lambda _{\phi } \lambda _{\phi \xi
   }^2 -48 \lambda _{\phi \xi}^3-12 \lambda _{\phi \chi }^3-144 \lambda _{\phi}^2 y_t^2-3 \lambda _{\phi } y_t^4+30 y_t^6-48
   \lambda _{\phi }^2 y_{\tau }^2-\lambda _{\phi }
   y_{\tau }^4+10 y_{\tau }^6,\\
\end{align*}
\begin{align*}
\beta_{\lambda_\xi}^{(0)} &=  +88 \lambda _{\xi }^2+3 g_2^4-24 g_2^2 \lambda _{\xi }+2 \kappa _1 \lambda_{\chi \xi }+\kappa _1^2+3\lambda _{\chi \xi }^2+2 \lambda _{\phi \xi }^2,\\[20pt]
\beta_{\lambda_\xi}^{(1)} &=-4416\lambda_{\xi }^3+2 \lambda _{\phi \xi } \Big(2 \lambda _{\phi \xi }
   \Big(-3 y_b^2-3y_t^2+g_1^2+3 g_2^2-4
   \lambda _{\phi \xi }-y_{\tau }^2\Big)+5 g_2^4-4
   \kappa _3^2\Big)\\
   &\quad +4 \kappa _1^2 \Big(2
   g_1^2+g_2^2-12 \lambda _{\xi }-8 \lambda _{\chi \xi}\Big)+4 \kappa _1 \Big(4 \lambda _{\chi \xi }
   \Big(g_1^2+2 g_2^2-5 \lambda _{\xi }\Big)+5
   g_2^4-6 \lambda _{\chi \xi }^2\Big)\\
   &\quad+\lambda _{\xi} \Big(\frac{470}{3}g_2^4+8 \kappa _3^2-80
   \lambda _{\phi \xi }^2\Big)+4 \Big(10
   g_2^4-\kappa _3^2\Big) \lambda _{\chi \xi }+24
   \lambda _{\chi \xi }^2 \Big(g_1^2+2 g_2^2-5
   \lambda _{\xi }\Big)\\
   &\quad+1024 g_2^2 \lambda _{\xi}^2-\frac{83 }{3}g_2^6-12 \kappa _1^3 -24 \lambda _{\chi \xi }^3,
\end{align*}
\begin{align*}
\beta_{\lambda_\chi}^{(0)} &=+28 \lambda _{\chi }^2+12 g_1^2
   \Big(g_2^2-\lambda _{\chi }\Big)-24 g_2^2
   \lambda _{\chi }+6 g_1^4+9 g_2^4+4 \kappa _1
   \lambda _{\chi \xi }+\kappa _1^2+\frac{1}{2}\kappa_2^2\\
   &\quad+6 \lambda _{\chi \xi}^2+16 \tilde{\lambda }_{\chi }^2+16 \lambda _{\chi }\tilde{\lambda }_{\chi }+2 \lambda _{\phi \chi }^2,\\[20pt]
   \beta_{\lambda_\chi}^{(1)} & =-384 \lambda _{\chi }^3 +2 g_2^2 \Big(16
   \Big(8 \lambda _{\chi } \tilde{\lambda }_{\chi }+5
   \tilde{\lambda }_{\chi }^2+11 \lambda _{\chi }^2+3
   \lambda _{\chi \xi }^2\Big)+32 \kappa _1 \lambda
   _{\chi \xi }+5 \kappa _1^2+6 \lambda _{\phi \chi
   }^2\Big)\\
   \nonumber
   &\quad+g_2^4 \Big(128
   \tilde{\lambda }_{\chi }+30 \kappa _1+\frac{638}{3}\lambda _{\chi }+80 \lambda _{\chi \xi }+20
   \lambda _{\phi \chi }\Big)+\frac{1}{3} g_1^2 \Big(6 g_2^2
   \Big(-80 \tilde{\lambda }_{\chi }+5 \kappa _2+68
   \lambda _{\chi }\Big)\\
   &\quad+3 \Big(4 \Big(32
   \tilde{\lambda }_{\chi }^2+32 \lambda _{\chi }
   \tilde{\lambda }_{\chi }+44 \lambda _{\chi
   }^2+\lambda _{\phi \chi }^2\Big)+\kappa
   _2^2\Big)-958 g_2^4\Big)+g_1^4 \Big(80
   \tilde{\lambda }_{\chi }-\frac{1138}{3}g_2^2\\
   &\quad+\frac{811}{3} \lambda _{\chi }+10 \lambda
   _{\phi \chi }\Big)-\kappa _2^2 \Big(3y_b^2+3y_t^2+y_{\tau}^2\Big)-\frac{598 }{3}g_1^6-29 g_2^6-2 \Big[2 \Big(80 \tilde{\lambda }_{\chi }^3+156
   \lambda _{\chi } \tilde{\lambda }_{\chi }^2\\
   &\quad +88
   \lambda _{\chi }^2 \tilde{\lambda }_{\chi }+3 y_b^2
   \lambda _{\phi \chi }^2+\kappa _3^2 \Big(-\lambda
   _{\chi }+2 \lambda _{\chi \xi }+2 \lambda _{\phi
   \chi }\Big)+15 \lambda _{\chi } \lambda _{\chi
   \xi }^2+12 \lambda _{\chi \xi
   }^3+5 \lambda _{\chi } \lambda _{\phi \chi }^2\\
   &\quad+2
   \lambda _{\phi \chi }^3+3 \lambda _{\phi \chi }^2
   y_t^2+\lambda _{\phi \chi }^2 y_{\tau }^2\Big)+8
   \kappa _1^2 \Big(\tilde{\lambda }_{\chi }+\lambda
   _{\chi }+3 \lambda _{\chi \xi }\Big)+\kappa _1
   \Big[\kappa _3^2+4 \lambda _{\chi \xi } (5
   \lambda _{\chi }+6 \lambda _{\chi \xi}\Big)\Big]\\
   &\quad+\kappa _2^2 \Big(3 \lambda _{\chi}+4 \lambda _{\phi \chi }\Big)+7 \kappa_1^3+\kappa _2 \kappa _3^2\Big],
   \end{align*}
\begin{align*}
\beta_{\tilde{\lambda}_\chi}^{(0)} &=+12\tilde{\lambda }_{\chi}^2+24 \tilde{\lambda }_{\chi } \Big(\lambda_{\chi }-g_2^2\Big)-12 g_1^2 \Big(\tilde{\lambda}_{\chi }+g_2^2\Big)+3g_2^4+\kappa _1^2-\frac{1}{2}\kappa _2^2,\\[20pt]
\beta_{\tilde{\lambda}_\chi}^{(1)} &=-16\tilde{\lambda }_{\chi }^3+\frac{1}{3} \Big[g_2^4 \Big(86 \tilde{\lambda
   }_{\chi }+30 \kappa _1+72 \lambda _{\chi }\Big)-6
   g_2^2 \Big[\kappa _1^2-48 \tilde{\lambda }_{\chi }
   \Big(\tilde{\lambda }_{\chi }+2 \lambda _{\chi
   }\Big)\Big]\\
   &\quad +g_1^2 \Big(-6 g_2^2 \Big(-84\tilde{\lambda }_{\chi }+5 \kappa _2+48 \lambda_{\chi }\Big)-3 \Big(48 \tilde{\lambda }_{\chi }
   \Big(\tilde{\lambda }_{\chi }-2 \lambda _{\chi}\Big)+\kappa _2^2\Big)+598 g_2^4\Big)\\
   &\quad+g_1^4
   \Big(331 \tilde{\lambda }_{\chi }+778
   g_2^2\Big)+6 \Big(-4 \kappa _1^2
   \Big(\tilde{\lambda }_{\chi }+\lambda _{\chi }+2
   \lambda _{\chi \xi }\Big)+\kappa _1 \Big(\kappa_3^2-20 \lambda _{\chi \xi } \tilde{\lambda }_{\chi}\Big)\\
   &\quad-2 \tilde{\lambda }_{\chi } \Big(4
   \Big(28 \lambda _{\chi } \tilde{\lambda }_{\chi
   }+28 \lambda _{\chi}^2\Big)-\kappa _3^2+15 \lambda _{\chi \xi }^2+5
   \lambda _{\phi \chi }^2\Big)+\kappa _2^2 \Big[3
   \tilde{\lambda }_{\chi }+2 \Big(\lambda _{\chi}+\lambda _{\phi \chi }\Big)\Big]\\
   &\quad-5 \kappa_1^3+\kappa _2 \kappa _3^2\Big)+3 \kappa _2^2
   \Big(3 y_b^2+3y_t^2+y_{\tau}^2\Big)-245 g_2^6\Big],
\end{align*}
\begin{align*}
\beta_{\kappa_1}^{(0)} &=+10\kappa_1^2-24 g_2^2 \kappa _1+6g_2^4+2 \kappa _1 \Big(4 \tilde{\lambda }_{\chi }-3 g_1^2+8 \lambda _{\xi }+2 \lambda _{\chi }+8\lambda _{\chi \xi }\Big)-2 \kappa _3^2,\\[20pt]
\beta_{\kappa_1}^{(1)} &=-22 \kappa _1^3+2 \kappa _1^2 \Big(-60
   \tilde{\lambda }_{\chi }+5 g_1^2+80 g_2^2-152
   \lambda _{\xi }-46 \lambda _{\chi }-66 \lambda
   _{\chi \xi }\Big) \\
   \nonumber
   &\quad +2 \kappa _3^2 \Big(4 \tilde{\lambda }_{\chi }+6y_b^2+6y_t^2+g_1^2+2 \kappa _2+8
   \lambda _{\xi }+2 \lambda _{\chi }+4 \lambda _{\chi
   \xi }+8 \lambda _{\phi \xi }+4 \lambda _{\phi \chi
   }+2 y_{\tau }^2\Big)\\
   &\quad +\frac{1}{6} \kappa _1 \Big[12
   \Big(-8 \tilde{\lambda }_{\chi } \Big(g_2^2+5
   \lambda _{\chi } +12 \lambda _{\chi \xi }\Big)-12
   \tilde{\lambda }_{\chi }^2+8 \lambda _{\chi \xi }
   \Big(13 g_2^2-36 \lambda _{\xi }-12 \lambda _{\chi}\Big)\\
   &\quad -4 \Big(4 \lambda _{\xi } \Big(g_2^2+13
   \lambda _{\xi }\Big)+g_2^2 \lambda _{\chi }+4
   \lambda _{\chi }^2\Big)-57 \lambda _{\chi \xi
   }^2\Big)+96 g_1^2 \Big(4 \tilde{\lambda }_{\chi
   }+2 \lambda _{\chi }+\lambda _{\chi \xi
   }\Big)\\
   &\quad+307 g_1^4+48 g_2^2 g_1^2+136 g_2^4+12
   \kappa _2^2+84 \kappa _3^2 -12 \Big(16 \lambda_{\phi \xi } \lambda _{\phi \chi }+4 \lambda _{\phi\xi }^2+\lambda _{\phi \chi}^2\Big)\Big]\\
   &\quad+\frac{2}{3} g_2^4 \Big(60
   \tilde{\lambda }_{\chi }-45 g_1^2-245 g_2^2+120
   \lambda _{\xi }+30 \lambda _{\chi }+12 \lambda
   _{\chi \xi }\Big),\\
\end{align*}
\begin{align*}
\beta_{\kappa_2}^{(0)} &=+\frac{3}{2} g_1^2 \Big(8 g_2^2-5 \kappa_2\Big)-\frac{33}{2} g_2^2 \kappa _2+4 \kappa_3^2\\
&\quad+2 \kappa _2 \Big(-4 \tilde{\lambda }_{\chi}+2\lambda _{\chi }+2\lambda _{\phi }+4 \lambda_{\phi \chi }+3 y_b^2+3 y_t^2+y_{\tau}^2\Big),\\[20pt]
\beta_{\kappa_2}^{(1)} &=+\frac{1}{2}\kappa _2^3+6 g_1^2 \kappa_2^2+\frac{1}{3} g_1^2 g_2^2 \Big(120 \lambda_{\chi }+120 \lambda _{\phi }-240 \tilde{\lambda }_{\chi }+24\lambda _{\phi \chi}+108 y_b^2+252 y_t^2\\
&\quad+132 y_{\tau }^2-643
   g_1^2-463 g_2^2\Big)+\kappa _3^2 \Big( 16
   \tilde{\lambda }_{\chi }-12 y_b^2+8 \kappa _1-8
   \lambda _{\chi }-16 \lambda _{\chi \xi } -24 \lambda_{\phi }\\
   &\quad -16 \lambda_{\phi \xi }-16 \lambda _{\phi
   \chi }-12 y_t^2-4y_{\tau }^2+g_1^2+47
   g_2^2\Big)+\frac{1}{48} \kappa _2 \Big[48 \Big(2 \lambda _{\phi
   \chi } \Big(16 \tilde{\lambda }_{\chi }-48 \lambda _{\chi }\\
   &\quad-12
   y_b^2-12 y_t^2-4y_{\tau
   }^2+5 g_1^2+23 g_2^2\Big) +8 \Big(4 g_1^2
   \Big(\lambda _{\chi }-2 \tilde{\lambda }_{\chi
   }\Big)+5 g_2^2 \Big(\lambda _{\chi }-2
   \tilde{\lambda }_{\chi }\Big)\\
   &\quad +9 \tilde{\lambda}_{\chi }^2+6 \lambda _{\chi } \tilde{\lambda
   }_{\chi }-4 \lambda _{\chi }^2\Big) +8 \lambda
   _{\phi } \Big(-3 y_b^2-3y_t^2+g_1^2-10
   \lambda _{\phi \chi }-y_{\tau }^2\Big)-28 \lambda
   _{\phi }^2\\
   &\quad -29 \lambda _{\phi \chi }^2\Big)+10
   g_1^2 \Big(10 y_b^2+489 g_2^2+34 y_t^2+30 y_{\tau
   }^2\Big) +180 g_2^2 \Big(3y_b^2+3y_t^2+y_{\tau }^2\Big)\\
   &\quad+24
   \Big(80 g_3^2 \Big(y_b^2+y_t^2\Big) -27
   \Big(y_b^2-y_t^2\Big)^2-9 y_{\tau
   }^4\Big)+96 \Big[-2 \kappa_1 \Big(\lambda _{\chi \xi }+4 \lambda _{\phi \xi}\Big)\\
   &\quad+3 \kappa _1^2-3 \Big(\lambda _{\chi \xi}^2+8 \lambda _{\chi \xi } \lambda _{\phi \xi}       
    +\lambda _{\phi \xi }^2\Big)\Big]+3121 g_1^4-911 g_2^4+192 \kappa_3^2\Big],\\[25pt]
\beta_{\kappa_3}^{(0)} &=-\frac{1}{2} \kappa _3 \Big(9 g_1^2+33g_2^2\Big)+ 2\kappa_3\Big( 3 y_b^2-\kappa_1+\kappa _2+2 \lambda _{\chi \xi }+2\lambda_{\phi }+4 \lambda _{\phi \xi }+2\lambda _{\phi \chi}+3 y_t^2+y_{\tau }^2\Big),\\
\end{align*}
\begin{align*}
\beta_{\kappa_3}^{(1)} &=+3 \kappa
   _3^3+\frac{1}{48} \kappa _3 \Big[1797 g_1^4-911 g_2^4-24 \Big(\kappa _2
   \Big(-16 \tilde{\lambda }_{\chi }+4 \Big(3y_b^2+3y_t^2+y_{\tau }^2\Big)-17g_1^2-47 g_2^2\\
   &\quad+8 \lambda _{\chi }+16 \lambda _{\chi\xi }+24 \lambda _{\phi }+16 \lambda _{\phi \xi}+16 \lambda _{\phi \chi }\Big)+2 \Big(-12\tilde{\lambda }_{\chi }^2+8 \tilde{\lambda }_{\chi} \Big(-\lambda _{\chi }+2 \lambda _{\chi \xi }+2\lambda _{\phi \chi }\Big)\\
   &\quad+\lambda _{\phi \xi }\Big(8 \Big(3 y_b^2+3y_t^2+y_{\tau}^2\Big)-2 g_1^2-46 g_2^2+160 \lambda _{\xi }+56
   \lambda _{\chi \xi }+24 \lambda _{\phi \chi}\Big)+4 \lambda _{\phi } \Big(6y_b^2+6y_t^2\\
   &\quad+g_1^2+20 \lambda _{\phi
   \xi }+10 \lambda _{\phi \chi }+2 y_{\tau
   }^2\Big)+12 y_b^2 \lambda _{\phi \chi }-4 g_1^2
   \lambda _{\chi \xi }-40 g_2^2 \lambda _{\chi \xi
   }-17 g_1^2 \lambda _{\phi \chi }-23 g_2^2 \lambda
   _{\phi \chi }\\
   &\quad+80 \lambda _{\xi } \lambda _{\chi \xi
   }+32 \lambda _{\chi } \lambda _{\chi \xi }-8
   \lambda _{\chi }^2-\lambda _{\chi \xi }^2+28
   \lambda _{\phi }^2+34 \lambda _{\phi \xi }^2+32
   \lambda _{\chi } \lambda _{\phi \chi }+40 \lambda
   _{\chi \xi } \lambda _{\phi \chi }+10 \lambda
   _{\phi \chi }^2\\
   &\quad+12 \lambda _{\phi \chi } y_t^2+4
   \lambda _{\phi \chi } y_{\tau }^2\Big)+4 \kappa
   _1 \Big(-4 \tilde{\lambda }_{\chi }+g_1^2+10
   g_2^2-2 \kappa _2+2 \lambda _{\chi }-7 \lambda
   _{\chi \xi }-4 \lambda _{\phi \xi }\Big)-28
   \kappa _1^2\\
   &\quad-2 \kappa _2^2\Big)+10 g_1^2 \Big(10
   y_b^2+45 g_2^2+34 y_t^2+30 y_{\tau }^2\Big)+180
   g_2^2 \Big(3 y_b^2+3y_t^2+y_{\tau
   }^2\Big)\\
   &\quad+24 \Big(80 g_3^2
   \Big(y_b^2+y_t^2\Big)-3
   \Big(y_b^2-y_t^2\Big)^2-y_{\tau
   }^4\Big)+3840 \lambda _{\xi }^2\Big],
\end{align*}
\begin{align*}
\beta_{\lambda_{\phi\xi}}^{(0)} &=+8\lambda _{\phi \xi }^2-\frac{33}{2} g_2^2 \lambda _{\phi \xi }+\frac{1}{2} \lambda _{\phi \xi } \Big(12y_b^2+80 \lambda _{\xi }+24 \lambda _{\phi }+12 y_t^2+4y_{\tau }^2-3g_1^2\Big)\\
&\quad +3g_2^4+2 \lambda _{\phi \chi } \Big(\kappa _1+3\lambda _{\chi \xi }\Big)+4 \kappa _3^2,\\[15pt]
\beta_{\lambda_{\phi\xi}}^{(1)} &=-46
   \lambda _{\phi \xi }^3+\frac{1609}{48} g_2^4
   \lambda _{\phi \xi }+640 g_2^2 \lambda _{\xi
   } \lambda _{\phi \xi }+100 g_2^4 \lambda _{\xi }+72
   g_2^2 \lambda _{\phi } \lambda _{\phi \xi }+24
   g_1^2 \lambda _{\phi } \lambda _{\phi \xi }+30
   g_2^4 \lambda _{\phi }\\
   &\quad+22 g_2^2 \lambda _{\phi \xi
   }^2+\frac{15}{8} g_1^2 g_2^2 \lambda _{\phi \xi }+2
   g_1^2 \lambda _{\phi \xi }^2+40 g_2^4 \lambda
   _{\phi \chi }+\frac{45}{4} g_2^2 \lambda _{\phi \xi
   } y_t^2+\frac{85}{12} g_1^2 \lambda _{\phi \xi }
   y_t^2+\frac{15}{4} g_2^2 \lambda _{\phi \xi }
   y_{\tau }^2\\
   &\quad+\frac{25}{4} g_1^2 \lambda _{\phi \xi }
   y_{\tau }^2+40 g_3^2 \lambda _{\phi \xi }
   y_t^2-8 \kappa _1^2 (\lambda _{\phi \xi }+2
   \lambda _{\phi \chi })-\frac{3}{2} \kappa
   _2^2 \lambda _{\phi \xi }-480 \lambda _{\xi }
   \lambda _{\phi \xi }^2-800 \lambda _{\xi }^2
   \lambda _{\phi \xi }\\
   &\quad-144 \lambda _{\phi } \lambda
   _{\phi \xi }^2-60 \lambda _{\phi }^2 \lambda _{\phi
   \xi }-12 \lambda _{\chi \xi }^2 (\lambda
   _{\phi \xi }+2 \lambda _{\phi \chi })-3
   \lambda _{\phi \xi } \lambda _{\phi \chi }^2-72 \lambda _{\phi } \lambda
   _{\phi \xi } y_t^2-24 \lambda _{\phi \xi }^2
   y_t^2\\
   &\quad-\frac{27}{2} \lambda _{\phi \xi } y_t^4-24
   \lambda _{\phi } \lambda _{\phi \xi } y_{\tau }^2-8
   \lambda _{\phi \xi }^2 y_{\tau }^2-\frac{9}{2}
   \lambda _{\phi \xi } y_{\tau }^4+\kappa _3^2 \Big(-4 \Big(3
   y_b^2+3y_t^2+y_{\tau }^2\Big)+17
   g_1^2\\
   &\quad+23 g_2^2-4 \kappa _2-80 \lambda _{\xi }-28
   \lambda _{\chi \xi }-40 \lambda _{\phi }-34 \lambda
   _{\phi \xi }-12 \lambda _{\phi \chi
   }\Big)+\frac{1}{48} \Big[689 g_1^4 \lambda
   _{\phi \xi }\\
   &\quad-4 g_2^4 \Big(12 \Big(3
   y_b^2+3y_t^2+y_{\tau }^2\Big)+45
   g_1^2-217 g_2^2\Big)\Big]+\frac{45}{4} g_2^2
   y_b^2 \lambda _{\phi \xi }+\frac{25}{12} g_1^2
   y_b^2 \lambda _{\phi \xi }-\frac{27}{2}
   y_b^4 \lambda _{\phi \xi }\\
  &\quad+40 g_3^2 y_b^2
   \lambda _{\phi \xi }-21 y_b^2 \lambda _{\phi \xi }
   y_t^2-72 y_b^2 \lambda _{\phi } \lambda _{\phi \xi
   }-24 y_b^2 \lambda _{\phi \xi }^2+3 \lambda _{\chi \xi }
   \Big(4 \lambda _{\phi \chi } \Big(4 g_1^2+8
   g_2^2-4 \lambda _{\phi \xi }\\
   &\quad-\lambda _{\phi \chi}\Big)+5 \Big(g_1^4+2 g_2^4\Big)-2 \kappa
   _2^2\Big)+\kappa _1 \Big[4 \lambda _{\phi \chi }
   \Big(4 g_1^2+8 g_2^2-4 \lambda _{\phi \xi
   }-\lambda _{\phi \chi }\Big)+5 \Big(g_1^4+2
   g_2^4\Big)\\
   &\quad-2 \kappa _2^2+4 \kappa _3^2-8 \lambda
   _{\chi \xi } \Big(\lambda _{\phi \xi }+2 \lambda
   _{\phi \chi }\Big)\Big],\\[15pt]
\beta_{\lambda_{\phi\chi}}^{(0)} &=+4 \lambda _{\phi \chi }^2+2 \lambda _{\phi \chi } \Big(4 \tilde{\lambda
   }_{\chi }+3 y_b^2+8 \lambda _{\chi }+6 \lambda
   _{\phi }+3 y_t^2+y_{\tau
   }^2\Big)+4 \lambda _{\phi \xi } \Big(\kappa _1+3
   \lambda _{\chi \xi }\Big)\\
   &\quad +2\kappa _2^2+4 \kappa
   _3^2-\frac{15}{2} g_1^2 \lambda _{\phi \chi
   }-\frac{33}{2} g_2^2 \lambda _{\phi \chi }+3
   g_1^4+6 g_2^4,
 \end{align*}
\begin{align*}
\beta_{\lambda_{\phi\chi}}^{(1)} &=-13
   \lambda _{\phi \chi }^3-\frac{1061 }{12}g_1^6+\frac{217 }{6}g_2^6+g_1^4\Big(-\frac{165 }{4}g_2^2+5
   y_b^2-19 y_t^2-25 y_{\tau }^2+30 \lambda _{\phi
   }+40 \lambda_{\chi }\\
   &\quad+20 \tilde{\lambda }_{\chi }+\frac{6121 }{48}\lambda _{\phi \chi }\Big)
   +g_2^4 \Big(-6 y_b^2-6 y_t^2-2 y_{\tau
   }^2+10 \kappa _1+60 \lambda _{\phi }+80 \lambda
   _{\phi \xi }+\frac{3529}{48} \lambda _{\phi \chi
   }\\
   &\quad+80 \lambda _{\chi }+30 \lambda _{\chi \xi
   }+40 \tilde{\lambda }_{\chi }\Big)
   +\frac{1}{24} g_1^2\Big[-900 g_2^4+\Big(96 \kappa
   _2+525 \lambda _{\phi \chi }\Big) g_2^2+60 \kappa
   _2^2+24 \kappa _3^2\\
   &\quad+2 \lambda _{\phi \chi }
   \Big(25 y_b^2+85 y_t^2+75 y_{\tau }^2+288 \lambda
   _{\phi }+60 \lambda _{\phi \chi }+1536
   \lambda _{\chi }+768\tilde{\lambda }_{\chi
   }\Big)\Big] +\frac{1}{4}
   g_2^2 \Big[46 \kappa _2^2\\
   &\quad+92 \kappa _3^2+256
   \lambda _{\phi \xi } \Big(\kappa _1+3 \lambda
   _{\chi \xi }\Big)+\lambda _{\phi \chi } \Big(45
   y_b^2+45 y_t^2+15 y_{\tau }^2+288 \lambda _{\phi
   }+44 \lambda _{\phi \chi }\\
    \end{align*}
\begin{align*}
   &\quad+1024 \lambda
   _{\chi }+512\tilde{\lambda }_{\chi
   }\Big)\Big]-6 y_b^2 \kappa _2^2-6 y_t^2
   \kappa _2^2-2 y_{\tau }^2 \kappa _2^2-12 y_b^2
   \kappa _3^2-12 y_t^2 \kappa _3^2-4 y_{\tau }^2
   \kappa _3^2\\
   \nonumber
   &\quad-8 \kappa _2 \kappa _3^2-16 \kappa _1
   \lambda _{\phi \xi }^2-12 y_b^2 \lambda _{\phi \chi
   }^2-12 y_t^2 \lambda _{\phi \chi }^2-4 y_{\tau }^2
   \lambda _{\phi \chi }^2-72 \lambda _{\phi } \lambda
   _{\phi \chi }^2-80 \lambda _{\phi \chi } \lambda
   _{\chi }^2\\
   \nonumber
   &\quad-48 \lambda _{\phi \xi } \lambda _{\chi
   \xi }^2-6 \lambda _{\phi \chi } \lambda _{\chi \xi
   }^2-120 \lambda _{\phi \chi } \tilde{\lambda
   }_{\chi }^2-20 \kappa _2^2 \lambda _{\phi }-40
   \kappa _3^2 \lambda _{\phi }-32 \kappa _1^2 \lambda
   _{\phi \xi }-24 \kappa _3^2 \lambda _{\phi \xi
   }\\
   &\quad-\frac{27}{2} y_b^4 \lambda _{\phi \chi
   }-\frac{27}{2} y_t^4 \lambda _{\phi \chi
   }-\frac{9}{2} y_{\tau }^4 \lambda _{\phi \chi }+40
   g_3^2 y_b^2 \lambda _{\phi \chi }+40 g_3^2 y_t^2
   \lambda _{\phi \chi }-21 y_b^2 y_t^2 \lambda _{\phi
   \chi }-4 \kappa _1^2 \lambda _{\phi \chi
   }\\
   &\quad-\frac{29}{2} \kappa _2^2 \lambda _{\phi \chi }-20
   \kappa _3^2 \lambda _{\phi \chi }-60 \lambda _{\phi
   }^2 \lambda _{\phi \chi }-6 \lambda _{\phi \xi }^2
   \lambda _{\phi \chi }-96 \lambda
   _{\phi \chi }^2 \lambda _{\chi }-72 y_b^2 \lambda _{\phi }
   \lambda _{\phi \chi }-72 y_t^2 \lambda _{\phi }
   \lambda _{\phi \chi }\\
   &\quad-24 y_{\tau }^2 \lambda _{\phi
   } \lambda _{\phi \chi }-16 \kappa _1 \lambda _{\phi
   \xi } \lambda _{\phi \chi }-24 \kappa _2^2 \lambda
   _{\chi }-32 \kappa _3^2 \lambda _{\chi }-40 \kappa _3^2
   \lambda _{\chi \xi }-48 \lambda _{\phi \xi }^2
   \lambda _{\chi \xi }-4 \kappa _1 \lambda
   _{\phi \chi } \lambda _{\chi \xi }\\
   &\quad-32 \kappa _1 \lambda _{\phi
   \xi } \lambda _{\chi \xi }-48 \lambda
   _{\phi \xi } \lambda _{\phi \chi } \lambda _{\chi
   \xi }+8 \tilde{\lambda
   }_{\chi }\Big[\kappa _2^2-2 \Big(\kappa _3^2+3
   \lambda _{\phi \chi }^2+5 \lambda _{\phi \chi }
   \lambda _{\chi }\Big)\Big] ,\\[20pt]
\beta_{\lambda_{\chi\xi}}^{(0)} &=+8\lambda _{\chi \xi }^2+2\lambda _{\chi \xi } \Big(4\tilde{\lambda }_{\chi }+20 \lambda _{\xi }+8\lambda _{\chi }-3g_1^2\Big)-24 g_2^2 \lambda _{\chi \xi }+6g_2^4+4 \kappa _1 \Big(2 \lambda _{\xi }+\lambda_{\chi }\Big)\\
&\quad+2\kappa _1^2+2\kappa _3^2+4 \lambda_{\phi \xi } \lambda _{\phi \chi },\\[15pt]
\beta_{\lambda_{\chi\xi}}^{(1)} &=-50 \lambda _{\chi \xi }^3-120 \lambda _{\chi \xi } \tilde{\lambda
   }_{\chi }^2+\frac{667}{6} g_1^4 \lambda _{\chi \xi}
  +10 g_1^4 \lambda _{\phi \xi }+20
   g_1^2 g_2^2 \lambda _{\chi \xi
   }+100
   g_2^4 \lambda _{\chi }+20 g_2^4 \lambda
   _{\phi \xi }\\
   &\quad+\frac{2}{3} g_2^4 \Big(-45 g_1^2+79
   g_2^2+360 \lambda _{\xi }\Big)+128
   g_1^2 \lambda _{\chi } \lambda _{\chi \xi }+8
   g_1^2 \lambda _{\chi \xi }^2+8 g_1^2
   \lambda _{\phi \xi } \lambda _{\phi \chi }+10 g_2^4 \lambda _{\phi \chi
   }\\
   &\quad+640 g_2^2 \lambda _{\xi } \lambda _{\chi \xi
   }+\frac{326 }{3}g_2^4
   \lambda _{\chi \xi }+256 g_2^2 \lambda _{\chi } \lambda _{\chi
   \xi }+32 g_2^2 \lambda _{\chi \xi }^2+24
   g_2^2 \lambda _{\phi \xi } \lambda _{\phi \chi
   }-\kappa _2^2 \lambda _{\chi \xi }-26 \kappa _1^3\\
   &\quad-4 \kappa _2^2
   \lambda _{\phi \xi }-480 \lambda
   _{\xi } \lambda _{\chi \xi }^2-800 \lambda _{\xi }^2
   \lambda _{\chi \xi }-192 \lambda _{\chi } \lambda
   _{\chi \xi }^2-80 \lambda _{\chi }^2 \lambda _{\chi
   \xi }-8 \lambda _{\chi \xi
   } \lambda _{\phi \xi }^2-32 \lambda _{\chi \xi }
   \lambda _{\phi \xi } \lambda _{\phi \chi }\\
   &\quad-8 \lambda
   _{\phi \xi } \lambda _{\phi \chi }^2-16 \lambda
   _{\phi \xi }^2 \lambda _{\phi \chi }-2 \lambda
   _{\chi \xi } \lambda _{\phi \chi }^2-24 y_b^2
   \lambda _{\phi \xi } \lambda _{\phi \chi }-24
   y_t^2 \lambda _{\phi \xi } \lambda _{\phi \chi
   }-8 y_\tau^2 \lambda _{\phi \xi } \lambda
   _{\phi \chi }\\
   &\quad+2 \kappa _1^2 \Big(-12 \tilde{\lambda }_{\chi
   }+g_1^2+4 g_2^2-56 \lambda _{\xi }-22
   \lambda _{\chi }-18 \lambda _{\chi \xi }\Big)+8
   \tilde{\lambda }_{\chi } \Big[2 \lambda _{\chi \xi
   } \Big(4 g_1^2+8 g_2^2-5 \lambda _{\chi
   }\\
   &\quad-6 \lambda _{\chi \xi }\Big)+5
   g_2^4\Big]-2 \kappa _3^2 \Big(4
   \tilde{\lambda }_{\chi }+g_1^2+2 \kappa _2+16
   \lambda _{\xi }+6 \lambda _{\chi }+\lambda _{\chi
   \xi }+12 \lambda _{\phi \xi }+8 \lambda _{\phi \chi
   }+2y_\tau^2\\
   &\quad+6y_b^2+6y_t^2\Big)+\kappa _1 \Big[-4 \Big(-12
   g_2^2 \tilde{\lambda }_{\chi }+8
   \tilde{\lambda }_{\chi }^2+8 \lambda _{\xi } \Big(4
   \lambda _{\chi \xi }-7 g_2^2\Big)+\lambda
   _{\chi } \Big(16 \lambda _{\chi \xi }-22
   g_2^2\Big)\\
   &\quad+3 \lambda _{\chi \xi } \Big(4
   g_2^2+\lambda _{\chi \xi }\Big)+32 \lambda
   _{\xi }^2+4 \lambda _{\chi }^2\Big)+20
   g_1^4+4 g_1^2 g_2^2+32 g_1^2
   \lambda _{\chi }+34 g_2^4-\kappa
   _2^2\Big]\,.
\end{align*}

For the sake of completeness, the beta functions of the dimensionful parameters at one-loop are given in Ref.~\cite{Blasi:2017xmc}. 
\section{Supplementary figures}\label{app:supp_figs}
\begin{figure}[!h]
     \centering
             \includegraphics[clip=true,width=0.85\columnwidth]{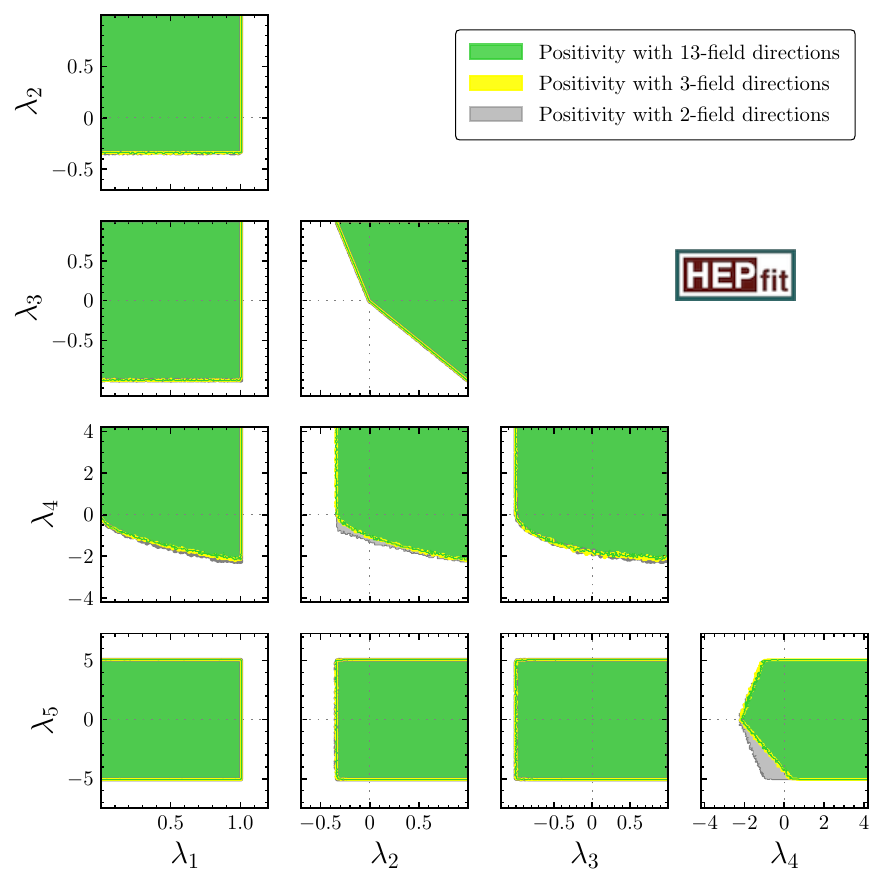}
     \caption{Allowed regions on the quartic coupling planes of the GM model with three different BFB constraints. The colors of the shaded regions  have the same meaning here as in Figure~\ref{fig:1}.}
    \label{fig:14}
\end{figure}
In this appendix, we present supplementary figures of the GM (eGM) model parameter space. 
The comparison of the three different BFB constraints in the quartic coupling planes of the GM model is shown in Figure~\ref{fig:14}. The colors in Figure~\ref{fig:14} have the same meaning as in Figure~\ref{fig:1}, and from this figure, we observe that the regions allowed from the 3-field BFB constraints largely overlap with the allowed regions considering all 13-field BFB constraints~\cite{Hartling:2014zca,Moultaka2020}.

\begin{figure}[!h]
     \centering
             \includegraphics[clip=true,width=0.85\columnwidth]{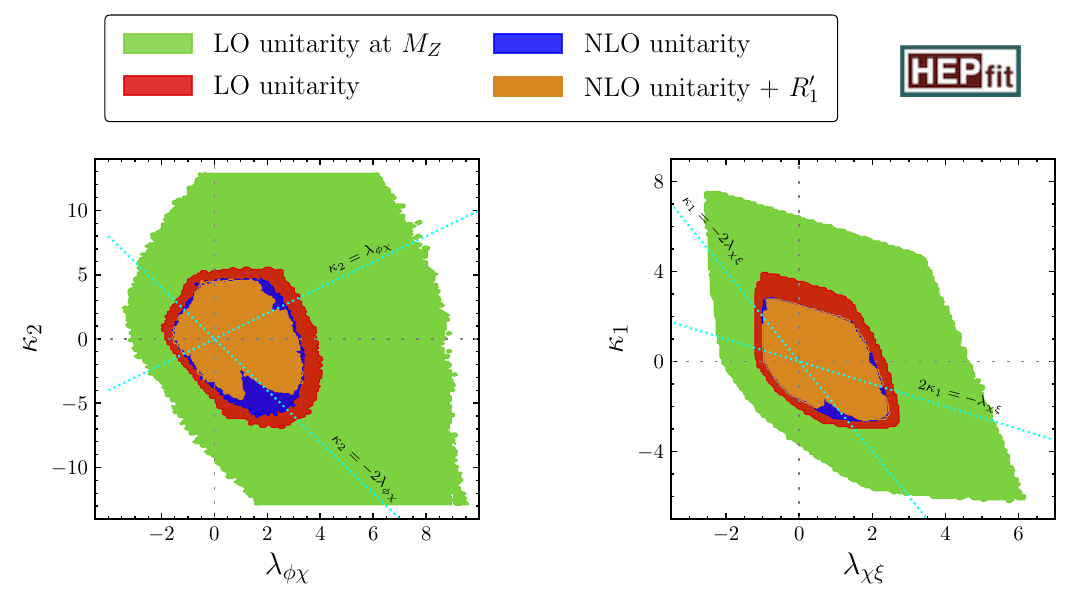}
     \caption{Allowed regions in the $\lambda_i$ vs.~$\kappa_j$ planes. The green, red,  and blue regions have the same meaning as in Figure~\ref{fig:2}. The brown areas show the parameter space without violating the perturbativity conditions ($R_1^\prime<1$) at the one-loop level. The cyan dashed lines represent directions along which the tree-level amplitudes vanish.\\[30pt]}
     \label{fig:11}
\end{figure}

The allowed regions with different  unitarity conditions in the $\lambda_{\phi\chi}$ vs.~$\kappa_2$ ($\lambda_{\chi\xi}$ vs.~$\kappa_1$) plane are shown in Figure~\ref{fig:11}. From Eq.~(\ref{eigenvalue_matrix}), the tree-level amplitudes vanish along the lines, $\kappa_2=\lambda_{\phi\chi},\;\kappa_2=-2\lambda_{\phi\chi}, \; \kappa_1=-2\lambda_{\chi\xi}$, and $\kappa_1=-\lambda_{\chi\xi}/2$, which are denoted by the cyan dotted lines in Figure~\ref{fig:11}. The perturbativity test ($R_1^\prime <1$) fails for coupling values lying in the neighborhood region of these lines. In Figure~\ref{fig:11}, the effects of such cancellations are visible in the NLO unitarity with $R_1^\prime$ contour. The colors in Figure~\ref{fig:11} have the same meaning as in Figure~\ref{fig:2}. 

\begin{figure}[h!]
     \centering
             \includegraphics[clip=true,width=0.85\columnwidth]{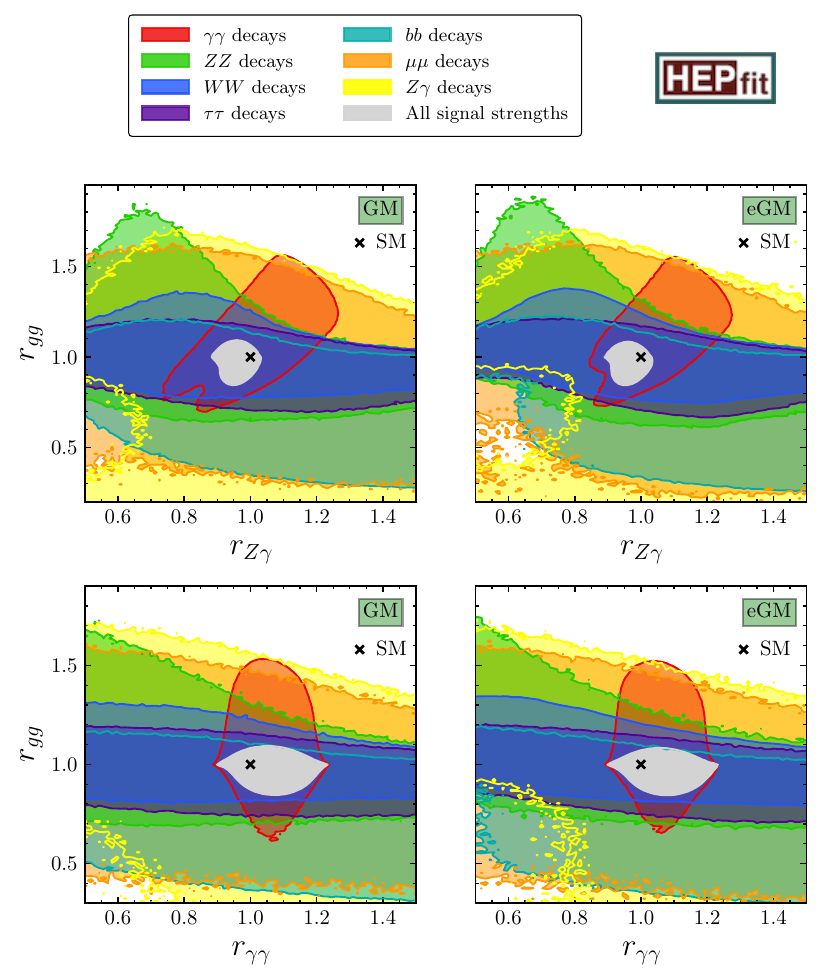}
     \caption{95.4\% probability regions show the impact of the $h$ signal strengths for different final states in the $r_{V\gamma}$ vs.~$r_{gg}$ ($V = \gamma, Z$) planes  for the GM model (left panel) and the eGM model (right panel). The individual fits to full Run 1 and Run 2 data from $h$ decays to $\gamma\gamma,\,ZZ,\,WW,\, \tau\tau\, , \,bb,\,\mu\mu\,,$ and $Z\gamma$ are displayed in red, green, blue, purple, cyan, orange, and yellow, respectively. The results from the combined fits are shown in the grey color shaded regions, where the cross indicates the expected value for the SM Higgs boson.\\[30pt]}
     \label{fig:15}
\end{figure}

In the top left (right) panel of Figure~\ref{fig:15}, we show the allowed ranges in $r_{Z\gamma}$ vs.~$r_{gg}$ plane for all different final states for the GM (eGM) model. The colors in Figure~\ref{fig:15} have the same meaning as in Figure~\ref{fig:5}. From Figure~\ref{fig:15}, we see that the most significant constraints in the $r_{Z\gamma}$ vs.~$r_{gg}$ plane come from  the $WW$, $bb$, and $\gamma\gamma$ decay channels. The SM predictions of $r_{Z\gamma}^{\text{SM}}=1$ and $r_{gg}^{\text{SM}}=1$ lie within the allowed regions (grey shaded area) of all signal strengths.  The maximal deviation of $r_{Z\gamma}$ ($r_{gg}$) from its SM value is roughly 10\% (15\%) in both models.
In the bottom left (right) panel of Figure~\ref{fig:15}, we show the allowed ranges in $r_{\gamma\gamma}$ vs.~$r_{gg}$ plane, for all different final states for the GM (eGM) model. The maximal deviation of $r_{\gamma\gamma}$ from its SM value ($r_{\gamma\gamma}^{\text{SM}}=1$) is roughly 25\% in both models.

\begin{figure}[h!]
     \centering
             \includegraphics[clip=true,width=0.85\columnwidth]{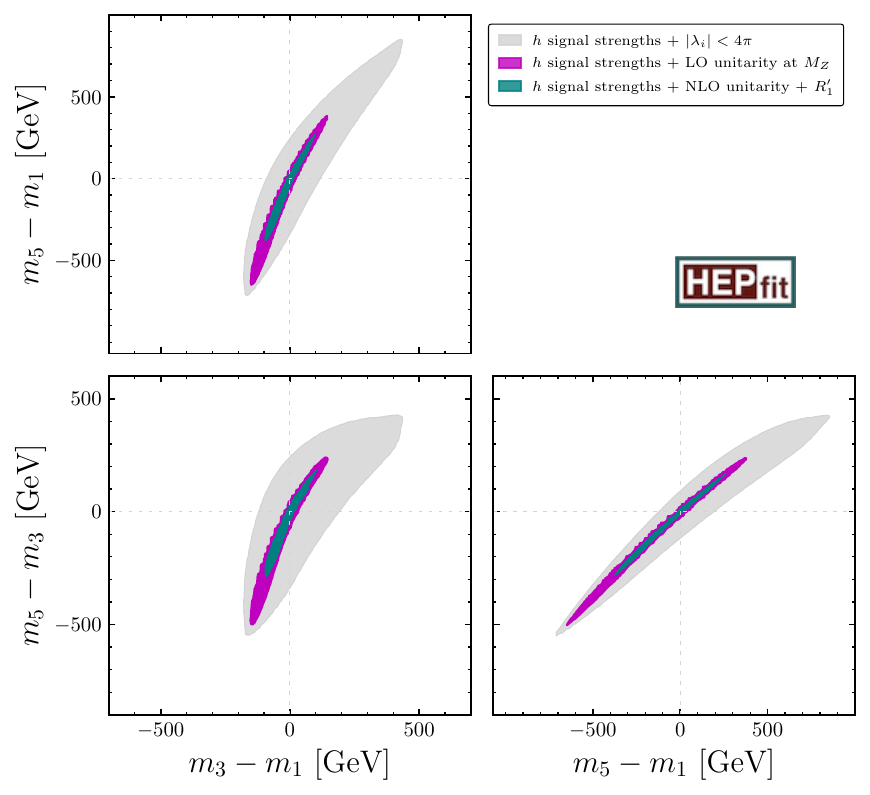}
     \caption{Allowed mass differences in the GM model constrained by the latest $h$ signal strength
data with various perturbativity and unitarity conditions. The light grey regions are allowed from the $h$ signal strengths data without violating the naive perturbativity limit on all the quartic couplings, $|\lambda_i| < 4\pi$. The magenta-shaded region represents the parameter space allowed by the combined constraints of $h$ signal strength data and tree-level (LO) unitarity. The teal-shaded regions denote the parameter space that remains viable when NLO unitarity conditions with $R_1^\prime$ are imposed in addition to the $h$ signal strength data. All shaded regions are shown at a $95.4\%$ probability.\\[110pt]}
     \label{fig:16}
\end{figure}

 The effects of various perturbativity and unitarity constraints on the fit to $h$ signal strength data are shown in Figure~\ref{fig:16}, where the allowed regions are obtained using only the mass-squared priors.
 
\begin{figure}[h!]
     \centering
             \includegraphics[clip=true,width=0.8\columnwidth]{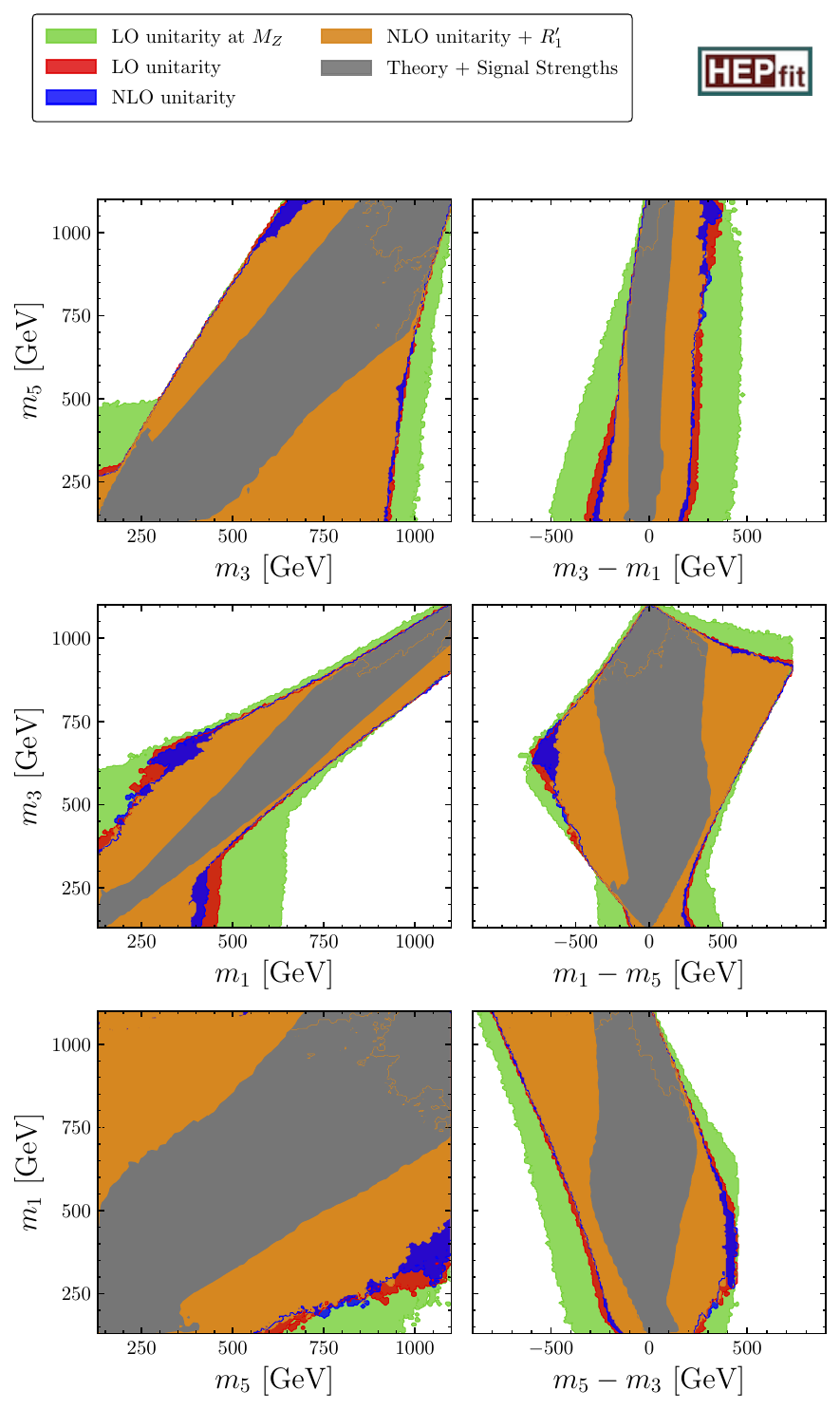}
     \caption{Allowed regions in the $m_i$ vs.~$m_j$ ($m_j-m_k$) planes in the GM model. The color shaded regions have the same meaning as in Figure~\ref{fig:6}. }
     \label{fig:12}
\end{figure}

\begin{figure}[h!]
     \centering
             \includegraphics[clip=true,width=0.8\columnwidth]{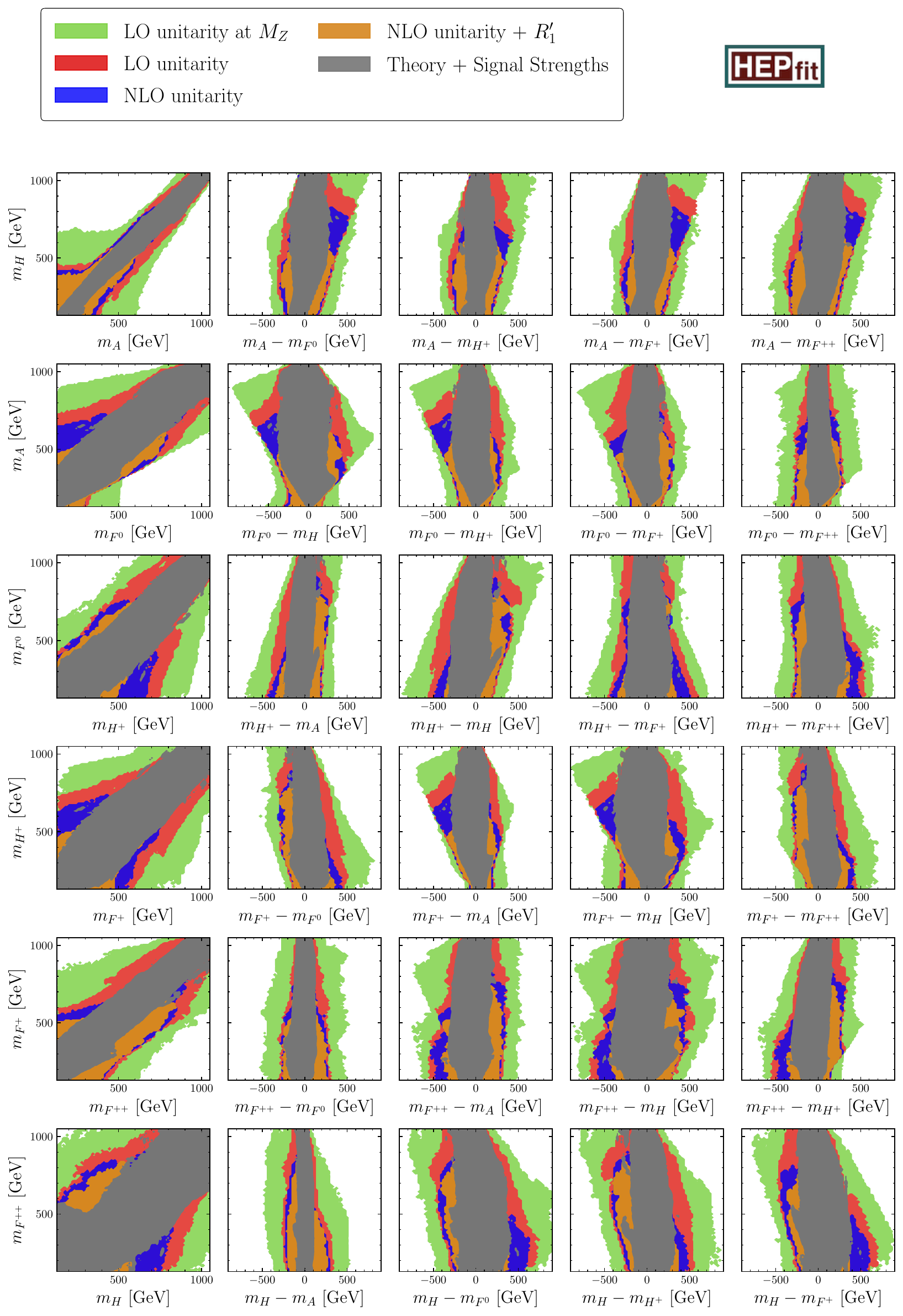}
     \caption{Allowed regions in the $m_i$ vs.~$m_j$ ($m_j-m_k$) planes in the eGM model. The color shaded regions have the same meaning as in Figure~\ref{fig:6}. }
     \label{fig:13}
\end{figure}

The masses of the heavy Higgs bosons and their mass differences are shown in Figures~\ref{fig:12} and \ref{fig:13}, for the GM and the eGM models, respectively. The colors in Figures~\ref{fig:12} and \ref{fig:13} have the same meaning as in Figure~\ref{fig:6}. 
\vspace*{1cm}

\pagebreak
\bibliography{NLOuni_v2}
\end{document}